\documentclass[twocolumn,eqsecnum,floatfix,aps,prc]{revtex4}
\usepackage{amsmath}  
\usepackage{epsfig} 
\usepackage{color} 
\usepackage{txfonts} 
 
\def\muvec{\mbox{\boldmath $\mu$}}

\begin{document}

\parindent=12pt
\title{Comprehensive study of muon-catalyzed nuclear reaction 
processes in the dt$\muvec$ molecule}

\author{M.\ Kamimura}
\email{mkamimura@a.riken.jp}
\affiliation{Meson Science Laboratory, RIKEN Nishina Center, RIKEN, Wako 351-0198, Japan}

\author{Y. Kino}
\email{yasushi.kino.e5@tohoku.ac.jp}
\affiliation{Department of Chemistry, Tohoku University, 
Sendai, 980-8578, Japan}

\author{T. Yamashita}
\email{tyamashita@tohoku.ac.jp}
\affiliation{Institute for Excellence in Higher Education
Division of Developmental Research in Education Programs,\\
Tohoku University, Sendai, 980-8576, Japan \\ 
\mbox{and Department of Chemistry,~Tohoku University,~Sendai, 980-8578, Japan}}

\date{\today}

\begin{abstract}
Muon catalyzed fusion ($\mu$CF) has recently regained 
considerable research interest owing to several 
new developments and applications.
In this regard, 
we have performed a comprehensive study of the most important fusion reaction, 
namely $(dt\mu)_{J=v=0} \to \alpha + n + \mu + 17.6\,{\rm MeV}\:$ or $ \;
(\alpha \mu)_{nl} + n + 17.6\,{\rm MeV}$.
For the first time, the coupled-channels
Schr\"{o}dinger equation for the reaction is solved, 
satisfying the boundary condition for the muonic 
molecule $(dt\mu)_{J=v=0}$
as the initial state and the outgoing wave in the $\alpha n \mu$ channel.
We employ the $dt\mu$- and $\alpha n \mu$-channel coupled three-body model.
All the nuclear interactions, the $d$-$t$ and $\alpha$-$n$ potentials, and 
the $d t$-$\alpha n$ channel-coupling nonlocal tensor potential 
are chosen to reproduce the observed 
low-energy ($1-300$ keV) astrophysical $S$-factor of 
the reaction $d+t\to \alpha+n + 17.6 \,{\rm MeV}$,
as well as the total cross section of the $\alpha+n$ reaction at 
the corresponding energies.
The resultant $dt\mu$ fusion rate is $1.15 \times 10^{12}\, {\rm s}^{-1}$.
Substituting the obtained total wave function into the \mbox{$T$-matrix} based 
on the Lippmann-Schwinger equation, we have calculated  {\it absolute} values of 
{the fusion rates $\lambda_{\rm f}^{\rm bound}$ and $\lambda_{\rm f}^{\rm cont.}$
}
going to the bound and continuum states 
of the outgoing $\alpha$-$\mu$ pair, 
{ 
respectively.
We then derived the initial $\alpha$-$\mu$ sticking
probability $\omega_S^0=\lambda_{\rm f}^{\rm bound}/
(\lambda_{\rm f}^{\rm bound}+\lambda_{\rm f}^{\rm cont.})=
0.857 \%$,} which is $\sim\!7\%$ smaller than 
the literature values
$(\simeq 0.91-0.93 \%)$ , and {can explain} the  recent 
\mbox{observations (2001)} at high D/T densities.  
{We have much improved the sticking-probability calculation
by employing the $D$-wave $\alpha$-$n$ outgoing channel with the non-local 
tensor-force $dt$-$\alpha n$ coupling
and by deriving $\omega_S^0$
based on the absolute values of 
the $\lambda_{\rm f}^{\rm bound}$ and $\lambda_{\rm f}^{\rm cont.}$.
}
We also calculate the  {\it absolute} values for the momentum and energy
spectra of the muon emitted during the fusion process. The most important 
result is that
the peak energy is 1.1 keV although the mean energy is 9.5 keV 
owing to the long higher-energy tail.
This is an essential result for the ongoing experimental project to 
realize the generation of an ultra-slow negative muon beam by utilizing 
the $\mu$CF
for various applications e.g., a scanning negative muon microscope 
and an injection source \mbox{for the muon collider.}
\end{abstract}


\maketitle

\section{INTRODUCTION}

\begin{figure} [t!]
\begin{center}
\epsfig{file=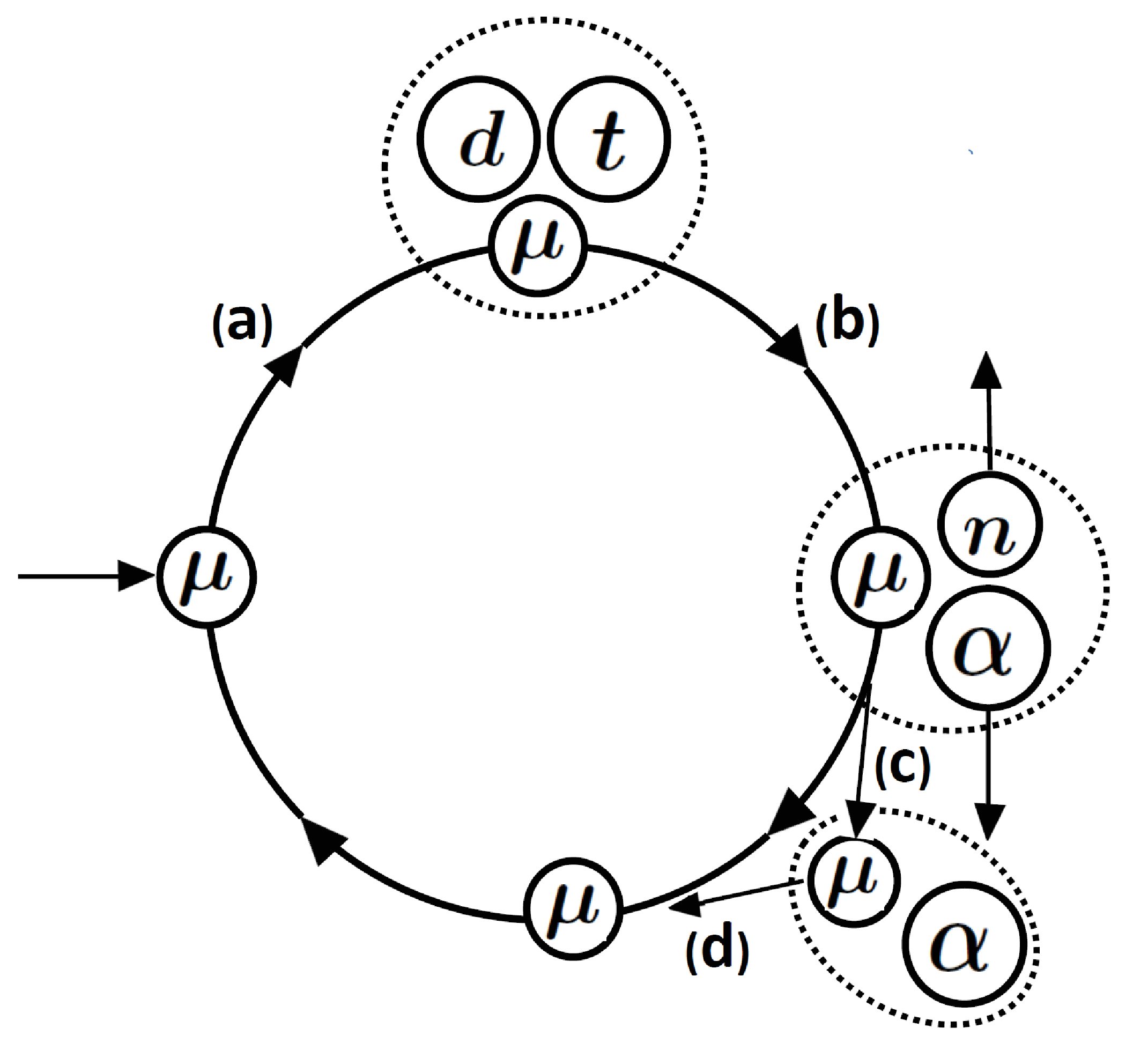,width=4.5cm}
\caption{ 
Schematic diagram of the $\mu$CF cycle by a muon injected into
the D-T mixture.
(a) Formation of $dt\mu$ molecule, (b) fusion reaction, 
(c) $\alpha$-$\mu$ initial sticking, and (d) muon reactivation.
} 
\label{fig:cycle}
\end{center}	
\end{figure}

In the mixture of deuterium (D) and tritium (T), 
an injected negatively charged muon ($\mu$) forms 
a muonic molecule with a deuteron ($d$) and a triton ($t$),
namely, $dt\mu$.
Since the mass of a muon is 207 times heavier than that of an electron, 
the nuclear wave funcitons of $d$ and $t$
overlap inside the $dt\mu$ molecule, which instantly 
results in an intramolecular nuclear fusion reaction
$d + t \to $ \mbox{$\alpha + n +  17.6 \,{\rm MeV}$}.
After this reaction, 
the muon \mbox{becomes} free and can \mbox{facilitate} 
another \mbox{fusion} reaction (Fig.~\ref{fig:cycle}). 
This cyclic reaction is called muon \mbox{catalyzed fusion ($\mu$CF).}
Among various isotopic species of muonic molecules ($pp\mu$, $pd\mu$, 
$dd\mu$, $dt\mu$ and $tt\mu$),
the $dt\mu$ has attracted particular attention in $\mu$CF with the
\mbox{expectation} that it may be exploited as a future energy source.

The $\mu$CF has been dedicatedly studied since 
1947~\cite{Frank1947,Sakharov1948}, and is reviewed 
in Refs.~\cite{Breunlich89,Ponomarev90,Froelich92,Nagamine98}.
Efficiency of the $\mu$CF has been discussed in the literature as following:
As seen in Fig.~\ref{fig:cycle}, 
the muon emitted after the \mbox{$d$-$t$} fusion sticks 
to the $\alpha$ particle
with a probability, $\omega_S^0$,  
and is lost from the cycle due to spending its lifetime 
($\tau_\mu=2.2 \times 10^{-6}$s) as a coupled entity, although the muon is 
reactivated (stripped) with a probability $R$ during
the collision of an $(\alpha \mu)^+$ ion with the D-T mixture.
The net loss-probability $\omega_S^{\rm eff}=\omega_S^0 (1-R)$
is called \mbox{{\it effective}} sticking probability, whereas $\omega_S^0$
is referred to as {\it initial} sticking probability.
The number of fusion events, $Y_\mathrm{f}$, catalyzed by one muon 
is essentially represented as~\cite{Ponomarev90}
\begin{equation}
Y_\mathrm{f} \simeq \left( \omega_S^{\rm eff} 
+ \lambda_0/\lambda_c\phi \right)^{-1}
\label{eq:cycle-rate}
\end{equation}
where $\lambda_0=1/\tau_\mu=0.455\times10^6$ s$^{-1}$,
$\lambda_c$ is a cycle rate, and $\phi$ is a target density relative 
to the liquid hydrogen ($4.25\times10^{22}$ atoms cm$^{-3}$).
A typical parameter set of $\omega_S^{\rm eff}\sim0.5\%$~\cite{Ishida2001} 
and $\lambda_c\sim1.1\times10^8$ s$^{-1}$~\cite{Kawamura2003} at $\phi=1.25$
results in $Y_\mathrm{f}\sim120$, which produces $\sim \! 2.1$ GeV  
per a muon. 
A literature reported $Y_\mathrm{f}\sim150$~\cite{Jones1986nature} which 
was the highest value known so far, and results in $\sim \! 2.6$ GeV,
whereas $\sim \! 5$ GeV of energy is required to generate 
a muon in accelerator; the efficiency of $\mu$CF 
for energy production is approximately half that required 
to achieve a scientific break-even.
If $\omega_S^{\rm eff}$ is omitted in Eq.~(\ref{eq:cycle-rate}), 
$Y_\mathrm{f}\sim \! 300$ cycles is obtained. On the other hand, if we omit 
$\lambda_0/\lambda_c\phi$ from Eq.~(\ref{eq:cycle-rate}),
we have $Y_\mathrm{f}\sim \! 200$. 
Therefore, the low \mbox{efficiency} of the $\mu$CF comes from both 
parameters of $\omega_S^{\rm eff}$ and $\lambda_c$;
it is desirable to examine $\omega_S^0$, $R$ and $\lambda_c$ carefully.

Although the fusion yield $Y_\mathrm{f}$ as well as the fundamental 
parameters $\omega_S^0$, $R$ and $\lambda_c$ 
have been often investigated in liquid/solid targets thus far, 
such a cold target $\mu$CF would not be realistic as a practical 
energy source due to
the low thermal efficiency of the Carnot cycle.
The experimental knowledge of $\mu$CF at high-temperature conditions, 
however, is limited.

The $\mu$CF has recently regained considerable interest
owing to several new developments and applications.
They are grouped into the following two types:

\begin{itemize}
\setlength{\parskip}{0.05cm} 
\setlength{\itemsep}{0.0cm}  
\item[I)] 
To realize the production of energy by the $\mu$CF using the high-temperature 
{\it gas} target of the D-T mixture with high thermal efficiency.
\item[II)] To realize  an ultra-slow negative muon 
beam by utilizing the $\mu$CF
for various applications e.g., a scanning negative muon microscope 
and an injection source for the muon collider.
\end{itemize}

\noindent
They are explained as follows:
 
Type I):  
The $\mu$CF kinetics model in high-temperature gas targets  
is re-examined, including the excited (resonant) muonic molecules 
and fusion in-flight 
processes~\cite{Iiyoshi2019,Yamashita2021submitted}.
Recent improvements in the energy resolution of X-ray detectors facilitate
the examination of the dynamics of muon atomic 
processes~\cite{Okada2020,Paul2021,Okumura2021},
and may allow for the detection of 
the resonance states of muonic molecules during the $\mu$CF cycle.
An intense muon beam~\cite{Miyake2014} also creates the upgraded 
conditions required to explore these $\mu$CF fundamental studies.
In parallel to the re-examination of the $\mu$CF kinetics model, 
there is a new proposal to 
strongly reduce the $\alpha$-$\mu$ sticking probability
by boosting the negative muon stripping 
using resonance radio-frequency acceleration of 
$(\alpha \mu)^+$ ions in a spatially located D-T mixture 
gas stream~\cite{Mori2021}. 
In addition to studies on the $\mu$CF as possible energy sources,
the 14.1 MeV neutron has been considered as a source for 
the mitigation of long-lived fission products (LLFPs) with nuclear 
transmutation~\cite{Yamamoto2021}. 
Since the mitigation of LLFPs requires a well-defined condition for a 
neutron beam, $\mu$CF-based monochromatic neutrons would be 
more suitable than those from a nuclear reactor and/or spallation 
neutron sources.

Type II): 
In general, muon beams generated by accelerators have $\sim$ MeV kinetic energies.
At present, the negative muon beam, with a size of a few tens of millimeters, 
has proven to be suitable for non-destructive elemental 
analysis~\cite{Rosen490,DANIEL1984,Kubo2016JPSJ}
in various research fields such as 
archeology, earth-and-planetary science, and industry. 
In contrast to the accelerator-based muon beam, the mean \mbox{kinetic} 
energy of the muon released after 
the $\mu$CF reaction is 
\mbox{$\sim \! 10$ keV} since the $dt\mu$ molecule
nearly takes the $(^5{\rm He} \mu)_{1s}$ configuration
at the instant of the fusion reaction.  
Therefore, 
the $\mu$CF can be utilized as 
a means for beam cooling ~\cite{Nagamine1989,NagamineMCF199091,
Nagamine1996Hyp103,Strasser1993,Strasser1996}.
Recently, the aim has been to produce a negative muon beam 
by reducing the beam size to an order of
10 $\mu$m using a set of beam optics, by utilizing the muons emitted by 
the $\mu$CF.
This beam is called an ultra-slow negative 
muon beam~\cite{Nagamine1989,Nagamine1996Hyp103,Natori2020},
which will facilitate various applications such as a scanning 
negative muon microscope, as well as
an injection source for a muon 
collider~\cite{Nagamine1996Hyp103}.
The scanning negative muon microscope that can utilize characteristic 
muonic X-rays 
will allow for three-dimensional analysis of elements and isotopes.
Owing to the high penetrability of muon, such a microscope can be applied to 
biological samples under the atmospheric environment.
Experiments for direct observation of 
the muon released after the $\mu$CF using a layered hydrogen thin disk 
target are in progress~\cite{Nagamine1989,NagamineMCF199091,
Strasser1993,Nagamine1996Hyp103,Strasser1996,okutsu,
YAMASHITA2021112580}.
%
\begin{figure}[t!]
\begin{center}
\epsfig{file=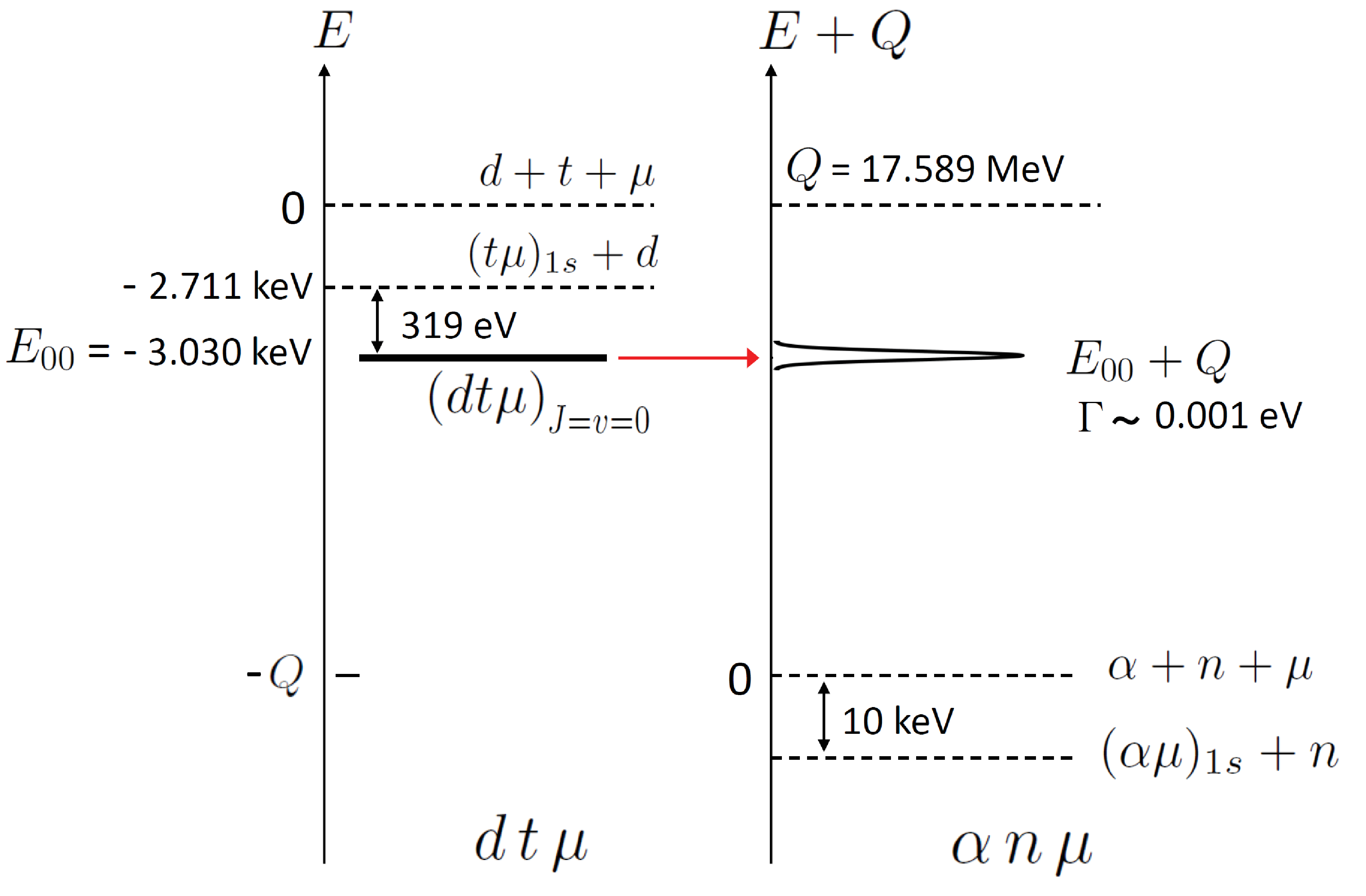,width=8.4cm,height=5.5cm}
\end{center}
\vskip -0.2cm
\caption{Schematic illustration of the energy relation between the $dt\mu$ and
$\alpha n \mu$ channels. Owing to the nuclear interaction, 
the muonic molecular 
bound state $(dt\mu)_{J=v=0}$ at $E_{00}=-3.030$ keV becomes
an extremely narrow Feshbach resonance that \mbox{decays}
into the $\alpha+n+\mu$ and $(\alpha \mu)_{nl} + n$ continuum states.
The width of the resonance ($\Gamma \sim 10^{-3}$ eV) 
was already derived, for example,
using the $d$-$t$ optical-potential-model calculation of the
reaction (\ref{eq:dt-reaction})~\cite{Kamimura89,
Bogdanova,Bogdanova89} 
and by the $R$-matrix method~\cite{Struensee88a,
Szalewicz90,Hale93,Hu1994,Cohen1996,Jeziorski91}.   
}
\label{fig:gainen-zu}
\end{figure}

Here, we note that
two of the present authors (Y.K. and T.Y.) 
have contributed to the aforementioned studies of \mbox{Type I} in 
Refs.~\cite{Iiyoshi2019,Yamashita2021submitted} 
and Type II in Refs.~\cite{okutsu,YAMASHITA2021112580}.

The purpose of the present paper is that considering the latest developments 
regarding the new $\mu$CF applications,
we thoroughly investigate the mechanism of
the nuclear reaction
\begin{subequations} 
\label{eq:mucf-reaction}
\begin{align}
(dt\mu)_{J=v=0} & \to \;\alpha \, + n \, + \mu  + 17.6 \,\mbox{MeV} 
\label{first equation}  \\
      & \searrow    \; (\alpha \mu)_{nl} + n  + 17.6 \,\mbox{MeV}, 
\label{second equation}  
\end{align}
\label{eq:mucf-reaction}
\end{subequations} 

by employing a sophisticated framework that has not been explored
in the literature work on this reaction.



The reaction~(\ref{eq:mucf-reaction}) 
is the most important  among the nuclear reactions in $\mu$CF,
but it has a complicated mechanism.
\mbox{Figure~\ref{fig:gainen-zu}}  
illustrates schematically the energy relation between the $dt\mu$ and
$\alpha n \mu$ channels;  the  molecular 
bound state $(dt\mu)_{J=v=0}$ becomes an extremely narrow Feshbach 
resonance with $\Gamma \! \sim \! 10^{-3}$ eV~\cite{Kamimura89,
Bogdanova,Bogdanova89, 
Struensee88a,Szalewicz90,Hale93,Hu1994,Cohen1996,Jeziorski91}
that decays into the $\alpha+n+\mu\,$ 
and $\,(\alpha \mu)_{nl} + n$ continuum states, owing to the nuclear 
interactions.

Therefore, due to the difficulty of the problem,
the reaction (\ref{eq:mucf-reaction})  has not been
studied  in the literature using sufficiently sophisticated methods.
\mbox{Another reason} is that 
such a precise calculation of the reaction
has not been required in previous $\mu$CF studies;
the required quantities were
the  fusion rate $\lambda_{\rm f} (=\Gamma/\hbar) $ and 
the $\alpha$-$\mu$ initial sticking probability $\omega_S^0$,
which were calculated using approximate models
(cf. Secs.~7 and 8 of the $\mu$CF review paper~\cite{Froelich92}), 
for example, in Ref.~\cite{Kamimura89} by one of the  
present authors (M.K.).

For the new situation of $\mu$CF mentioned
in Types I and II,
it is desirable to precisely calculate the following quantities of
the reaction (\ref{eq:mucf-reaction}): 
\setlength{\parskip}{-0.1cm} 
\begin{itemize}
\setlength{\itemsep}{0.2cm}  
\item[a)]  Reaction rates, in {\it absolute} values,
going to the individual bound 
and continuum states of the outgoing $\alpha$-$\mu$ pair,
together with the $\alpha$-$\mu$ sticking probability $\omega_S^0$ 
based on those reaction rates.  
\item[b)] Momentum and energy spectra, in {\it absolute} values,
of the muon emitted by the fusion reaction.
\end{itemize}

%
\begin{figure}[b]
\begin{center}
\epsfig{file=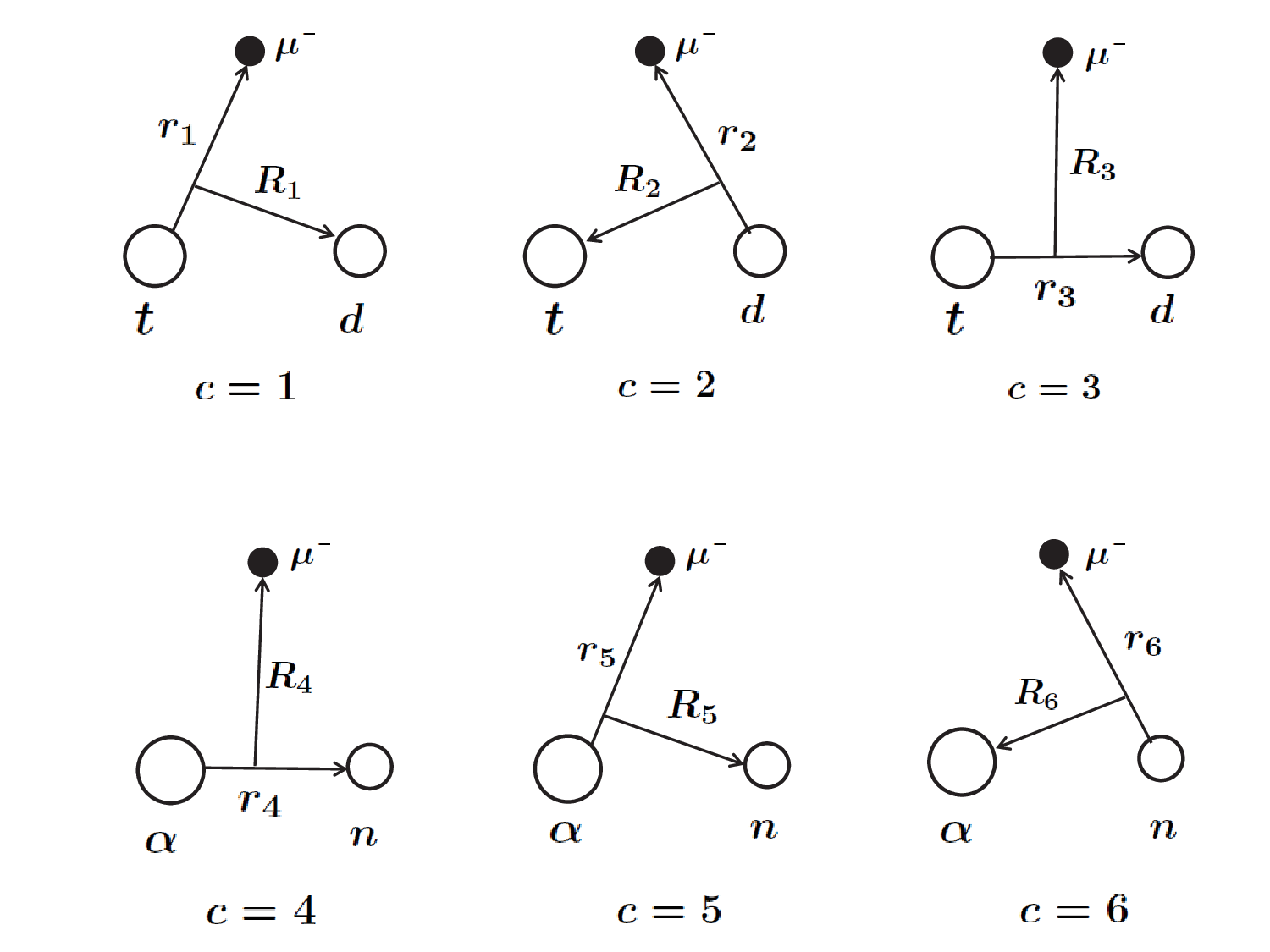,width=9.0cm,height=6.8cm}
\end{center}
\vskip -0.2cm
\caption{All the sets of Jacobi coordinates in the $dt\mu$ 
system $(c=1,2$ and 3)
and in the $\alpha n \mu$ system $(c=4,5$ and 6). 
}
\label{fig:jacobi-dtmu}
\end{figure}

\noindent
To calculate these quantities and conduct additional analyses,
we employ the $dt\mu$- and $\alpha n \mu$-channel coupled three-body model
and perform the following:
\begin{itemize}
\setlength{\itemsep}{0.1cm}  
\item[i)] We first determine all the nuclear interactions 
(the \mbox{$d$-$t$} 
and $\alpha$-$n$ potentials and the $d t$-$\alpha n$ 
channel coupling potential)
to \mbox{reproduce} 
the observed \mbox{astrophysical} \mbox{$S$-factor} of 
the reaction 
\begin{equation}
d + t \to \alpha + n + 17.6\: \mbox{MeV}
\label{eq:dt-reaction}
\end{equation}
at the low-energies of $1-300$ keV~\cite{Sfactor-exp} 
as well as the total cross section of the $\alpha+n$ reaction at the
corresponding energies~\cite{Haesner}  (see Sec.~II).

\item[ii)] We solve a coupled-channels three-body Schr\"{o}dinger equation  
on the Jacobi coordinates in Fig.~\ref{fig:jacobi-dtmu},
satisfying the boundary condition to have the muonic molecular 
bound state $(dt\mu)_{J=v=0}$
as the initial state (as the source term of Schr\"{o}dinger equation)
and the outgoing wave in the $\alpha n \mu$ channel
with the 17.6-MeV $D$-state $\alpha$-$n$ relative motion based on the 
observation (see Sec.~III).

\item[iii)] We calculate  the quantities a) and b) 
by substituting the obtained total wave function into the 
\mbox{$T$-matrix} elements for a) and b) based on the 
Lippmann-Schwinger equation~\cite{Lippmann} (see Secs.~IV,V and VI).
\end{itemize}

\setlength{\parskip}{0.0cm} 

The reliability of these 
calculations shall be carefully examined as follows:
The fusion rate $\lambda_{\rm f}$ 
(the number of fusions per second) is calculated 
in the three different prescriptions: 
\begin{itemize}
\setlength{\parskip}{-0.1cm} 
\setlength{\itemsep}{0.1cm} 
\item[A)] $\lambda_{\rm f}$ from the $S$-matrix of the 
asymptotic amplitude of the total wave function solved using 
the coupled-channels Schr\"{o}dinger equation (see Sec.~III).
\item[B)] $\lambda_{\rm f}$ from the \mbox{$T$-matrix} calculation 
of the reaction rates mentioned in item a) (see Sec.V).  
\item[C)] $\lambda_{\rm f}$ from the $T$-matrix calculation of 
the muon spectra mentioned
in item b) (see Sec.~VI).
\end{itemize}
%

\noindent
If our total wave function is the  exact rigorous  solution of
the coupled-channels Schr\"{o}dinger equation,
the fusion rates $\lambda_{\rm f}$ obtained by A)-C) should be {\it equal}
according to the Lippmann-Schwinger equation that is
equivalent to the Schr\"{o}dinger equation.
However, since the function space employed in our total wave function 
is not complete, the resultant $\lambda_{\rm f}$ are not equal, but
they should be {\it consistent} with each other.
This check of $\lambda_{\rm f}$ is 
one of the highlights of the present paper.

The authors possess their own 
three methods to solve the present coupled-channels three-body 
Schr\"{o}dinger equation.
Namely, the Kohn-type variational method for the reactions 
between composite particles~\cite{Kamimura77}, 
the Gaussian expansion method (GEM) for 
few-body systems~\cite{Kamimura88,Kameyama89,Hiyama03} 
here for describing the $(dt\mu)$ molecule 
\mbox{nonadiabatically}~\cite{Kamimura88},
and the continuum-discretized coupled-channels (CDCC) 
method~\cite{Kamimura86, Austern, Yahiro12}  
here for discretizing the $\alpha$-$\mu$ and \mbox{$\alpha$-$n$} 
continuum states.

This paper is organized as follows. In Sec.~II \mbox{we will}
determine  all the nuclear interactions used in this work 
to reproduce the observed   $S$-factor of 
the reaction (\ref{eq:dt-reaction}) and the total cross section of 
the $\alpha + n$ reaction.
In Sec.~III we solve the
coupled-channels Schr\"{o}dinger equation for the
reaction (\ref{eq:mucf-reaction}).
Section IV is devoted to providing an overview of the $T$-matrix framework
based on the Lippmann-Schwinger equation.
In Sec.~V, based on the \mbox{$T$-matrix} calculation, we derive
the reaction rates going to the $\alpha$-$\mu$ bound and
continuum states.
In Sec.~VI,  we also  derive
the momentum and energy spectra of the muon ejected from the fusion reaction.
Finally, a summary is given in Sec.~VII.

\section{Nuclear interactions  for low-energy  
{\mbox{\boldmath  $\lowercase{d}+\lowercase{t} 
\to \alpha+\lowercase{n}$ }} reaction}
%

\begin{figure}[b!]
\begin{center}
\epsfig{file=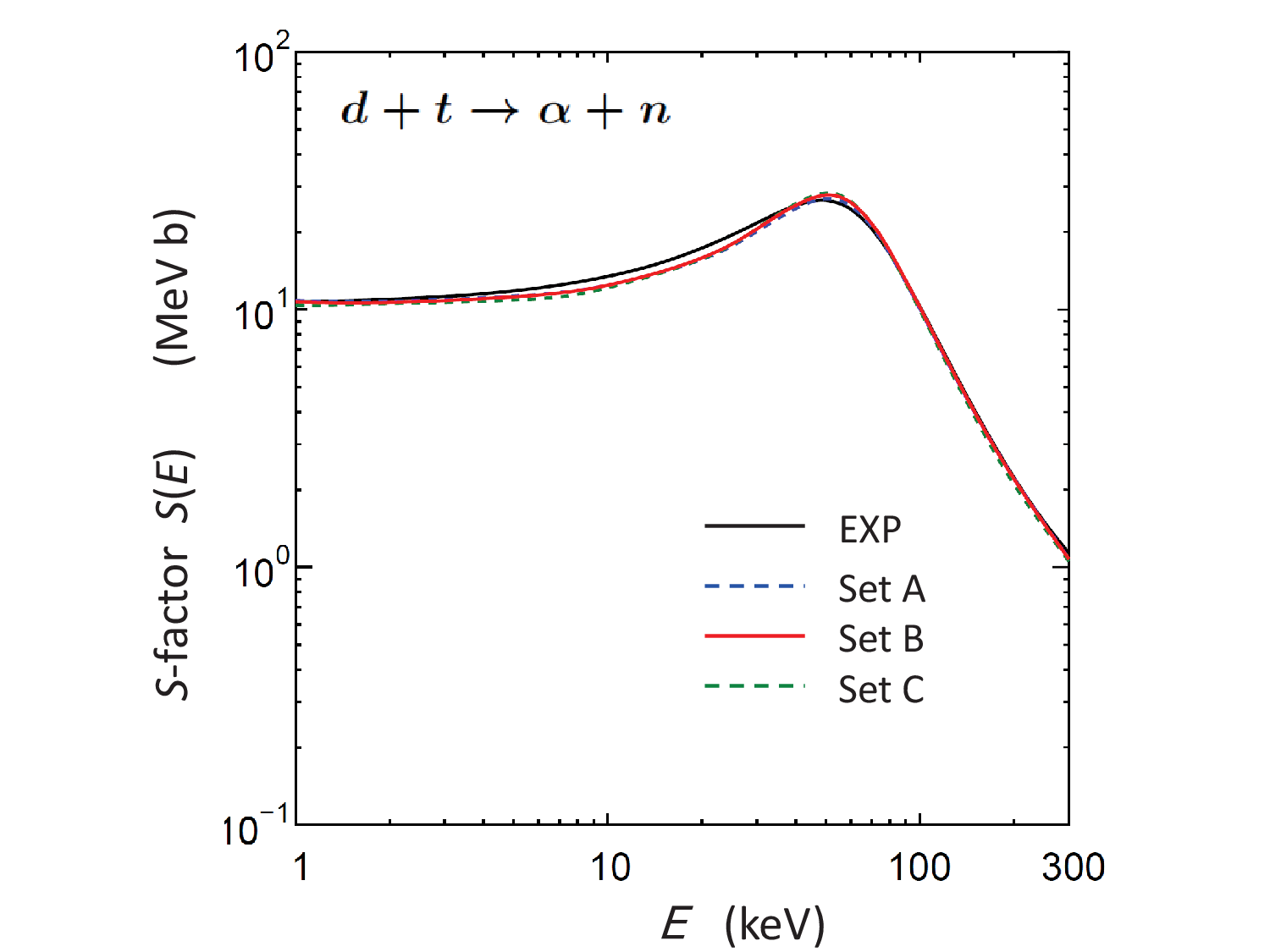,width=8.9cm,height=7.0cm}
\end{center}
\vskip -0.2cm
\caption{
Calculated and observed $S$-factor $S(E)$ of 
the \mbox{$ d+t \to \alpha + n$} reaction
with respect to the c.m. energy $E$ of the incoming $d$-$t$ wave.
The three curves for Sets A, B, and C are obtained using  
the nuclear interactions listed in Table~I. 
The black curve (EXP) is taken from a review paper~\cite{Sfactor-exp};
it fits the literature data  using  the function 
$S(E)=(26 - 0.361E + 248 E^2)/(1 + ((E - 0.0479)/0.0392)^2) 
{\rm  MeV}\,{\rm b}$ ($E$ in MeV).
The calculated curves lie within the error range of
the data.
}
\label{fig:cal-cc-dt-an}
\end{figure}

\begin{figure}[b!]
\begin{center}
\epsfig{file=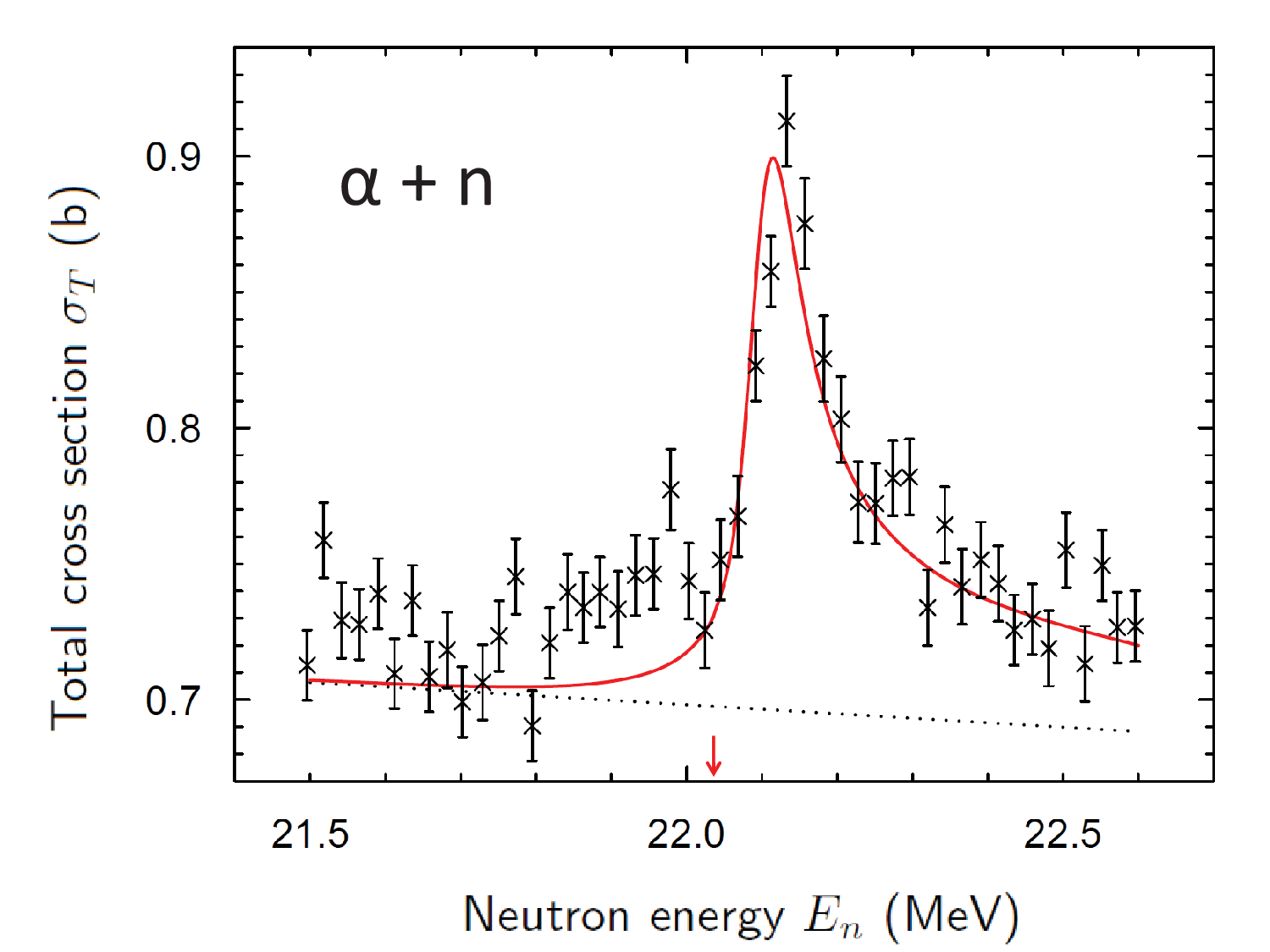,width=7.0cm,height=5.8cm}
\end{center}
\vskip -0.2cm
\caption{The total cross section $\sigma_T(E_n)$ 
of the $\alpha+n$ reaction based on the observation~\cite{Haesner} and
the present calculation 
with the nuclear interactions Set B 
(the red curve), where 
$E_n$ is the neutron incident energy.
The dotted line denotes the background cross section based on partial 
waves other than the $D_{3/2^+}$ wave~\cite{Haesner,Barker}.
The calculated red curve for the $I={\frac{3}{2}}^+$ state is piled on the
background dotted line.  The peak at $E_n \approx 22.1 $MeV corresponds 
to the peak in Fig.~\ref{fig:cal-cc-dt-an} for the $d$-$t$ channel. 
The $d$-$t$ threshold 
\mbox{$(E=0)$} corresponds to $E_n=22.03$ MeV (at the red arrow)
in the $\alpha$-$n$ channel.
}
\label{fig:sigma-an}
\end{figure}

To investigate the mechanism of the $\mu$CF 
reaction~(\ref{eq:mucf-reaction})
based on the $dt\mu$- and $\alpha n \mu$-channels coupled three-body model,
it is necessary to use the nuclear interactions (the \mbox{$d$-$t$} 
and $\alpha$-$n$ 
potentials and the $d t$-$\alpha n$ coupling potential)
that reproduce the observed 
low-energy astrophysical $S$ factor of the reaction
(\ref{eq:dt-reaction}) (cf. Fig.~\ref{fig:cal-cc-dt-an})
as well as the total cross section of the $\alpha+n$ reaction 
at the corresponding energies (cf. Fig.~\ref{fig:sigma-an}).

Considering that a similar framework for channel coupling will also be 
used in the study of $\mu$CF reaction
(\ref{eq:mucf-reaction}), we apply the same coordinates ${\bf r}_3$
and ${\bf r}_4$ in Fig.~\ref{fig:jacobi-dtmu} 
to the $d$-$t$ and $\alpha$-$n$ relative coordinates, respectively,
in this section.

The total wave function of the system 
has spin-parity $I=\frac{3}{2}^+$
in the energy region of the resonance and below.
It is known that the $d$-$t$ channel has $S$-wave and spin $\frac{3}{2}$
and the \mbox{$\alpha$-$n$} channel  has $D$-wave with spin $\frac{1}{2}$ 
coupled to \mbox{$I=\frac{3}{2}^+$}.
Let $E$ denote the c.m. energy of the $d$-$t$ relative motion. 
The total wave function is written as 
\begin{eqnarray}
\!\!\!\!\!\! \Phi_{\frac{3}{2} M}(E) & = &  
      \phi_{0}({\bf r}_3)\, \chi_{\frac{3}{2} M}(dt) +
   \big[ \psi_2 ({\bf r}_4)\,\chi_{\frac{1}{2}}
      (\alpha n) \big]_{\frac{3}{2} M} \nonumber \\
  &=&   \frac{\bar{\phi}_{0}(r_3)}{r_3} \, Y_{00}(\widehat{{\bf r}}_3)\, 
        \chi_{\frac{3}{2} M}(dt) \nonumber \\
\!\!\!\!\!\!   &+&  \frac{\bar{\psi}_{2}(r_4)}{r_4} \,
   \big[ Y_2 (\widehat{{\bf r}}_4) \,
\chi_{\frac{1}{2}}(\alpha n) \big]_{\frac{3}{2} M}\: ,
\label{eq:total-wf-dt-an}
\end{eqnarray} 
which has the asymptotic behavior
\begin{eqnarray}
\label{eq:dt-boundary-1}
&&  \bar{\phi}_0(r_3) \stackrel{r_3 \to \infty}{\longrightarrow} 
    \!\!\!  U_0^{(-)}(k_3, r_3) - S_0^{(dt)} \,U_0^{(+)}(k_3, r_3), \;\;  \\
&& \bar{\psi}_2(r_4) \stackrel{r_4 \to \infty}{\longrightarrow} 
     \quad  \quad   - \, \sqrt{\frac{v_3}{v_4}}\, S_2^{(dt,\alpha n)} \,
                        U_2^{(+)}(k_4, r_4), \;\;    \\
     && U_L^{(\pm)}(k, r)= G_L(k, r) \mp i F_L(k, r) ,
\label{eq:dt-boundary-2}
\end{eqnarray} 
where  $k_3 (k_4)$ and $v_3 (v_4)$ are 
the wave number and the velocity of  relative motion 
along ${\bf r}_3 ({\bf r}_4)$, respectively, and
$F_L$ and $G_L$ are the regular and
irregular Coulomb functions, respectively.



We assume that the $S$- and $D$-state wave functions are 
\mbox{coupled} to each other by the following 
tensor force, which is nonlocal between ${\bf r}_3$ and~${\bf r}_4$;
\begin{eqnarray}
\label{eq:tensor-1} 
\!\!\!\!\!\!\!\!\!
  V_{dt, \alpha n}^{({\rm T})}({\bf r}_3, {\bf r}_4)
            &\!\!\! =\!\!\!&v^{({\rm T})}(r_{34},R_{34}) 
          \big[ Y_2({\widehat {\bf r}}_{34})\, 
          {\cal S}_2(dt,\alpha n) \big]_{00},   \\ 
\!\!\!\!\!\!\!\!\! 
 v^{({\rm T})}(r_{34},R_{34}) &\!\!=\!\!& v_0^{({\rm T})} r_{34}^2\, 
          e^{-\mu \,r_{34}^2- \mu' R_{34}^2},  
\label{eq:tensor-2}
\end{eqnarray}
where  ${\bf r}_{34}= {\bf r}_3 - {\bf r_4}$ and 
${\bf R}_{34}= {\bf r}_3 + {\bf r_4}$. 
In Eq.~(\ref{eq:tensor-1}), ${\cal S}_2(dt,\alpha n)$ is a spin-tensor operator
composed of spins of  $dt$- and $\alpha n$-pairs.
However, it is {\it not} necessary to know the
explicit form of ${\cal S}_2(dt,\alpha n)$ in the present work,
as will be explained in the paragraph below Eq.~(\ref{eq:spin-fac}).

%



The coupled-channels Schr\"{o}dinger equation required to solve 
$\phi_{0}({\bf r}_3)$ and  $\psi_2 ({\bf r}_4)$  is written as
\begin{subequations} 
\label{eq:schr-dt}
\begin{align}
  \!\!\!\!   ( H_{dt} - E )\,  \phi_{0}({\bf r}_3) \,
               \chi_{\frac{3}{2} M}(dt)  
  = -V^{({\rm T})}_{dt, \alpha n} \: 
   \big[ \psi_2 ({\bf r}_4)\,\chi_{\frac{1}{2}}(\alpha n) 
   \big]_{\frac{3}{2} M},\!\!\! &
\label{first equation}     \\
 \!\!\!\!\!\!   \left( H_{\alpha n} - (E + Q) \right)
  \,   \big[ \psi_2 ({\bf r}_4)\,
      \chi_{\frac{1}{2}}(\alpha n) \big]_{\frac{3}{2} M} 
\qquad \qquad \qquad \;\;\; & \: \nonumber \\
  \!\!\!\!\!\!  = -V^{({\rm T})}_{\alpha n, dt} \:
      \phi_{0}({\bf r}_3)\, \chi_{\frac{3}{2} M}(dt) ,\quad
\label{second equation}     
\end{align}
\end{subequations}
where  $Q=17.589$ MeV and 
\begin{eqnarray}
  H_{dt}&=& T_{{\bf r}_3}  
             + V_{dt}^{({\rm N})}(r_3) + V_{dt}^{({\rm C})}(r_3) , \;\;   \\ 
  H_{\alpha n }&=& T_{{\bf r}_4} 
           + V_{\alpha n}^{({\rm N})}(r_4) , \\ 
  V^{({\rm T})}_{dt, \alpha n}&=& \int d{\bf r}_4 \:
           V_{dt, \alpha n}^{({\rm T})}({\bf r}_3, {\bf r}_4).
\label{eq:tensor-int-2}
\end{eqnarray}

The coupled-channels Schr\"{o}dinger 
equation~(\ref{eq:schr-dt}) 
with the scattering boundary 
condition~(\ref{eq:dt-boundary-1})-(\ref{eq:dt-boundary-2})  
can be accurately solved by using the couple-channels 
Kohn-type variational method  for composite-particle
reactions that was proposed by one of the authors (M.K.)~\cite{Kamimura77} 
and has been employed in the literature
in three-body transfer reactions, for example, 
in Refs.~\cite{Kawai86,Kino93a,Kamimura09}.

When the matrix element of the tensor force is calculated,
the spin part $ S^{({\rm T})}_0$ is factored out as follows:
\begin{eqnarray}
&&\!\!\!\!\! \!\!\!\!\!\!\! \!\!\!
      \langle \,\big[ \phi_2 ({\bf r}_4)\,\chi_{\frac{1}{2}}(\alpha n) 
           \big]_{\frac{3}{2} M}
     \,|\, V^{({\rm T})}_{\alpha n, dt}
     \,|\,\big[ \phi_0 ({\bf r}_3)\,\chi_{\frac{3}{2}}(dt)
           \big]_{\frac{3}{2} M} \rangle        \nonumber \\
&&\!\!\!\!\!\!\!\!\!\!\!\!\!\!\!\!  = v^{({\rm T})}_0 S^{({\rm T})}_0 
   \langle \, \phi_{2 m} ({\bf r}_4)
     |\,r_{34}^2\, e^{-\mu \,r_{34}^2- \mu' R_{34}^2}
     |\,\big[ Y_2({\widehat {\bf r}}_{34}) \phi_0 ({\bf r}_3) \big]_{2 m} 
     \rangle , \;
\label{eq:tensor-mat}
\end{eqnarray}
where
\begin{eqnarray}
\!\!\!\!\!\!\!  S^{({\rm T})}_0= \frac{1}{\sqrt{10}}
      \langle \,\chi_{\frac{1}{2} m_s}(\alpha n) \,|\, 
 \big[ {\cal S}_2(\alpha n, dt) 
\chi_{\frac{3}{2}}(dt) \big]_{\frac{1}{2} m_s} 
      \rangle \:.
\label{eq:spin-fac}
\end{eqnarray}
The R.H.S. of Eqs.~(\ref{eq:tensor-mat}) and (\ref{eq:spin-fac}) are 
independent of $m$ and $m_s$, respectively, 
and hence, the L.H.S. of Eqs.~(\ref{eq:tensor-mat})
does not depend on $M$.
We shall verify, in   Secs.~III, V and VI,
that the same $v^{({\rm T})}_0 S^{({\rm T})}_0$ as above is factored out
when calculating the three-body matrix elements of the tensor force 
and the \mbox{$T$-matrix} elements due to the same force. 
Consequently, we can treat the tensor force consistently throughout 
the present work without knowing the explicit forms of 
${\cal S}_2(dt,\alpha n)$ and $S^{({\rm T})}_0$.
It is sufficient to search for the optimum value of the product 
$v^{({\rm T})}_0 S^{({\rm T})}_0$ when producing the observed data.



\begin{figure}[t!]
\begin{center}
\epsfig{file=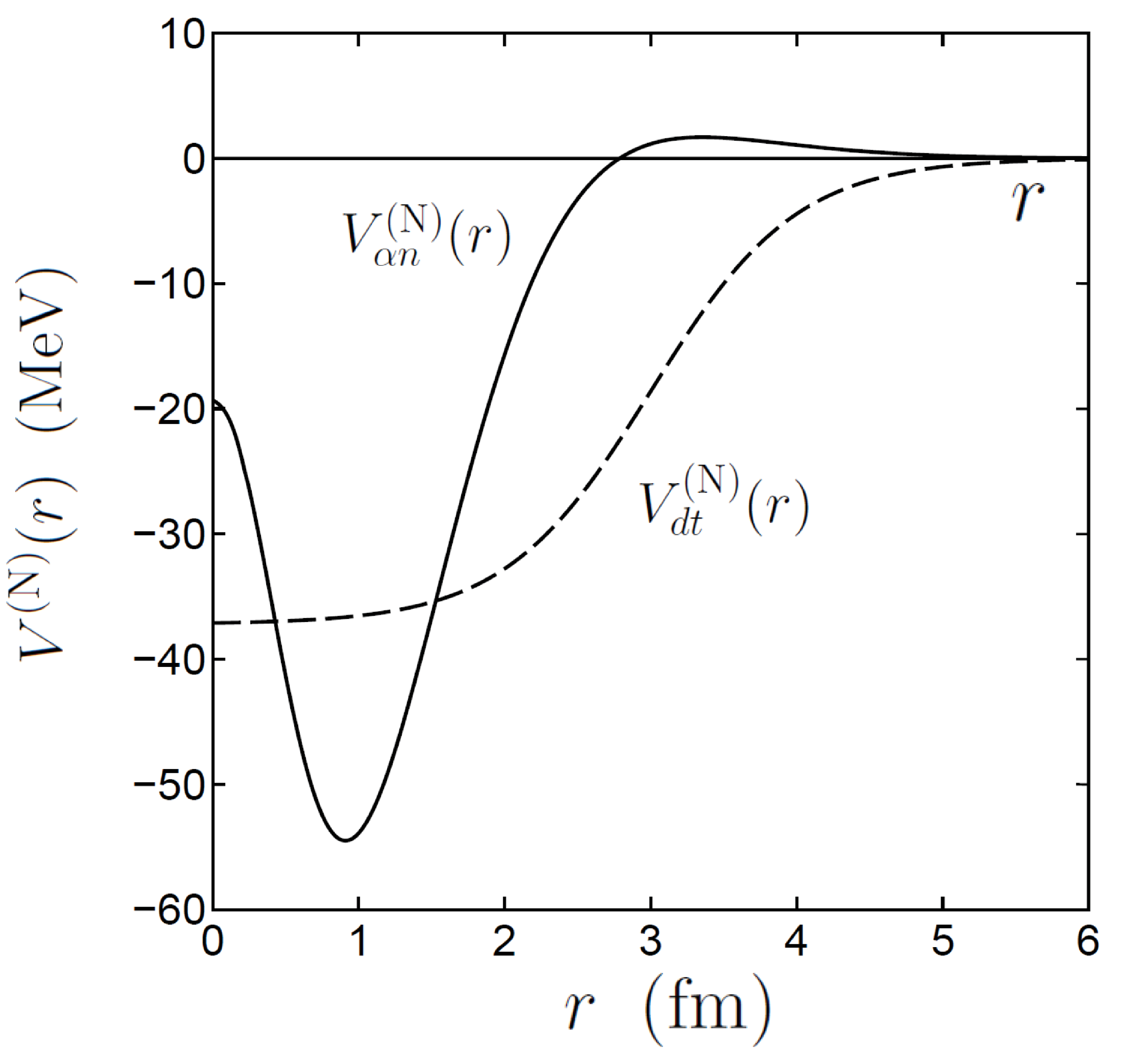,width=6.0cm,height=5.5cm}
\end{center}
\vskip -0.4cm
\caption{The nuclear $d$-$t$ potential $V_{dt}^{({\rm N})}(r)$ 
of Set B
(the dashed curve) and the Kanada-Kaneko $\alpha$-$n$ potential 
$V_{\alpha n}^{({\rm N})}(r)$~\cite{Kaneko,PTP-suppl-68-III}
for the even-parity case (the solid curve).
}
\label{fig:vpot-dt-an}
\end{figure}

The nuclear $d$-$t$ potential and accompanied Coulomb potential are
employed, respectively, in the form
\begin{eqnarray}
\label{eq:opt-real}
   &&  \; V_{dt}^{{\rm (N})}(r_3)=V_0/\{1+e^{(r_3-R_0)/a}\} , \\   
\label{eq:opt-coul}
&& V_{dt}^{({\rm C})}(r_3)=\begin{cases}
           \frac{e^2}{R_{\rm c}}(\frac{3}{2}-\frac{r_3^2}{2R_{\rm c}^2}) 
       &  \text{$(r_3 < R_{\rm c}\!=\!R_0)$},  \\
          \frac{e^2}{r_3}  &  \text{$(r_3 \geq  R_{\rm c}\!=\!R_0)$}.
\end{cases}
\end{eqnarray}
As the $\alpha$-$n$ potential $V_{\alpha n}^{({\rm N})}(r_4)$, 
we employ 
the  Kanada-Kaneko $\alpha$-$n$ potential~\cite{Kaneko,PTP-suppl-68-III}
(see Fig.~\ref{fig:vpot-dt-an}),
which is  derived based on an equivalent local potential
to the nonlocal kernel of the resonating-group method for the
$\alpha$-$n$ system, and is often used in the
\mbox{$\alpha$-cluster-model} calculations of 
light nuclei~\cite{PTP-suppl-68-III}.
We then fix the potential as $V_{\alpha n}^{({\rm N})}(r_4)$
throughout this work.



\begin{table}[t!]
{
\caption{Parameter Sets A, B, and C of the nuclear interactions 
$V_{dt}^{\rm (N)}(r_3)$ in Eq.~(\ref{eq:opt-real}) and 
$V_{dt,\alpha n}^{\rm (T)}({\bf r}_3, {\bf r}_4)$ 
in Eq.~(\ref{eq:tensor-1}) for reproducing the 
observed data in Figs.~\ref{fig:cal-cc-dt-an} and~\ref{fig:sigma-an}.  
}} 
{
\begin{center}
\begin{tabular}{cccccccccccccc} 
\hline \hline
\noalign{\vskip 0.1 true cm} 
  &  & $V_0$  & $\;\;$  & $R_0$ & $\;\;$ & $a_0$ & & 
              $v_0^{({\rm T})} S_0^{({\rm T})}$ & & $\mu$ & & $\mu'$  \\
\noalign{\vskip 0.0 true cm} 
            &  &  (MeV) & &   (fm)  & &   (fm)  &  & 
$({\rm MeV}\,{\rm fm}^{-5})$ && $({\rm fm}^{-2})$ && $({\rm fm}^{-2})$   \\
\noalign{\vskip 0.1 true cm} 
\hline
\noalign{\vskip 0.15 true cm} 
   Set A  &  & $-57.0$ & & $2.5$ & & $ 0.3$ & & $7.04$& & 
          $1/1.16^2$ && $1/6.0^2$\\ 
   Set B  &  & $-37.2$ & & $3.0$ & & $ 0.5$ & & $5.43$& & 
          $1/1.16^2$ && $1/6.0^2$\\ 
   Set C  &  & $-27.7$ & & $3.0$ & & $ 1.0$ & & $4.76$& & 
          $1/1.20^2$ && $1/6.0^2$\\ 
\noalign{\vskip 0.15 true cm} 
\hline
\hline
\end{tabular}
\end{center}
}
\label{table:3-set-interactions}
\end{table}



%
%


As the nuclear interactions $V_{dt}^{({\rm N})}(r_3)$ and
$V^{({\rm T})}_{dt, \alpha n}({\bf r}_3, {\bf r}_4)$ introduced above are
phenomenological ones, it is desirable to be
shown that the calculated results for the reaction (\ref{eq:mucf-reaction})
are independent of the interaction details 
as long as they reproduce the observed data
in Figs.~\ref{fig:cal-cc-dt-an} and~\ref{fig:sigma-an}.
Concequently, three sets of the interactions, 
Sets A, B, and C  shown in Table~I,  are examined.
$V_{dt}^{\rm (N)}(r_3)$ is acquired from the real part of the $d$-$t$ 
optical potentials A, B, and C in Ref.~\cite{Kamimura89}  
with a slight change in $V_0$ while $R_0$ and $a_0$ are the same;
in the study, the fusion rate $\lambda_{\rm f} =
\mbox{(1.22 - 1.28)} \times 10^{12} {\rm s}^{-1}$ was derived
using the optical-potential model.  
The potentials  $V_{dt}^{({\rm N})}(r_3)$ of Set B and 
$V_{\alpha n}^{({\rm N})}(r_4)$ are illustrated in Fig.~\ref{fig:vpot-dt-an}.

The reaction cross section $\sigma_{dt \to \alpha n}$ is given by
\begin{eqnarray}
\sigma_{dt \to \alpha n}(E)=\frac{2I+1}{(2I_d+1)(2I_t+1)}
\frac{\pi}{k_3^2}\, |S_2^{(dt,\alpha n)}|^2 ,
\end{eqnarray}
and the  $S$-factor $S(E)$ is derived from 
\begin{equation}
\sigma_{dt \to \alpha n}(E)=S(E)\, e^{-2\pi \eta(E)}/E,
\label{eq:S-factor}
\end{equation}
where $\eta(E)$ is the Sommerfeld parameter.  

Calculated $S$-factor $S(E)$ using the nuclear interactions of
Sets A, B, and C is illustrated in Fig.~\ref{fig:cal-cc-dt-an}.
The black curve represents the observed data summarized in a review paper
~\cite{Sfactor-exp}.  All the calculated curves well reproduce the observed data
within the error range of the data that are not shown here. 
 
We then discuss the observed total cross section 
$\sigma_T(E_n)$  of the 
$\alpha+n $ reaction for the $I={\frac{3}{2}}^+$ state~\cite{Haesner}
given in Fig.~\ref{fig:sigma-an}. The nuclear interactions of Set B is used.
The data should be explained by our calculation simultaneously as
the $S$-factor $S(E)$ of the $d +t \to \alpha + n$ reaction.
The total cross section  
can be expressed using our model as the sum of elastic and reaction
cross sections, with the spin factor $\frac{2I+1}{(2I_\alpha +1)(2I_n+1)}=2$,
%
\begin{eqnarray}
\!\!\!\!\!\!\sigma_T(E_n) = \sigma_{\rm el.}(E_n)  +  
    \sigma_{\rm re.}(E_n) 
    =\frac{4\pi}{k_4^2}\, [1-{\rm Re}(S_2^{(\alpha n)})], 
\label{eq:sig-total}
\end{eqnarray}
%
%
with
\begin{eqnarray}
\label{eq:sig-elastic}
\!\!\!\!\!\sigma_{\rm el.}(E_n)= \frac{2\pi}{k_4^2} |1-S_2^{(\alpha n)}|^2  ,  \;\;\;
\sigma_{\rm re.}(E_n)=\frac{2\pi}{k_4^2} |S_2^{(\alpha n,dt)}|^2, 
\label{eq:sig-reaction}
\end{eqnarray}
and the unitarity $|S_2^{(\alpha n)}|^2 +  |S_2^{(\alpha n,dt)}|^2=1.$
The $S$-matrix elements are determined by the 
the  asymptotic behavior, similarly to Eqs.~(\ref{eq:dt-boundary-1})
and (\ref{eq:dt-boundary-2}),
\begin{eqnarray}
\label{eq:dt-boundary-3}
\!\!\! \bar{\phi}_0(r_3) \stackrel{r_3 \to \infty}{\longrightarrow} 
 && \!\! \quad \;\,\;\, - \, \sqrt{\frac{v_4}{v_3}}\, S_2^{(\alpha n,dt)} \,
                     U_0^{(+)}(k_3, r_3), \;\;  \\
\!\!\!  \bar{\psi}_2(r_4) \stackrel{r_4 \to \infty}{\longrightarrow} 
   && \!\!\!\!\!  U_2^{(-)}(k_4, r_4) 
- S_2^{(\alpha n)} \,U_2^{(+)}(k_4, r_4). \;\;  
\label{eq:dt-boundary-4}
\end{eqnarray}



Recall that when tuning the potential
parameters to reproduce 
the \mbox{$S$-factor $S(E)$} in Fig.~\ref{fig:cal-cc-dt-an},
we fixed the $\alpha$-$n$ potential $V_{\alpha n}^{({\rm N})}(r_4)$
and did not include the observed  $\alpha+n $ data for fitting.
Therefore, it is impressive to see in Fig.~\ref{fig:sigma-an} that 
our result (the red curve) agrees well with the  data.

However, this is not unexpected because of the following reason: 
The reaction  $d+t \to \alpha + n$ in Fig.~\ref{fig:cal-cc-dt-an} 
is known as one of the most strongly channel-coupled \mbox{reactions}
in nuclear physics~\cite{Hale1987PRL,Bogdanova1991}.
Actually, at the peak, our calculation shows  
$|S_2^{(dt, \alpha n)}| \approx 1$ and
\mbox{$S_0^{(dt)}\approx 0$}. 
Accordingly, we see that at the peak of the $\alpha+n $ reaction, 
$|S_2^{(\alpha n, dt)}|=|S_2^{(dt, \alpha n)}| \approx 1$
and $S_2^{(\alpha n)}\approx 0$ 
and hence $\sigma_T(I=\frac{3}{2}^+) \approx 4\pi/k_4^2=0.19$ b 
with $k_4=0.824 \, {\rm fm}^{-1}$.
This is the peak value of
$\sigma_T(I=\frac{3}{2}^+)$ of the calculation (observation)
in Fig.~\ref{fig:sigma-an} 
piled on the dotted  background line.  
It is asserted that 
even if we employ another $\alpha$-$n$ potential instead of the above
$V_{\alpha n}^{({\rm N})}(r_4)$, the same situation will be repeated
as long as the observed data of the $d +t \to \alpha + n$ reaction
are reproduced using the replaced $\alpha$-$n$ potential.

\section{Coupled-channels Schr\"odinger 
equation for 
\mbox{\boldmath {\large $\lowercase{d} \lowercase{t} \mu$}}
fusion reaction} 
%

In this section, we formulate and solve the Schr\"{o}dinger equation
for the fusion reaction (\ref{eq:mucf-reaction})
using the nuclear interactions that were
determined in the previous section and
satisfying the boundary condition to have the muonic 
molecule $(dt\mu)_{J=v=0}$
as the initial state and the outgoing wave in the $\alpha n \mu$ channel.
We first divide the fusion
decay process  into the following two steps:
\begin{description}
 \item[\mbox{{\rm Step 1)}}] Construction of the  nonadiabatic 
wave function, denoted as  
$\mathring{\Phi}_{0}^{({\rm C})}({{\bf r}_3, {\bf R}_3})$
(cf. Eq.~(\ref{eq:psi3})), 
of the initial $(dt\mu)_{J=v=0}$ state 
using the Coulomb potentials only~\cite{Kamimura88}.  
The eigenenergy of the state, $E_{00}$, is given by 
\begin{equation}
  E_{00}=-0.003030 \,{\rm MeV} 
\label{eq:E00-0303}
\end{equation}
with respect to the $d+t+\mu$ threshold (cf. Fig.~\ref{fig:gainen-zu}).
 \item[\mbox{{\rm Step 2)}}]  Decay of the $(dt\mu)_{J=v=0}$ state 
(now, a Feshbach resonance after the nuclear interactions are switched on)
into the outgoing wave due to the nuclear $d$-$t$, $\alpha$-$n$ 
and $dt$-$\alpha n$ interactions. 
The kinetic energy of the outgoing wave with the wave number $k_4$ 
is given by
\begin{equation}
    \frac{\hbar^2}{2\mu_{{\bf r}_4}} k_4^2 = E_{00} +  Q,  
    \quad Q=17.589 \,{\rm MeV}\,.
\label{eq:conserv-k4}
\end{equation} 
\end{description}

By employing the \mbox{Step-1} wave function 
$\mathring{\Phi}_{0}^{({\rm C})}({{\bf r}_3, {\bf R}_3})$ 
as the fixed {\it source term} of the coupled-channels 
Schr\"{o}dinger equation, it becomes possible to \mbox{impose} 
the outgoing-wave boundary condition upon the $\alpha n \mu$ channel
with no incoming wave for the entire system.
The symbol ($\,\mathring{\:}\,$) placed on the top of 
$\mathring{\Phi}_{0}^{({\rm C})}({{\bf r}_3, {\bf R}_3})$
is to show 'given' before solving the Schr{\" o}dinger equation
of \mbox{Step 2.}
$\mathring{\Phi}_{0}^{({\rm C})}({{\bf r}_3, {\bf R}_3})$ 
will be calculated in Sec.~III\,A.

The total angular momentum of the entire system is 
$\frac{3}{2}$ with its $z$-component $M$
similarly to the $dt$-$\alpha n$ case in Sec.~II\,B (muon spin is neglected).
We then describe the total wave function 
in term of the three parts based on the aforementioned 
two steps as follows:
\begin{equation}
 \Psi_{\frac{3}{2} M}^{(+)}(E)
  = \mathring{\Psi}_{\frac{3}{2} M}^{({\rm C})}(dt\mu)  
   + \Psi_{\frac{3}{2} M}^{({\rm N})}(dt\mu)  
   + \Psi_{\frac{3}{2} M}^{(+)}(\alpha n \mu), 
\label{eq:total-wf}
\end{equation}
with
\begin{eqnarray}
\label{eq:wf-spin-1}
&&\mathring{\Psi}_{\frac{3}{2} M}^{({\rm C})}(dt\mu) 
=\mathring{\Phi}_{0}^{({\rm C})}({{\bf r}_3, {\bf R}_3})\,
\chi_{\frac{3}{2} M}(dt),   \\
\label{eq:wf-spin-2}
&&\Psi_{\frac{3}{2} M}^{({\rm N})}(dt\mu) 
=\Phi_{0}^{({\rm N})}({{\bf r}_3, {\bf R}_3}) \,
\chi_{\frac{3}{2} M}(dt), \\
\label{eq:wf-spin-3}
&&\Psi_{\frac{3}{2} M}^{(+)}(\alpha n\mu) 
=\big[\Phi_2^{(+)}({{\bf r}_4, {\bf R}_4})\,\chi_{\frac{1}{2}}(\alpha n)
\big]_{\frac{3}{2} M}\,.
\end{eqnarray}
Here, $\mathring{\Phi}_{0}^{({\rm C})}, \Phi_{0}^{({\rm N})}$ and
$\Phi_2^{(+)}$ are spatially $S\!$-, $S\!$- and $D$-wave 
functions, respectively. 
The spin functions $\chi_{\frac{3}{2} M}(dt)$ and
$\chi_{\frac{1}{2} M}(\alpha n)$ were introduced 
in Eq.~(\ref{eq:total-wf-dt-an}).
However, the explicit form of the spin functions is not necessary 
because the phenomenological
spin-tensor operator ${\cal S}_2(dt,\alpha n)$ and 
its matrix element $S_0^{(T)}$ of Eq.~(\ref{eq:spin-fac}) are used.

In Eq.~(\ref{eq:total-wf}),  the first component
$\mathring{\Psi}_{\frac{3}{2} M}^{({\rm C})}(dt\mu)$
becomes the fixed source term of the coupled-channels 
Schr\"{o}dinger equation.
The second component
$\Psi_{\frac{3}{2} M}^{({\rm N})}(dt\mu)$ is introduced to describe
the $d$-$t$ relative motion due to the  {\it nuclear} interactions.
The third term $\Psi_{\frac{3}{2} M}^{(+)}(\alpha n \mu)$
is for the outgoing $\alpha n \mu$ channel.
We can then derive the fusion rate from 
the asymptotic behavior of 
$\Psi_{\frac{3}{2} M}^{(+)}(\alpha n\mu)$.

The coupled-channels Schr\"{o}dinger equation required to solve  
$\Psi_{\frac{3}{2} M}^{({\rm N})}(dt\mu)$ and 
$\Psi_{\frac{3}{2} M}^{(+)}(\alpha n \mu)$ can be written  as
\begin{subequations} 
\label{eq:schr-eq}
\begin{align}
   \!\!\!\!  \!\!\!\!  \!\!\!\!      \big( H_{dt\mu} - E_{00} \big) 
   \left[  \mathring{\Psi}_{\frac{3}{2} M}^{({\rm C})}(dt\mu)  
   + \Psi_{\frac{3}{2} M}^{({\rm N})}(dt\mu) \right] \qquad & \nonumber \\ 
    \qquad \quad = -V^{({\rm T})}_{dt, \alpha n} \:
\label{first equation}     
  \Psi_{\frac{3}{2} M}^{(+)}(\alpha n \mu), & \\
    \big( H_{\alpha n \mu} - (E_{00} + Q) \big)
             \, \Psi_{\frac{3}{2} M}^{(+)}(\alpha n \mu) 
\qquad \qquad \;\;  & \nonumber \\
\quad \quad \qquad \qquad   = -V^{({\rm T})}_{\alpha n, dt} 
     \:\left[\mathring{\Psi}_{\frac{3}{2} M}^{({\rm C})}(dt\mu) + 
     \Psi_{\frac{3}{2} M}^{({\rm N})}(dt\mu) \right], \!\!\!\!\!\!\!\! 
\label{second equation}     
\end{align}
\end{subequations}
where
\begin{eqnarray}
\label{eq:hamil-dtm}  
\!\!\!\!  H_{dt\mu}&=& T_{{\bf r}_c} + T_{{\bf R}_c} 
           + V_{t \mu}^{({\rm C})}(r_1) + V_{d \mu}^{({\rm C})}(r_2) \nonumber \\ 
             &+& V_{dt}^{({\rm C})}(r_3) + V_{dt}^{({\rm N})}(r_3) ,
                  \;\;   (c=1,2 \: \mbox{or}\: 3),   
 \\
\label{eq:ham-anm}
\!\!\!\!  H_{\alpha n \mu}&=& T_{{\bf r}_4} + T_{{\bf R}_4} 
           + V_{\alpha n}^{({\rm N})}(r_4) + V_{\alpha \mu}^{({\rm C})}(r_5), \\ 
\label{eq:tensor-nonloc-34}
\!\!\!\!  V^{({\rm T})}_{dt, \alpha n}&=& \int d{\bf r}_4
           V_{dt, \alpha n}^{({\rm T})}({\bf r}_3, {\bf r}_4). 
\end{eqnarray}
{
An outline of the manner in which the three-body coupled-channels 
Schr\"{o}dinger equation~(\ref{eq:schr-eq})  
is solved is provided based on the following senario presented in i) to iv): 

i)
It is to be noted that  
the  kinetic energy (17.6 MeV) of the \mbox{$\alpha$-$n$}
relative motion is 
much larger than the potential energy (in the order of 10 keV)
of $V_{\alpha \mu}^{({\rm C})}(r_5)${, which can be neglected};
the muon is located nearly around the $({\rm He}\mu)_{1s}$ orbital
with the Bohr radius $\sim\!$ 130 fm
when the fusion takes place in the $dt\mu$ molecule.
Therefore, when solving the Schr\"{o}dinger equation~(\ref{eq:schr-eq}),
the Coulomb potential $V_{\alpha \mu}^{({\rm C})}(r_5)$
{is} {\it omitted} from $H_{\alpha n \mu}$
(\ref{eq:ham-anm}) for the $\alpha n \mu$ channel. 
The contribution of the potential $V_{\alpha \mu}^{({\rm C})}(r_5)$
to the fusion rate $\lambda_{\rm f}$ {is} afterwards  
estimated by calculating the related
$T$-matrix elements, which {is} then  added 
to $\lambda_{\rm f}$ as a correction. 
A similar methodology can be followed for the contribution to
the energy and momentum spectra of the emitted muons.
Those contributions {are} found to be small (cf. Sec.~VI).

ii) {First,}  Eq.~(3.7a) 
{is} solved with {\it switching off} the coupling to the
$\alpha n \mu$ channel and treating 
$\mathring{\Phi}_{0}^{({\rm C})}({\bf r}_3, {\bf R}_3)$ as the given 
source term.    
The resulting nuclear-correlated amplitude, say 
${\widehat \Phi}_{0}^{({\rm N})}({\bf r}_3, {\bf R}_3)$ (here, instead of
$\Phi_{0}^{({\rm N})}({\bf r}_3, {\bf R}_3)$
in Eq.~(\ref{eq:wf-spin-2})), {is} found to be very well separated 
in the form
(cf. Eqs.~(\ref{eq:Phi(N)-hat-eq})-(\ref{eq:Psi-separation-amb2}) and 
Figs.~\ref{fig:psi-1s-1} and \ref{fig:phi4-3cases})
\begin{equation}
{\widehat \Phi}_{0}^{({\rm N})}({\bf r}_3, {\bf R}_3)=
{\widehat \varphi}_0^{\,({\rm N})}(r_3) \, \psi_0^{({\rm N})}(R_3) .
\label{eq:Phi(N)hat-separation}
\end{equation}
This separation can be attributed to the fact that the nuclear \mbox{$d$-$t$} 
interaction is very short ranged, whereas the $d$-$\mu$ and \mbox{$t$-$\mu$}
Coulomb potentials are quite long ranged. Further,
the muon is located far away approximately in the
$1s$ orbital  around the $^5$He nucleus when the nuclear fusion takes place.
Therefore, the Coulomb potentials {do} not affect the
nuclear part of the \mbox{$d$-$t$} motion.
Moreover, as $\int |{\widehat \Phi}_{0}^{({\rm N})}({\bf r}_3, {\bf R}_3)|^2
{\rm d}{\bf r}_3 {\rm d}{\bf R}_3 \ll 
\int |\mathring{\Phi}_{0}^{({\rm C})}({\bf r}_3, {\bf R}_3)|^2
{\rm d}{\bf r}_3 {\rm d}{\bf R}_3=1 $ (cf. Fig.~\ref{fig:density-dtmu-rs3}),
the renormalization of  
$\mathring{\Phi}_{0}^{({\rm C})}({\bf r}_3, {\bf R}_3)$
due to ${\widehat \Phi}_{0}^{({\rm N})}({\bf r}_3, {\bf R}_3)$ 
does not need to  be considered.

iii) The short-range nuclear interaction
$V_{\alpha n, dt}^{({\rm T})}({\bf r}_4, {\bf r}_3)$ to couple  
$\Phi_{0}^{({\rm N})}({\bf r}_3, {\bf R}_3)$ and 
$\Phi_2^{(+)}({\bf r}_4, {\bf R}_4)$
does not affect  the muon motion $\psi_0^{({\rm N})}(R_3)$, 
and  the $\alpha$-$\mu$ Coulomb potential {is} omitted.

Therefore, it is trivial that, after the $dt$-$\alpha n$ coupling {is} switched on,
${\widehat \Phi}_{0}^{({\rm N})}({\bf r}_3, {\bf R}_3)$ {changes} into
$\Phi_{0}^{({\rm N})}({\bf r}_3, {\bf R}_3)$ taking the form
\begin{equation}
\Phi_{0}^{({\rm N})}({\bf r}_3, {\bf R}_3)= 
\varphi_0^{({\rm N})}(r_3) \, \psi_0^{({\rm N})}(R_3),
\label{eq:Phi(N)-separation}
\end{equation}
and the $\alpha n \mu$ outgoing wave function
{is generated} in the form \mbox{(note ${\bf R}_4={\bf R}_3$)}
\begin{equation}
\Phi_{2m}^{(+)}({\bf r}_4, {\bf R}_4)
= \varphi^{(+)}_{2m}(k_4,{\bf r}_4) \, \psi_0^{({\rm N})}(R_4). 
\label{eq:Phi(+)-separation}
\end{equation}

iv) The unknown functions $\varphi_0^{({\rm N})}(r_3)$ and 
$\varphi^{(+)}_{2m}(k_4,{\bf r}_4) $ {are} determined by solving
the Schr\"{o}dinger equation~(\ref{eq:schr-eq}) 
\mbox{without} $V_{\alpha \mu}^{({\rm C})}(r_5)$
in a straightforward manner. 
The fusion rate of the $dt\mu$ molecule {is} derived from the
asymptotic behavior of $\varphi_{2m}^{(+)}(k_4,{\bf r}_4)$
(cf. Eq.~(\ref{eq:outgoing})).
}

\vskip -0.2cm
\subsection{Structure of 
$\mathring{\Psi}_{\frac{3}{2} M}^{({\rm C})}(dt\mu)$, 
$\Psi_{\frac{3}{2} M}^{({\rm N})}(dt\mu)$, and
$\Phi_{\frac{3}{2} M}^{(+)}(\alpha n \mu)$}

\vskip -0.2cm
First, we calculate the initial-state ($dt\mu$) wave function 
$\mathring{\Phi}_{0}^{({\rm C})}({\bf r}_3, {\bf R}_3)$ 
of  Eq.~(\ref{eq:wf-spin-1}) and its energy $E_{00}$ 
by solving  the Coulomb-three-body Schr\"{o}dinger equation
\begin{eqnarray}
( H_{dt\mu}^0 - E_{00} )\, \mathring{\Phi}_0^{({\rm C})}
({\bf r}_3, {\bf R}_3)=0 ,  
\label{eq:sch-dtm-00}
\end{eqnarray}
where $H_{dt\mu}^0$ is given by Eq.~(\ref{eq:hamil-dtm})
omitting  $V_{dt}^{({\rm N})}(r_3)$.
Since, we study fusion decay starting from
one $dt\mu$ molecule, we normalize 
$\mathring{\Phi}_0^{({\rm C})}$ as 
{
$\langle \mathring{\Phi}_0^{({\rm C})} \,|\, 
\mathring{\Phi}_0^{({\rm C})} \rangle = 1.$
}

We solve Eq.~(\ref{eq:sch-dtm-00}) 
by using the Gaussian expansion method (GEM). 
This method was proposed by one of the present 
authors (M.K.) to accurately solve the $dt\mu$ molecule~\cite{Kamimura88}
and has been applied to variou few-body systems (cf. its review 
papers~\cite{Hiyama03,Hiyama09,Hiyama12FEW,Hiyama18}). 

The Coulomb three-body wave function 
$\mathring{\Phi}_0^{({\rm C})}({\bf r}_3, {\bf R}_3)$ is
described as a sum of the amplitudes of three rearrangement channels
$c=1, 2$ and $3$ (cf. Fig.~\ref{fig:jacobi-dtmu}):
\begin{equation} 
\mathring{\Phi}_0^{({\rm C})}({\bf r}_3, {\bf R}_3)
\!=\!\Phi_{IM}^{(1)}({\bf r}_1, {\bf R}_1)
+\Phi_{IM}^{(2)}({\bf r}_2, {\bf R}_2)
+\Phi_{IM}^{(3)}({\bf r}_3, {\bf R}_3). 
\label{eq:psi3}  
\end{equation} 

Each amplitude is expanded in terms of 
the Gaussian basis functions
of the Jacobi coordinates ${\bf r}_c$ and ${\bf R}_c$  $(c=1-3)$,:
\begin{eqnarray}
&&\!\!\!\!\!\!\!\!\!\! \Phi_{IM}^{(c)}({\bf r}_c, {\bf R}_c)  
=\!\!\!\!\! \sum_{n_cl_c,N_cL_c} \!\!\!\!\!\! A^{(c)}_{n_cl_c,N_cL_c} 
\big[\phi_{n_cl_c}({\bf r}_c)\:
\psi_{N_cL_c}({\bf R}_c)
\big]_{IM},  \\ 
\label{eq:gauss-expansion-1}
%
%
\label{eq:3gauss-1}
&&\!\!\!\!\!\!\!\!\!\!  \phi_{nlm}({\bf r}) =
 N_{nl}\,r^l\:e^{-\nu_n r^2}\:Y_{lm}({\widehat {\bf r}}) , \quad (n=1-n_{\rm max}) \\
\label{eq:3gaussa}
&&\!\!\!\!\!\!\!\!\!\!  \psi_{NLM}({\bf R}) = 
 N_{NL}\,R^L\:e^{-\lambda_N R^2}\:Y_{LM}({\widehat {\bf R}}),
\;\; (N\!=\!1\!-\!N_{\rm max}) \; 
\label{eq:3gauss}
\end{eqnarray}
where $N_{nl}$ and $N_{NL}$ are the normalization constants.
The Gaussian ranges are postulated to lie in  geometric progression:
\begin{eqnarray}
\label{eq:prog-r}
&& \!\!\!\!  \!\!\!\! \!\!\!\! \nu_n=1/r_n^2, \quad \;\; r_n=r_1\, a^{n-1},\; 
(n=1-n_{\rm max}), \\
&& \!\!\!\! \!\!\!\! \!\!\!\! \lambda_N\!=\!1/R_N^2, 
\quad R_N\!=\!R_1\, A^{N-1},\; 
(N\!=\!1\!-\!N_{\rm max}).
\label{eq:progR}
\end{eqnarray}
 $l_{c}$ and $L_{c}$ are restricted 
to $0 \leq l_{c} \leq l_{{\rm max}}$ and
$| I-l_{c} | \leq L_{c} \leq I+l_{c}$.

Eigenenergy $E_{00}$ and coefficients $A^{(c)}_{n_cl_c,N_cL_c}$
are determined by the Rayleigh-Ritz variational principle.
In the precise calculation~\cite{Kamimura88} of the eigenenergies of
the $(dt\mu)_{J,v}$ molecule,
we took $l_{{\rm max}}=4$, but $l_{{\rm max}}=1$ is sufficient
in the present fusion reaction problem.
The Gaussian-basis parameters employed are the same as those
in Ref.~\cite{Kamimura88}.
Although the large-size calculation~\cite{Kamimura88} 
of the $J\!=\!v\!=\!0$ with $l_{{\rm max}}=4$
gave an eigenenergy of  $-319.14$ eV 
with respect to the $(t\mu)_{1s}+d$
threshold, the case of $l_{{\rm max}}=1$ results in 
$-319.12$ eV, which is sufficient
in the present reaction calculations. 
We have $E_{00}=-0.003030 \; \mbox{MeV}$as mentioned in 
Eq.~(\ref{eq:E00-0303}) (cf. Fig.~\ref{fig:gainen-zu}).

We then derive the nuclear-part amplitude
in Eq.~(\ref{eq:Phi(N)hat-separation}) for the case of 
the $dt\mu$-$\alpha n\mu$ coupling switched off;
namely, we solve the linear equation 
\begin{eqnarray}
\!\!\!\!\!\!( H_{dt\mu} - E_{00} )\, 
{\widehat \Phi}_{0}^{({\rm N})}({\bf r}_3, {\bf R}_3)
=-( H_{dt\mu} - E_{00} )\, 
\mathring{\Phi}_{0}^{({\rm C})}({\bf r}_3, {\bf R}_3),\:
\label{eq:Phi(N)-hat-eq}
\end{eqnarray}
where $E_{00}$ and the source term 
$\mathring{\Phi}_0^{({\rm C})}({\bf r}_3, {\bf R}_3)$  
were  given above.
$\widehat{\Phi}_0^{({\rm N})}({\bf r}_3, {\bf R}_3)$ is expanded 
in terms of the Gaussian basis functions of the $c=3$ channel  
as  in  Eqs.~(\ref{eq:3gauss-1})-(\ref{eq:progR}); namely,

\noindent
{
the Gaussian-range parameters with $l=\!L\!=0$ lie 
in the geometric progressions of 
$(n_{\rm max}, r_1, r_{n_{\rm max}})=(20, 0.5\, {\rm fm}, 20\, {\rm fm})$
and $(N_{\rm max}, R_1, R_{N_{\rm max}})
=(15, 10\, {\rm fm}, 1500 \,{\rm fm})$,
}
which are suitable for correlating with the 
nuclear interactions. 
The bases with $l=L>0$ 
are not employed for the nuclear-interaction region 
since they have negligible contributions.

\begin{figure}[t!]
\begin{center}
\epsfig{file=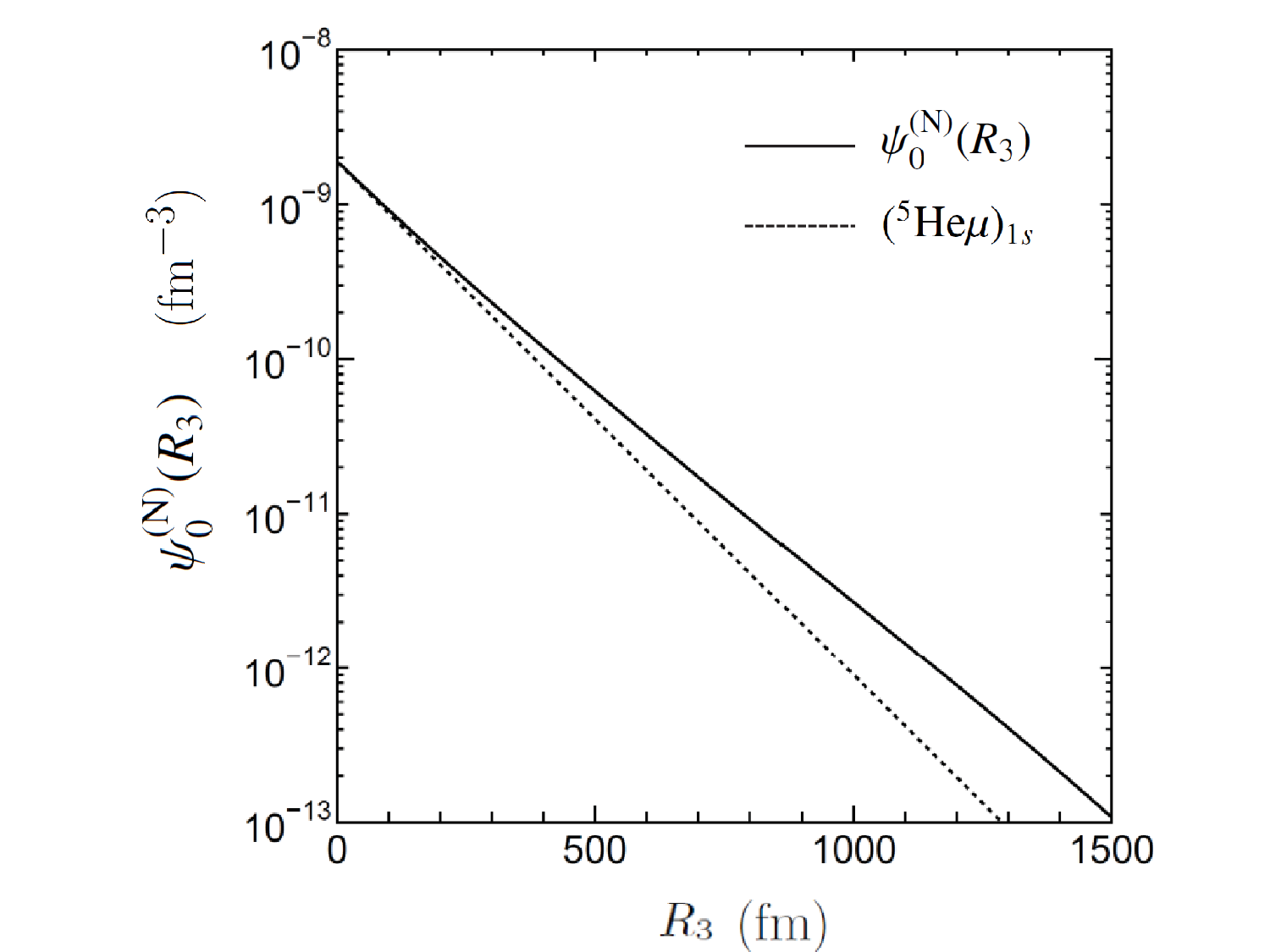,width=8.0cm,height=6.0cm}
\end{center}
\vskip -0.2cm
\caption{Amplitude  {$\psi_0^{({\rm N})}(R_3)$} of the $(dt)$-$\mu$ 
relative motion in the nuclear-correlated term
{${\widehat \Phi}_{0}^{({\rm N})}({\bf r}_3, {\bf R}_3)$} 
is represented by the solid curve, whereas the dotted curve represents the
wave function of the $(^5{\rm He}\mu)_{1s}$ atom, 
$\propto e^{-R_3/a_0}$ with $a_0=131$ fm, 
normalized to {$\psi_0^{({\rm N})}(R_3)$} at $R_3=0$ for comparison.
The solid curve is well simulated by the function 
$\propto e^{-R_3/a}$ with $a=154$ fm; this will be used 
in Eqs.~(\ref{eq:rNK-simulate}) and (\ref{eq:rNE-simulate}).
Here, the nuclear interactions of Set B is employed.
}
\label{fig:psi-1s-1}
\end{figure}

\begin{figure}[h!]
\begin{center}
\epsfig{file=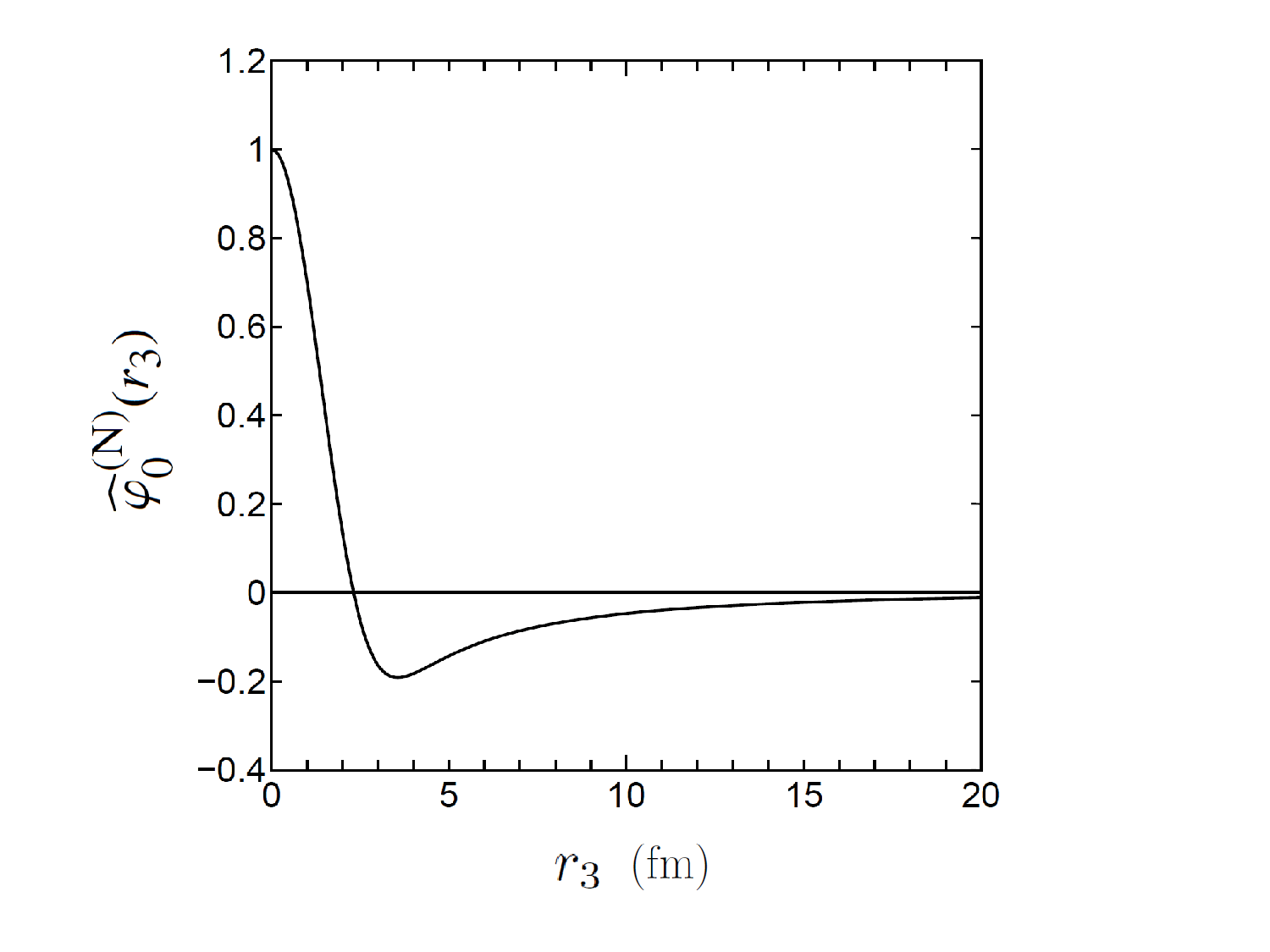,width=8.8cm,height=6.8cm}
\end{center}
\vskip -0.6cm
\caption{
Amplitude of  the $d$-$t$ relative motion 
{${\widehat \varphi}_0^{\,({\rm N})}(r_3)$}
in the nuclear-part wave function
{${\widehat \Phi}_{0}^{({\rm N})}({\bf r}_3, {\bf R}_3)$} 
without the coupling of the $\alpha n \mu$ channel 
is illustrated by the solid curve.
The curves have one node at $r_3 \sim 2$ fm due to the orthogonality to the
Pauli-forbidden (spurious) $0s$ wave function for the $^5$He nucleus 
of the $d$-$t$ potential model.
Here, the nuclear interactions of Set B is employed.
\label{fig:phi4-3cases}
}
\end{figure}

{

The resulting  
${\widehat \Phi}_0^{({\rm N})}({\bf r}_3, {\bf R}_3)$,
is found to be  well expressed in the separated form of 
Eq.~(\ref{eq:Phi(N)hat-separation}) with
\begin{eqnarray}
\psi_0^{({\rm N})}(R_3)&=& 
{\widehat \Phi}_0^{({\rm N})}(0, {\bf R}_3), 
\label{eq:Psi-separation-amb1}  \\
{\widehat \varphi}_0^{\,({\rm N})}(r_3) &=&
 {\widehat \Phi}_0^{({\rm N})}({\bf r}_3, 0)
         /{\widehat \Phi}_0^{({\rm N})}(0,0).
\label{eq:Psi-separation-amb2}
\end{eqnarray}
As the separation ambiguity  
$[\gamma {\widehat \varphi}_0^{\,({\rm N})}(r_3)][\frac{1}{\gamma} 
\psi_0^{({\rm N})}(R_3)]$ does not affect the expression 
(\ref{eq:lambda-muon-out}) for
$\lambda_{\rm f}$, $\gamma=1$ is taken here.

$\psi_0^{({\rm N})}(R_4)$ and ${\widehat \varphi}_0^{\,({\rm N})}(r_3)$
} 
are illustrated by the solid curves
in Figs.~\ref{fig:psi-1s-1} and \ref{fig:phi4-3cases}, respectively. 
The dotted curve in Fig.~\ref{fig:psi-1s-1} is the wave function of
the $(^5{\rm He}\mu)_{1s}$ atom, normalized to 
{
$\psi_0^{({\rm N})}(R_3)$ 
}  
at $R_3=0$ for comparison.
The less-steep slope of the solid curve is due to that
the charge density of the \mbox{$d$-$t$} pair \mbox{along ${\bf r}_3$}
spreads up to $r_3 \sim 20$ fm as shown in Fig.~\ref{fig:phi4-3cases}. 
The r.m.s. radius $\langle R_3^2\, \rangle^{1/2}$ of 
the solid- and dotted-curve wave functions in Fig.~\ref{fig:psi-1s-1}  are 
\mbox{260 and 227 fm}, respectively.

%

Thus, we have reached a step closer to solving  
the coupled-channel Schr\"{o}dinger equation
(\ref{eq:schr-eq}) with the Coulomb force $V_{\alpha \mu}^{({\rm C})}(r_5)$ 
omitted.  As shown in Eqs.~(\ref{eq:Phi(N)-separation}) and (\ref{eq:Phi(+)-separation}),
the problem is deduced to solve the unknown functions
$\varphi_0^{({\rm N})}(r_3)$ and 
$\varphi_{2m}^{(+)}(k_4, {\bf r}_4)$. 
We expand 
$\varphi_0^{({\rm N})}(r_3)$ in terms of the Gaussian basis functions
in Eq.~(\ref{eq:3gauss-1}) with the parameters 
$(n_{\rm max}, r_1, r_{n_{\rm max}})=(20, 0.5\, {\rm fm}, 20\, {\rm fm})$
as used in the case of ${\widehat \varphi}_0^{\,({\rm N})}(r_3)$ 
in Eq.~(\ref{eq:Psi-separation-amb2}).

We rewrite $\varphi_{2m}^{(+)}(k_4, {\bf r}_4)$ as
\begin{eqnarray}
\varphi^{(+)}_{2m} (k_4, {\bf r}_4)= \frac{u^{(+)}_2 (k_4, r_4)}{r_4} \,
Y_{2m}({\widehat {\bf r}}_4)
\end{eqnarray}
and impose the outgoing boundary condition as
\begin{eqnarray}
    u^{(+)}_{2}(k_4, r_4)   
      \stackrel{r_4 \to \infty}{\longrightarrow} 
     \: S_2 \,e^{i k_4  r_4} . 
\label{eq:outgoing}
\end{eqnarray}
The amplitude $S_2$ of the outgoing wave~(\ref{eq:outgoing}) 
is slightly different 
from the usual definition of $S$-matrix in the scattering 
with an  {\it incoming} wave. 
In the present case, the dimension of $S_2$ is $L^1$
owing to the dimension $L^{-3}$ of the initial $(dt\mu)$ bound state 
$\mathring{\Phi}_{\frac{3}{2} M}^{({\rm C})}(dt\mu)$ 
in Eq.~(\ref{eq:total-wf}).  

The flux of the
$\alpha$-$n$ relative motion at $r_4 \to \infty$ 
into the full direction ($4\pi$ sr)
in a unit time is given as 
(note ${\widehat {\bf r}}_4={\widehat {\bf k}}_4$ )
\begin{eqnarray}
v_4 N_\mu \int \left|  S_2 \big[ Y_2({\widehat {\bf k}_4}) 
           \chi_{\frac{1}{2}}(\alpha n) \big]_{\frac{3}{2}} \right|^2 
     {\rm d}{\widehat {\bf k}}_4\, 
     = v_4 N_\mu |S_2|^2   \:,   \quad 
\label{eq:rate-definition}
\end{eqnarray} 
where $v_4= \hbar k /\mu_{{\bf r}_4}$ 
is the velocity of the $\alpha$-$n$ relative motion, 
and $N_\mu$ is given by
\begin{eqnarray}
    N_\mu = \int |\, \psi_0^{({\rm N})}(R_4)\, |^2 \: {\rm d}{\bf R}_4.
\label{eq:N_mu}
\end{eqnarray}
The R.H.S. of Eq.~(\ref{eq:rate-definition}) gives the fusion rate, 
namely, the number of $\alpha$-$n$ pair (number of muon)
outgoing from {\it one} $(dt\mu)_{J=v=0}$ molecule:
\begin{equation}
    \lambda_{\rm f} = v_4 N_\mu |S_2 |^2  \,  . \quad 
\label{eq:lambda-muon-out}
\end{equation}

\subsection{Results}

According to the above procedure, 
the coupled-channels Schr\"{o}dinger 
equation~(\ref{eq:schr-eq}) 
with the outgoing boundary condition~(\ref{eq:outgoing})  
can be precisely solved  by using the couple-channels 
Kohn-type variational method  for composite-particle
scattering~\cite{Kamimura77}, which was  used in Sec.~II.

The $S$-matrix $S_2$ in Eq.~(\ref{eq:outgoing}) is obtained as 
\begin{eqnarray}
S_2= (  0.5934 +  0.1985\, i) \: {\rm fm}.
\end{eqnarray}
Therefore,  using 
$|S_2|^2=3.915 \times 10^{-1} \,{\rm fm}^2$,
$v_4=6.50 \times 10^{22} \,{\rm fm \, s}^{-1}$ and
$N_\mu=4.045 \times 10^{-11} {\rm fm}^{-3}$ in Eq.~(\ref{eq:N_mu}),
\mbox{the calculated} fusion rate is given as
\begin{equation}
    \lambda_{\rm f} = 1.03 \times 10^{12} \: {\rm s}^{-1}.
\label{eq:lambda-rate}
\end{equation}
{
This result is obtained 
by  using the nuclear interaction \mbox{Set B.}
Table~II lists the fusion rates
for three Sets A, B, and C; the rates imply 
that the present fusion-rate calculation is independent of the
details of the employed nuclear interactions that
reproduced the observed data in \mbox{Figs.~\ref{fig:cal-cc-dt-an}
and \ref{fig:sigma-an}.}
Therefore, only Set B is used in \mbox{Secs.~V and VI.}

A correction to $\lambda_{\rm f}$ owing to the 
\mbox{$\alpha$-$\mu$} Coulomb potential, which is omitted when solving 
the Schr\"{o}dinger equation, is discussed  in Sec.~VI\,A.

}
\begin{table}[h!]
{
\caption{
Calculated fusion rate $\lambda_{\rm f}$ of  $(dt\mu)_{J=v=0}$, 
Eq.~(\ref{eq:lambda-muon-out}), 
using the nuclear-interaction parameter sets A, B, and C in 
Table~I. 
}
\begin{center}
\begin{tabular}{ccccccc} 
\hline \hline
\noalign{\vskip 0.1 true cm} 
  & $\quad$ &   Set A &$\quad$ &  Set B &$\quad$ &  Set C   \\
\noalign{\vskip 0.1 true cm} 
\hline
\noalign{\vskip 0.15 true cm} 
$\lambda_{\rm f}$  (s$^{-1}$) &  &  $1.03 \times 10^{12}$ 
& &   $1.03 \times 10^{12}$   & &   $1.02 \times 10^{12}$    \\
\noalign{\vskip 0.2 true cm} 
\hline
\hline
\end{tabular}
\end{center}
}
\label{table:fusion-rate-setABC}
\end{table}

{
The above value of $\lambda_{\rm f}$ in (\ref{eq:lambda-rate}) supports 
the literature results $\mbox{(1.0 - 1.3)} \times 10^{12} {\rm s}^{-1}$
(cf. Table 8 in the $\mu$CF review paper~\cite{Froelich92})
obtained  in 1980's - 90's by using the $d$-$t$ optical-potential 
model~\cite{Kamimura89,
Bogdanova,Bogdanova89} and by the $R$-matrix method~\cite{Struensee88a,
Szalewicz90,Hale93,Hu1994,Cohen1996,Jeziorski91}.   

It is found  that $\varphi_0^{({\rm N})}(r_3)$ is well simulated by
\begin{equation}
\varphi_0^{({\rm N})}(r_3) \simeq ( 1.67 + 0.59\,i )\:
{\widehat \varphi}_0^{\,({\rm N})}(r_3).
\label{eq:enhancement}
\end{equation}
}
This enhancement in Eq.~(\ref{eq:enhancement}) by the
$dt\mu$-$\alpha n \mu$ coupling
will play an important 
role in the analysis in Secs.~V and VI with the use of the
$T$-matrix calculational method based on the Lippmann-Schwinger equation
to be introduced in Sec.~IV.

\begin{figure}[t!]
\begin{center}
\epsfig{file=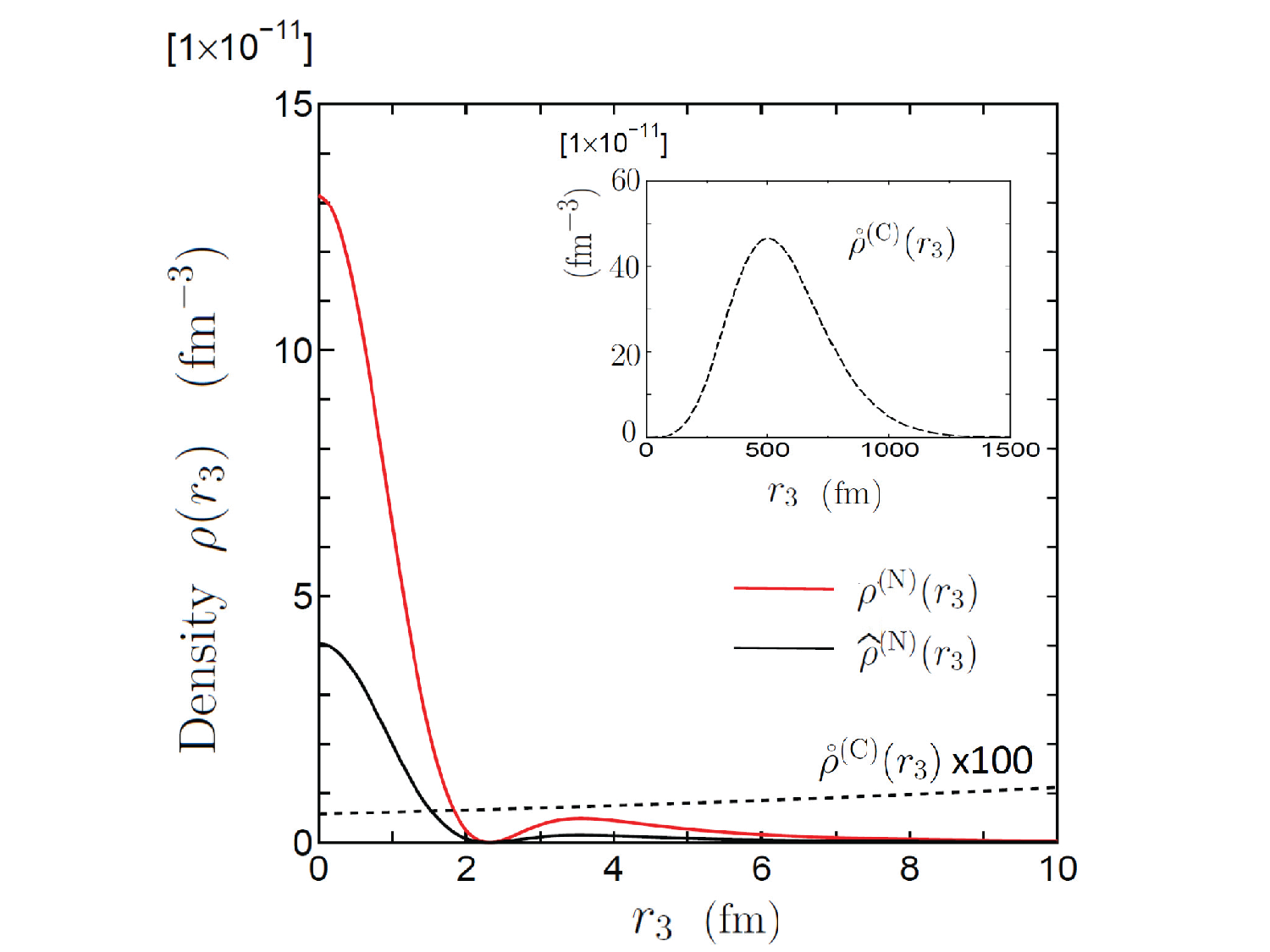,width=8.0cm,height=6.0cm}
\end{center}
\hskip -1.5cm
\vskip -0.5cm
\caption{
Three types of the densities of the $d$-$t$ relative motion 
associated with ${\bf r}_3$,  which are defined in 
Eqs.~(\ref{eq:density-1})-(\ref{eq:density-3}).
The dashed curve illutrates ${\mathring \rho}^{\rm (C)}(r_3)$
multiplied by 100 for $r_3 < 10$ fm, whereas
the entire behavior of ${\mathring \rho}^{\rm (C)}(r_3)$
is given in the inserted figure.
The red curve shows $\rho^{\rm (N)}(r_3)$ 
for the nuclear part of the $d$-$t$ relative motion 
when the $dt\mu$-$\alpha n \mu$ coupling is switched on.
The black solid curve for ${\widehat \rho}^{\rm (N)}(r_3)$ 
represents the case for which the $dt\mu$-$\alpha n \mu$ 
coupling is switched off. 
{
We have 
$\int {\mathring \rho}^{\rm (C)}(r_3) \,{\rm d}{\bf r}_3 =1$
($\langle \mathring{\Phi}_0^{({\rm C})} \,|\, 
\mathring{\Phi}_0^{({\rm C})} \rangle = 1$),
$\int \rho^{\rm (N)}(r_3) \,{\rm d}{\bf r}_3 
= 7.5 \times  10^{-9}$
and 
$\int{\widehat \rho}^{\rm (N)}(r_3)\,{\rm d}{\bf r}_3 
= 2.4 \times  10^{-9}$.
Here, the nuclear interactions of Set B is used.
}  
}
\label{fig:density-dtmu-rs3}
\end{figure}

Figure~\ref{fig:density-dtmu-rs3} illustrates the three types of densities
of the $d$-$t$ relative motion along ${\bf r}_3$;
\begin{eqnarray}
\label{eq:density-1}
\rho^{\rm (N)}(r_3)\!\!&=&\!\!\! \int \big| \Phi_0^{({\rm N})}
                    ({\bf r}_3, {\bf R}_3)\big|^2 \,{\rm d}{\bf R}_3 ,  \\
\label{eq:density-2}
{\widehat \rho}^{\rm (N)}(r_3)\!\!&=&\!\!\! \int 
                    \big|{\widehat \Phi}_0^{({\rm N})}
                    ({\bf r}_3, {\bf R}_3) \big|^2 \,{\rm d}{\bf R}_3 
           \\
{\mathring \rho}^{\rm (C)}(r_3)\!\!&=&\!\!\! \int 
                     \big|{\mathring \Phi}_0^{({\rm C})}
                    ({\bf r}_3, {\bf R}_3)\big|^2 \,{\rm d}{\bf R}_3 ,  
\label{eq:density-3}
\end{eqnarray}
where  the integration
over the muon-coordinate ${\bf R}_3$ is performed.
In Fig.~\ref{fig:density-dtmu-rs3}, 
${\mathring \rho}^{\rm (C)}(r_3)$ multiplied by 100 is illustrated
by the dashed curve for $r_3 < 10$ fm, whereas
the \mbox{inserted} \mbox{figure} shows  ${\mathring \rho}^{\rm (C)}(r_3)$ 
for the entire region.
The red curve gives $\rho^{\rm (N)}(r_3)$ 
for the $d$-$t$ relative motion when all the nuclear interactions
are employed.
It should be emphasized that ${\mathring \rho}^{\rm (C)}(r_3)$ is 
significantly smaller
than $\rho^{\rm (N)}(r_3)$ in the nuclear interaction region.
Therefore, ${\mathring \Phi}_0^{({\rm C})}({\bf r}_3, {\bf R}_3)$
is expected to play a minor role compared to  
$\Phi_0^{({\rm N})}({\bf r}_3, {\bf R}_3)$ 
in the estimation of the  $\alpha$-$\mu$ sticking and 
the ejected muon's spectrum after the fusion;
this will be discussed in detail in Secs. V and VI.

\vspace{0.0cm}
\section{Use of Lippmann-Schwinger 
\vskip 0.05cm
equation for 
\mbox{\boldmath {\large $\lowercase{d} \lowercase{t} \mu$}}
fusion reaction}

\subsection{$T$-matrix element}

In this section, we propose the use of the Lippmann-Schwinger 
equation~\cite{Lippmann}
as another method for studying the fusion reaction (\ref{eq:mucf-reaction}), 
in particular, to calculate  the initial sticking of 
a muon to an $\alpha$ particle 
and the momentum and energy spectra of the emitted muon,
respectively in Secs.~V and~VI.   

Let's suppose that the reaction from a plane-wave 
initial \mbox{$\alpha$-channel} 
is outgoing to the final $\beta$-channels assuming the following wave functions: 
\begin{eqnarray}
&& \mbox{Total wave function} : \Psi^{(+)}_\alpha (E_\alpha), \nonumber \\
  && \mbox{Initial $\alpha$-channel wave function} :  
         e^{i\, {\bf K}_\alpha\cdot{\bf R}_\alpha} 
        \phi_\alpha(\xi_{\alpha}) \nonumber ,  \\
   && \mbox{Asymptotic form of final $\beta$-channel wave function} : \nonumber \\ 
   &&    \quad  f_{\beta \alpha}(\Omega_\beta)  
         \frac{e^{i\,K_\beta  R_\beta}}{R_\beta}\, \nonumber 
        \phi_\beta(\xi_{\beta}), \nonumber  
\end{eqnarray}
where $\phi_\alpha(\xi_{\alpha})$ and 
$\phi_\beta(\xi_{\beta})$ are
ortho-normalized intrinsic wave functions of 
the $\alpha$- and $\beta$-channels, respectively,
and $f_{\beta \alpha}(\Omega_\beta)$ is the scattering amplitude
to be determined.

One can calculate the transition matrix elements $T_{\beta \alpha}$
from the well-known integral formula~\cite{Lippmann,Gell-Mann}
based on the Lippmann-Schwinger equation as
\begin{eqnarray}
T_{\beta \alpha} = \langle \,e^{i\, {\bf K}_\beta\cdot{\bf R}_\beta} 
        \phi_\beta(\xi_{\beta}) 
      \,|\, V_\beta \,|\,\Psi^{(+)}_\alpha (E_\alpha) \,\rangle , 
\label{eq:T-mat}
\end{eqnarray}
where $V_\beta$ is the interaction in the $\beta$-channel.
Using this $T_{\beta \alpha}$,
the $\beta$-channel asymptotic form of the total wave function 
$\Psi^{(+)}_\alpha (E_\alpha)$  is written as 
\begin{eqnarray}
 \langle \, \phi_{\beta}(\xi_\beta) \,|\,
      \Psi^{(+)}_{JM}(E) \,\rangle 
         \stackrel{R_\beta \to \infty}{\longrightarrow} 
      -\frac{\mu_{R_\beta}}{2 \pi \hbar^2}\,
           T_{\beta \alpha}\, \frac{e^{i K_\beta R_\beta}}{R_\beta} ,
\end{eqnarray}    
where $\mu_{R_\beta}$ is the reduced mass associated with ${\bf R}_\beta$.
Therefore, we have 
\begin{equation}
f_{\beta \alpha}(\Omega_\beta) =
-\frac{\mu_{R_\beta}}{2 \pi \hbar^2}\, T_{\beta \alpha} \,.
\end{equation}

The reaction cross section $\sigma_{\beta \alpha}$ is usually defined by
the flux of the outgoing wave with a velocity 
$v_\beta (=\hbar K_\beta/\mu_{{\bf R}_\beta})$
into the full direction ($4\pi$ st)
in a unit time divided by the flux of the incident wave $(= v_\alpha)$ 
as 
\begin{eqnarray}
\!\!\!   \sigma_{\beta \alpha}=
     \frac{v_\beta}{v_\alpha}
    \!  \int |f_{\beta \alpha}(\Omega_\beta)|^2 \, 
          {\rm d}\Omega_\beta 
     = \frac{v_\beta}{v_\alpha}
      \left( \frac{\mu_{R_\beta}}{2 \pi \hbar^2} \right)^2
  \!\!   \int  |T_{\beta \alpha}|^2 \,{\rm d}{\widehat {\bf K}}_\beta  .  
\label{eq:sigma-a-b}
\end{eqnarray}
The preceding expressions are exact provided that the total wave function
$\Psi^{(+)}_\alpha (E_\alpha)$ is rigorously exact. 
However, for typical reaction calculations,
$\Psi^{(+)}_\alpha (E_\alpha)$ in Eq.~(\ref{eq:T-mat})
is replaced by an approximate wave function.

In the present fusion reaction (\ref{eq:mucf-reaction}), 
however, the initial \mbox{$\alpha$-channel} 
is not the plane wave $e^{i\, {\bf K}_\alpha \cdot {\bf R}_\alpha} 
        \phi_\alpha(\xi_{\alpha})$,
but the $dt\mu$ bound state 
${\mathring \Phi}_{\frac{3}{2} M}^{({\rm C})}(dt\mu)$
in Eq.~(\ref{eq:total-wf}). 
Therefore, in Eq.~(\ref{eq:sigma-a-b}),
we omit the incoming-channel information 
and replace  the total wave function 
by our $\Psi_{\frac{3}{2} M}^{(+)}(E)$  of Eq.~(\ref{eq:total-wf}),
and introduce the `reaction rate', $r_{\beta \alpha}$, as
\begin{eqnarray}
r_{\beta \alpha}= v_\beta
      \left( \frac{\mu_{R_\beta}}{2 \pi \hbar^2} \right)^2
     \int  |T_{\beta \alpha} |^2 \,{\rm d}{\widehat {\bf K}_\beta} \, .  
\label{eq:reaction-rate-def}
\end{eqnarray}  

Since the initial-state wave function  
${\mathring \Phi}_{\frac{3}{2} M}^{({\rm C})}(dt\mu)$ is normalized to unity
(namely, starting with one molecule),
$r_{\beta \alpha}$ represents the number (probability) of a molecule
decaying into the $\beta$ channel per unit time.
Therefore, the sum of $r_{\beta \alpha}$ over $\beta$ becomes 
{
the $T$-matrix expression of 
}
the fusion rate $\lambda_{\rm f}$:
\begin{eqnarray}
   \lambda_{\rm f} = \sum_\beta \; r_{\beta \alpha} \,.
\end{eqnarray}
Note that we call $r_{\beta \alpha}$ the reaction rate and
$\lambda_{\rm f}$ the fusion rate throughout this work.

The definition of reaction rate $r_{\beta \alpha}$ 
is applied to the study
of $\alpha$-$\mu$ sticking in Sec.~V
and to the momentum and energy spectra of the emitted muon in Sec.~VI.
In these applications,  
{
the fusion rates 
$\lambda_{\rm f}$  are calculated 
using $r_{\beta \alpha}$; 
these two types of additional 
calculations of the fusion rate
}
should be consistent with the value already obtained 
in Eq.~(\ref{eq:lambda-rate}), 
which will be a significant test  of the reliability of 
the present calculations.

%
\subsection{Discretization of two-body continuum states}

In the definition of the   $T$-matrix (\ref{eq:T-mat}), 
$\phi_\beta(\xi_{\beta})$ are treated as ortho-normalized
states. As the $\beta$-channel intrinsic states, however,
we will consider the $\alpha$-$\mu$ continuum states
associated with ${\bf r}_5$ (Sec.~V)
and the $\alpha$-$n$ continuum states
associated with ${\bf r}_4$ (Sec.~VI)

It is difficult to directly treat the continuum states   
in the $T$-matrix calculation.
Instead, we discretize these states and 
construct the ortho-normalized discretized continuum states, 
such as ${\widetilde \phi}_{ilm}({\bf r}_3)$ using $i$ to number  
the discretized states; we then consider their convergence 
back to the continuum. 
This discretization is performed by employing the CDCC 
(Continuum-Discretized-Coupled Channels) method that was developed
by one of the present authors (M.K.) and his collaborators
(for example, see review papers~\cite{Kamimura86,Austern,Yahiro12})
for the study of projectile-breakup reactions.
At present, this is one of the standard methods for investigating
various reactions using light- and heavy-ion projectiles.

The discretization of continuum states is performed as follows:
Let $\phi_{lm}(k,{\bf r})$ denote $k$-continuum states with 
angular-momentum $lm$ that satisfies the Schr\"{o}dinger equation
\begin{eqnarray}
\Big[ -\frac{\hbar^2}{2\mu_r} \nabla^2_r +V(r) 
    -\varepsilon \,\Big] \phi_{lm}(k,{\bf r})=0,
\quad \varepsilon = \frac{\hbar^{2}\,  k^2}{2\mu}. \quad
\end{eqnarray}
We confine the range of momentum
as  $ 0<k<k_N$ and divide it based on the interval
$\mathit{\Delta}k_i =k_{i-1} - k_i \:  (i=1,...,N)$;
usually, $\mathit{\Delta}k_i$ is taken to be independent of $i$.
We then take an average of the continuum wave functions in each 
momentum bin as
\begin{eqnarray}
{\widetilde \phi}_{i l m}({\bf r})
 =\frac{1}{\sqrt{\mathit{\Delta}k_i}}
\int_{k_{i-1}}^{k_{i}} \!\! \phi_{lm}(k,{\bf r})\, dk , \;
  (i=1,...,N)
\label{eq:bin}
\end{eqnarray}
where the integration is performed numerically with
the \mbox{required} accuracy.

Since $\phi_{lm}(k,{\bf r})$ is normalized as
\begin{eqnarray}
   \langle \, \phi_{lm}(k,{\bf r})\,|\, \phi_{lm}(k',{\bf r})
   \, \rangle    =\delta(k-k') , 
\end{eqnarray}
the discretized-continuum wave functions ${\widetilde \phi}_{i lm}({\bf r})$
have the ortho-normal relation 
\begin{eqnarray}
\langle\, {\widetilde \phi}_{i lm}({\bf r}) \,|\,  
          {\widetilde \phi}_{i' l m}({\bf r}) \rangle
=\delta_{i  i'} .
\end{eqnarray}
Namely, ${\widetilde \phi}_{i l m}({\bf r})$ becomes an $L^2$-integrable
function because of the cancelation between the asymptotic (oscillating)
amplitudes of $\phi_{lm}(k,{\bf r})$ during the $k$-integration.
A typical example of such a damping in the asymptotic region 
by averaging oscillating functions is as follows:

\begin{equation}
\int_{k_i}^{k_i+\mathit{\Delta}k_i}\!\!\! {\rm sin}\, kr\, dk
=\frac{2 \,{\rm sin} \frac{\mathit{\Delta}k_i}{2}r}{r}\, 
{\rm sin}\,(k_i\!+\!\frac{\mathit{\Delta}k_i}{2})r\,.
\end{equation}
Each $\widetilde{\phi}_{ilm}$ can be regarded as if it were a discrete 
excited-state wave function with energy $\widetilde{\varepsilon}_{i l}$
given by the expectation value of the Hamiltonian as
\begin{eqnarray}
  \widetilde{\varepsilon}_{i}=
\frac{\hbar^{2}\, {\widetilde k}_i^2}{2\mu}, \qquad
    {\widetilde k_i}^2=\Big(\frac{k_i+k_{i-1}}{2}\Big)^2 
+ \frac{\mathit{\Delta}k_i^2}{12}, \;
\label{eq:average-energy}
\end{eqnarray}

Convergence of the calculated results
with $\mathit{\Delta}k_i \to 0$ is well discussed
in review papers of the CDCC method~\cite{Kamimura86,Austern,Yahiro12}.
Therefore, we can treat the $T$-matrix elements
of three-body break-up systems similar to those of `two-body' systems
with many `discrete excited' states.

\section{Muon sticking to ${\mbox{\boldmath {\large $\alpha$}}}$ particle}

After the fusion reaction (\ref{eq:mucf-reaction})
takes place, the emitted muon 
sticks to the $\alpha$ particle
or goes to the $\alpha$-$\mu$ continuum states.
The probability that this muon sticks to the bound state is referred to
as the initial sticking probability, $\omega_S^{0}$,
and is one of the most important parameters for determining 
fusion efficiency,
since this muon is not available for further $\mu$CF cycles.
However, as summarized in a previously published review 
paper~\cite{Froelich92},
the $\alpha$-$\mu$ sticking is not yet completely understood.

In this section, we study the muon sticking problem
in a much more sophisticated manner
than that in the literature.
We derive, \mbox{for the first time,}
\mbox{{\it absolute}} values of the reaction (\ref{eq:mucf-reaction})
going to the $\alpha$-$\mu$ bound  and continuum states.
This is performed by calculating the $T$-matrix (\ref{eq:T-mat}) for 
the reaction rate (\ref{eq:reaction-rate-def}),   
in which the {\it exact} total wave function
$\Psi^{(+)}_\alpha (E_\alpha)$ is approximated by
$\Psi_{\frac{3}{2} M}^{(+)}(E)$ of Eq.~(\ref{eq:total-wf}) that was 
already obtained  
by solving Schr\"{o}dinger 
equation (\ref{eq:schr-eq}).
{
In this Section, Set B is employed for the nuclear interactions.
}

\subsection{$T$-matrix calculation of fusion rate}

We calculate the $T$-matrix~(\ref{eq:T-mat}) for the 
reaction rate~(\ref{eq:reaction-rate-def}) of
the fusion reaction
\begin{eqnarray}
\label{eq:mucf-sticking}
(dt\mu)_{J=v=0}  &\to& (\alpha \mu)_{il} + n  + 17.6 \,\mbox{MeV} \nonumber \\
      &\searrow&     (\alpha \mu)_{nl} + n  + 17.6 \,\mbox{MeV}, 
\end{eqnarray}
where $(\alpha \mu)_{nl}$ denotes the $n$-th
$(\alpha \mu)$ bound state with $l$ presented by 
${\phi}_{nlm}({\bf r}_5)$ with the eigenenergy $\varepsilon_{nl}$,
whereas $(\alpha \mu)_{il}$ describes 
the $i$-th discretized $\alpha$-$\mu$ continuum state
that is obtained by discretizing 
$\phi_{lm}(k, {\bf r}_5)$ into \{${\widetilde \varepsilon}_{il},
{\widetilde \phi}_{ilm}({\bf r}_5), i=1,...,N\}$
by performing Eqs.~(\ref{eq:bin}) and (\ref{eq:average-energy}).
We take \mbox{$l=0$ to 25,} $N=200$ and the maximum momentum
$\hbar k_N=10 \,{\rm MeV}/c$ ($\varepsilon_N \simeq 487$ keV)
in this section.

In $T$-matrix~(\ref{eq:T-mat}), we replace 
the exact total wave function $\Psi^{(+)}_\alpha (E_\alpha)$  
by  $\Psi_{\frac{3}{2} M}^{(+)}(E)$ that was
given in Eq.~(\ref{eq:total-wf}) as the sum of the three components.
Correspondingly, we divide the  \mbox{$T$-matrix~(\ref{eq:T-mat})}  into 
three parts employing channel $c=5$ with the Jacobi coordinates 
$({\bf r}_5, {\bf R}_5)$ as 
\begin{eqnarray}
&& \!\!\!\!\!\! T^{({\rm C})}_{nl, m m_s}\!\!  = \! \langle \, 
   e^{ i {\bf K}_n \cdot {\bf R}_5 } \,\phi_{nlm}({\bf r}_5) 
       \chi_{\frac{1}{2} m_s}\!(n)\,  
    |\,  V^{({\rm T})}_{\alpha n, dt} \,|   
 \mathring{\Psi}_{\frac{3}{2} M}^{({\rm C})}(dt\mu) \,\rangle ,\nonumber  \\ 
&& \!\!\!\!\!\!  T^{({\rm N})}_{nl, m m_s}\!\!  = \! \langle \, 
   e^{ i {\bf K}_n \cdot {\bf R}_5 } \,\phi_{nlm}({\bf r}_5) 
       \chi_{\frac{1}{2} m_s}\!(n)\,  
    |\,  V^{({\rm T})}_{\alpha n, dt} \,|   
  \Psi_{\frac{3}{2} M}^{({\rm N})}(dt\mu) \,\rangle , \nonumber \\ 
&& \!\!\!\!\!\!  T^{(+)}_{nl, m m_s}\!\!  = \! \langle \, 
           e^{ i {\bf K}_n \cdot {\bf R}_5 } \, \phi_{nlm}({\bf r}_5) 
       \chi_{\frac{1}{2} m_s}\!(n)\,  
    |\,  V_{\alpha n} \,|   
 \Psi_{\frac{3}{2} M}^{(+)}(\alpha n \mu) \,\rangle , \nonumber  \\
\label{eq:three-Tmat-0}
\end{eqnarray}
for the $\alpha$-$\mu$ bound states
$\phi_{nlm}({\bf r}_5)$  with the energy $\varepsilon_{nl}$, and
\begin{eqnarray}
&& \!\!\!\!\!\! {\widetilde T}^{({\rm C})}_{il, m m_s}\!\!  = \! \langle \, 
   e^{ i {\widetilde {\bf K}}_i \cdot {\bf R}_5 } \,{\widetilde \phi}_{ilm}({\bf r}_5) 
       \chi_{\frac{1}{2} m_s}\!(n)\,  
    |\,  V^{({\rm T})}_{\alpha n, dt} \,|   
 \mathring{\Psi}_{\frac{3}{2} M}^{({\rm C})}(dt\mu) \,\rangle ,\nonumber  \\ 
&& \!\!\!\!\!\! {\widetilde T}^{({\rm N})}_{il, m m_s}\!\!  = \! \langle \, 
   e^{ i {\widetilde {\bf K}}_i \cdot {\bf R}_5 } \,{\widetilde \phi}_{ilm}({\bf r}_5) 
       \chi_{\frac{1}{2} m_s}\!(n)\,  
    |\,  V^{({\rm T})}_{\alpha n, dt} \,|   
  \Psi_{\frac{3}{2} M}^{({\rm N})}(dt\mu) \,\rangle , \nonumber \\ 
&& \!\!\!\!\!\! {\widetilde T}^{(+)}_{il, m m_s}\!\!  = \! \langle \, 
           e^{ i {\widetilde {\bf K}}_i \cdot {\bf R}_5 } \, 
         {\widetilde \phi}_{ilm}({\bf r}_5) 
       \chi_{\frac{1}{2} m_s}\!(n)\,  
    |\,  V_{\alpha n} \,|   
 \Psi_{\frac{3}{2} M}^{(+)}(\alpha n \mu) \,\rangle , \nonumber  \\
\label{eq:three-Tmat-1}
\end{eqnarray}
\noindent
for the $\alpha$-$\mu$ discretized continuum states
${\widetilde \phi}_{ilm}({\bf r}_5)$ with ${\widetilde \varepsilon}_{il}$. 
 
 In Eqs.~(\ref{eq:three-Tmat-0}) and (\ref{eq:three-Tmat-1}),
the plane-wave  momenta  ${\bf K}_n$ and ${\widetilde {\bf K}}_i$
are determined, respectively, as
\begin{eqnarray}
\label{eq:E-conservation-2}
&&\frac{\hbar^2}{2\mu_{R_5}} K_n^2 = E_{00} + Q - {\varepsilon}_{nl} , \\
&&\frac{\hbar^2}{2\mu_{R_5}} {\widetilde K}_i^2 = E_{00} + Q - 
{\widetilde \varepsilon}_i, \qquad 
{\widetilde \varepsilon}_i
=\frac{\hbar^2}{2\mu_{r_5}} {\widetilde k}_i^2 . \quad 
\label{eq:E-conservation}  
\end{eqnarray}
The reaction rate (\ref{eq:reaction-rate-def}) is written as  
\begin{eqnarray}
&&\!\!\!\!\!\!\!\!\!\!\!  {r}_{nl}= {v}_{nl}
      \left( \frac{\mu_{R_5}}{2 \pi \hbar^2} \right)^2    
 \!  \sum _{m,m_s} \!
   \int \! \big| {T}^{({\rm C})}_{nl,m m_s}
     \!\!+\!{T}^{({\rm N})}_{nl,m m_s} 
     \!\!+\!{T}^{(+)}_{nl,m m_s}    \big|^2 \,
     \!{\rm d}{\widehat {\bf K}_n},    \\
&&\!\!\!\!\!\!\!\!\!\!\!{\widetilde r}_{il}= {v}_{il}
      \left( \frac{\mu_{R_5}}{2 \pi \hbar^2} \right)^2   
   \!    \sum _{m,m_s} \!  
   \int \! \big|  {\widetilde T}^{({\rm C})}_{il,m m_s}
     \!\!+\!{\widetilde T}^{({\rm N})}_{il,m m_s} 
     \!\!+\!{\widetilde T}^{(+)}_{il,m m_s}    \big|^2 \,
     \!{\rm d}{\widehat {\widetilde {\bf K}}_i},  
\end{eqnarray}  
respectively, for the bound state $(n l)$ and for the discretized 
continuum state $(i l)$.
$v_{il} =\hbar {\widetilde K}_i/\mu_{R_5}$ is the velocity of the
\mbox{$(\alpha \mu)_{il}$-$n$} relative motion associated with ${\bf R}_5$,
and similarly for ${v}_{nl}=\hbar K_n/\mu_{R_5}$.
Since the reaction rates ${r}_{nl} \,({\widetilde r}_{il}) $ 
do not depend on the
$M$ \mbox{($z$-component} of the total angular momentum $\frac{3}{2}$),
it is not necessary to take the average with respect to $M$.

The components of the $T$-matrix elements are   explicitly expressed 
to identify the dominant contribution to the reaction rates 
$r_{nl} ({\widetilde r}_{il})$.
This is a new approach for analyzing
the initial sticking probability $\omega_S^{0}$ in Sec.~V\,B.

We then transform the summation  $\sum_{i} {\widetilde r}_{il}$
into the integration of a smooth continuum function   $r_l(k)$ of $k$ 
as
\begin{eqnarray}
\sum_{i=1}^{K_N} {\widetilde r}_{il}
  =\!\sum_{i=1}^{K_N} \Big( \frac{{\widetilde r}_{il}}{\mathit{\Delta}k} \Big)
 \mathit{\Delta}k
 \stackrel{\mathit{\Delta}k \to 0}{\longrightarrow} 
 \! \int_0^{k_N}\! r_{l}(k)\, {\rm d} k \equiv r_l^{\rm cont.}. \;
\label{eq:k-convergence}
\end{eqnarray}  
Test of this procedure is well explained in the review papers 
of the CDCC method~\cite{Kamimura86, Austern, Yahiro12}

%
\begin{figure}[h]
\begin{center}
\epsfig{file=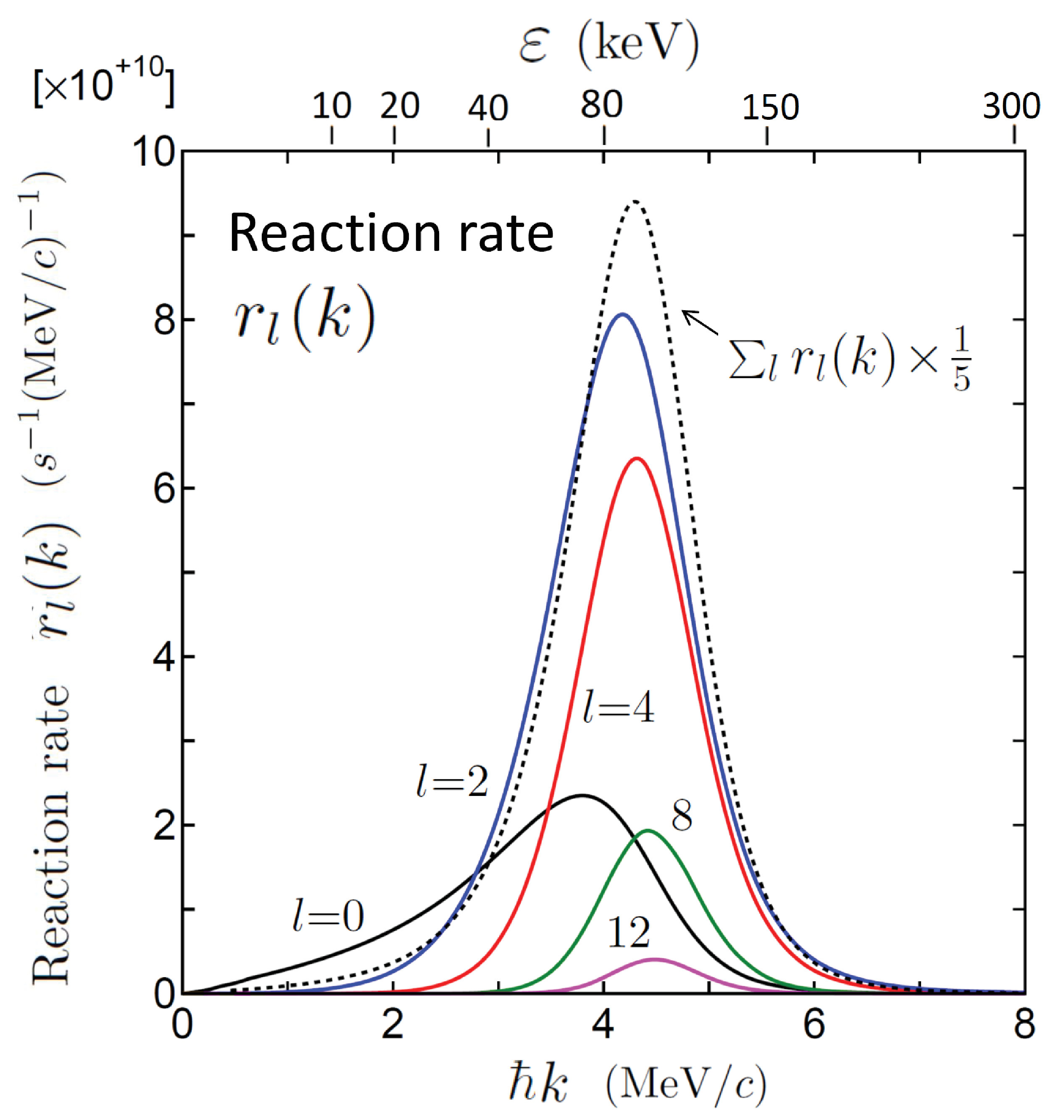,width=7.0cm,height=7.0cm}
\end{center}
\vskip -0.5 cm
\caption{Calculated reaction rates ${r}_{l}(k)$ going 
to the $\alpha$-$\mu$ continuum states with the angular momentum $l$
defined in Eq.~(\ref{eq:k-convergence}).
The dotted black curve represents 
$\sum_{l=0}^{25} r_l(k)$ multiplied by $\frac{1}{5}$.
The rates are given in units of ${\rm s}^{-1} ({\rm MeV}/c)^{-1}$.
}
\label{fig:k-continuum-rate}
\end{figure}
%

\begin{figure}[h]
\begin{center}
\epsfig{file=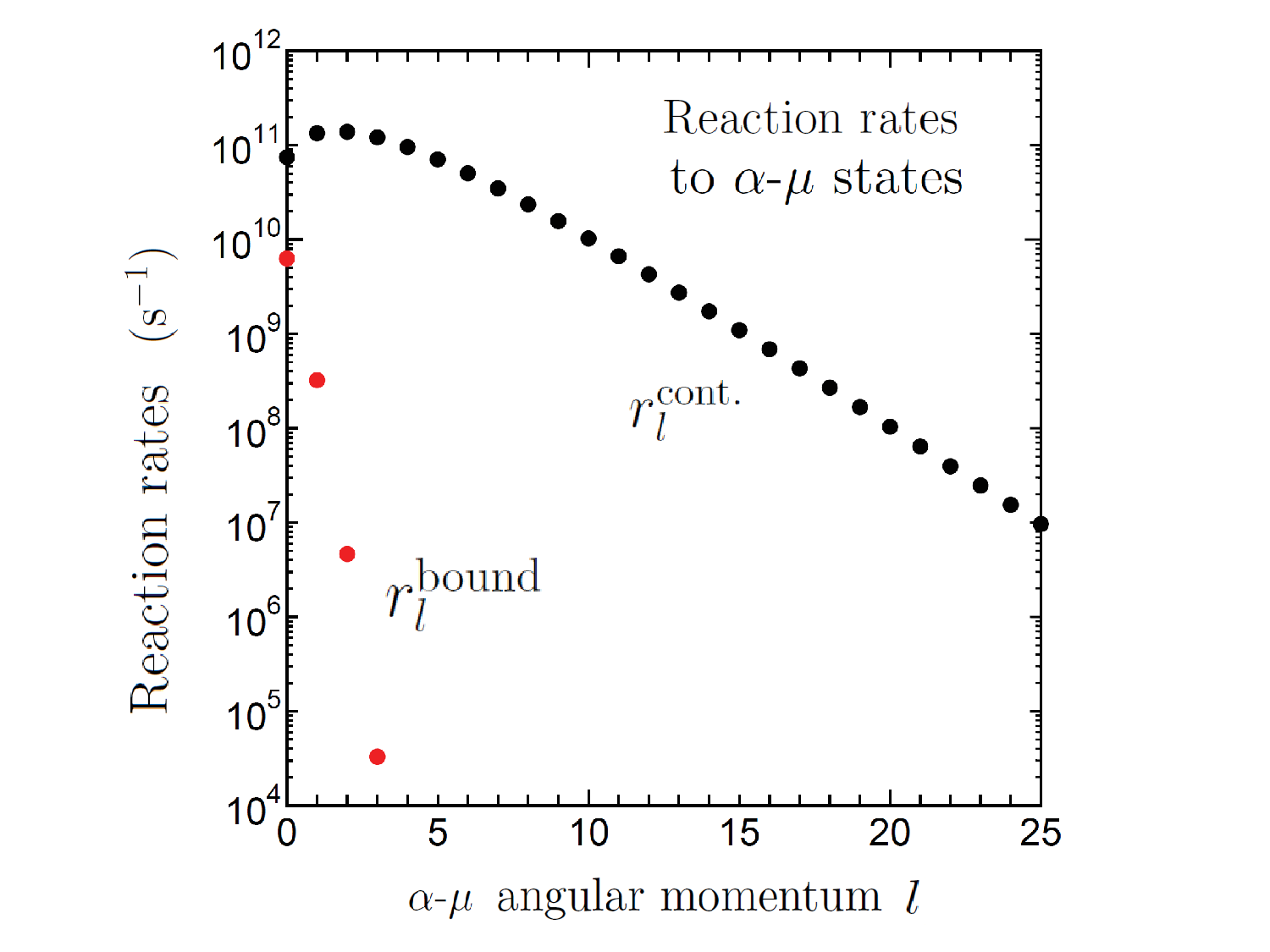,width=9.5cm,height=7.0cm}
\end{center}
\vskip -0.5cm
\caption{Calculated reaction rates $r_l^{\rm bound}$ (red circles) to 
the $\alpha$-$\mu$ bound states 
in Eq.~(\ref{eq:rate-bound})
and $r_l^{\rm cont.}$ (black circles) to the $\alpha$-$\mu$ continuum states 
in Eq.~(\ref{eq:k-convergence})
with respect to the angular \mbox{momentum $l$}. 
They are given in units of ${\rm s}^{-1}$.
}
\label{fig:fusion-rate-l5}
\end{figure}

The calculated reaction rates ${r}_{l}(k)$ 
are shown in Fig.~\ref{fig:k-continuum-rate}
for angular momenta $l$ between $\alpha$ and $\mu$.
The dotted black curve represents the summed rates
$\sum_{l=0}^{25}\,r_l(k)$ multiplied by $\frac{1}{5}$.
We see that the peak of the dotted curve is at 
$\hbar k \sim 4.3 \, {\rm MeV}/c$ ($\varepsilon \sim 88$ keV).
This is understood as follows: 
With the kinetic energy $3.5$ MeV (with speed $v_\alpha/c=0.043$),
the $\alpha$ particle  escapes
from the muon cloud which has approximately the $(^5{\rm He} \mu)_{1s}$ 
wave function of ${\bf R}_4$.
Conversely, the muon cloud is moving with respect to 
the $\alpha$  particle with the 
same speed $v_\alpha/c$, 
namely $\hbar k \sim 4.3$ MeV/c. 
The width of the peak of the dotted curve 
corresponds to the width of the momentum distribution {\it within} 
the muon cloud.

Furthermore, the reason why so many angular momenta $l$ appear 
in $r_l(k)$ in Fig.~\ref{fig:fusion-rate-l5} 
is as follows: In the \mbox{$T$-matrix} elements (\ref{eq:three-Tmat-1}), 
the component '$|V|\Psi(dt\mu)\rangle$' is composed of  
very short-range functions of ${\bf r}_4$ and long-range functions of
${\bf R}_4$. Therefore, many angular momenta $l$ are necessary to expand
this unique function of $({\bf r}_4, {\bf R}_4)$ 
in terms of the functions $e^{ i {\widetilde {\bf K}}_i \cdot {\bf R}_5 } 
\,{\widetilde \phi}_{ilm}({\bf r}_5)$ on the different Jacobi coordinates
$({\bf r}_5, {\bf R}_5)$.

For comparison with $r_l^{\rm cont.}$ in Eq.~(\ref{eq:k-convergence}),
we introduce
\begin{eqnarray}
    r_l^{\rm bound}= \sum_n r_{nl}
\label{eq:rate-bound}
\end{eqnarray}
for the transition to all the $\alpha$-$\mu$ 
bound states with $l$.
\mbox{Figure ~\ref{fig:fusion-rate-l5}} illustrates
how $r_l^{\rm bound}$ and $r_l^{\rm cont.}$ 
depend on the angular momentum $l$
of $0 \leq l \leq 25$. 
$r_l^{\rm bound}$ are given
by the red circles and $r_l^{\rm cont.}$ by the black ones.
The rates $r_l^{\rm bound}$ 
decrease quickly with \mbox{increasing $l$}, whereas $r_l^{\rm cont.}$ 
changes slowly with respect to $l$.
The ratio ($\sim 1 \% $) of the strength of the red-circle group to 
that of the black-circle group is the essence of the
initial $\alpha$-$\mu$ sticking probability, which is discussed 
in the next subsection.

\begin{table}[t!]  
\caption{ Fusion rates $\lambda_{\rm f}^{\rm cont.}$ and 
$\lambda_{\rm f}^{\rm bound}$ of the reaction 
$(dt\mu)_{J=v=0} \to (\alpha \mu)_{\rm cont.} + n\:$ \mbox{or} $\: 
(\alpha \mu)_{\rm bound}+n$, respectively, and their sum 
$\lambda_{\rm f}$, defined by 
Eqs.~(\ref{eq:lambda-bound})-(\ref{eq:lambda-sum})
on the Jacobi-coordinate channel $c=5$.
Contribution to them from the full $T$-matrix and from 
the individual $T$-matrix elements 
$T^{({\rm C})}, T^{({\rm N})}$ and $T^{(+)}$ are listed. 
The initial sticking probability $\omega_S^{0}
(=\lambda_{\rm f}^{\rm bound}/\lambda_{\rm f})$
is given in the last line. Since the contribution from 
$T^{({\rm C})}$ and $T^{(+)}$ are minor,  
$\omega_S^{0}$ is not calculated there.
}
\begin{center}
\begin{tabular}{ccccccccc} 
\hline \hline
\noalign{\vskip 0.2 true cm} 
\noalign{\vskip 0.1true cm} 
  &    &$ \!\!\!\!\!\!\!\!\!$  $|T^{({\rm C})}\!+\!T^{({\rm N})}\!+\!T^{(+)}|^2$   
           &    &    $|T^{({\rm C})}|^2$  &   
         & $|T^{({\rm N})}|^2$   &    & $|T^{({\rm +})}|^2$    \\ 
\noalign{\vskip 0.2true cm} 
\hline 
\noalign{\vskip 0.1 true cm} 
 $\lambda_{\rm f}\,$ $({\rm s}^{-1})$ & &    $8.05 \times 10^{11} $  &     
     &  $3.70 \times 10^{8} $ &   &   $7.73 \times 10^{11} $ &   
     &   $1.18 \times 10^{6} $    \\ 
 $\lambda_{\rm f}^{\rm cont.}$ $({\rm s}^{-1})$ & &    $7.98 \times 10^{11} $  &     
     &  $3.67 \times 10^{8} $ &   &   $7.66 \times 10^{11} $ &   
     &   $1.17 \times 10^{6} $    \\ 
\noalign{\vskip 0.1 true cm} 
 $\lambda_{\rm f}^{\rm bound}$ $({\rm s}^{-1})$ & &    $\!\!6.90 \times 10^{9} $  &     
     &  $3.14 \times 10^{6} $ &   &   $\!\!6.63 \times 10^{9} $ &   
     &   $9.41 \times 10^{3} $    \\ 
\noalign{\vskip 0.1 true cm} 
\hline
\noalign{\vskip 0.1 true cm} 
 $\omega_S^{0} \:(\%)$  & &    $0.857 $  &     
     & ---  &   &   $ (0.857)  $ &   
     & ---    \\ 

\noalign{\vskip 0.1 true cm} 
\noalign{\vskip 0.2 true cm} 
\hline
\hline
\end{tabular} 
\label{table:fusion-rate-sticking}
\end{center}
\end{table}

Finally, we define the fusion rates to all the bound states 
and to all the continuum states, respectively as
\begin{eqnarray}
 \lambda_{\rm f}^{\rm bound}=\sum_{l=0}^{5}\: r_l^{\rm bound}, 
\quad  \lambda_{\rm f}^{\rm cont.}=\sum_{l=0}^{25}\: r_l^{\rm cont.}
\label{eq:lambda-bound} 
\end{eqnarray}
and the sum of them as
\begin{eqnarray} 
 \lambda_{\rm f}&=& \lambda_{\rm f}^{\rm bound}
                 +\lambda_{\rm f}^{\rm cont.}.
\label{eq:lambda-sum}
\end{eqnarray}
This $\lambda_{\rm f}$ is the fusion rate of
the reaction (\ref{eq:mucf-reaction}) defined by using 
the $T$-matrix in channel $c=5$.

In Table~III,    
the calculated results of
$\lambda_{\rm f}, \lambda_{\rm f}^{\rm bound}$ and 
$\lambda_{\rm f}^{\rm cont.}$ are listed together with the 
contributions from  the individual
$T$-matrix elements $T^{({\rm C})}, T^{({\rm N})}$ and $T^{(+)}$.
We see that the fusion rate $\lambda_{\rm f}$ is obtained as
\begin{equation}
\lambda_{\rm f}=8.05 \times 10^{11}\, {\rm s}^{-1}.
\label{eq:lambda-stick-sum}
\end{equation}
We consider that this value is consistent with  
$1.03 \times 10^{12}\, {\rm s}^{-1}$ in Eq.~(\ref{eq:lambda-rate}) 
obtained by solving the couple-channels Schr\"{o}dinger equation 
with  the outgoing wave in channel \mbox{$c\!=\!4$,}
taking into account the significant difference in the 
calculational methods and the fact that the $c\!=\!5$ channel component
is not included in the total wave function $\Psi_{\frac{3}{2} M}^{(+)}(E)$.

Another important result in Table~\ref{table:fusion-rate-sticking}
is that the contribution to the fusion rates from $T^{({\rm C})}$ 
is much smaller than that from $T^{({\rm N})}$;
this was expected from Fig.~\ref{fig:density-dtmu-rs3}
since ${\mathring \rho}^{\rm (C)}(r_3)$ is much smaller
than $\rho^{\rm (N)}(r_3)$ in the nuclear interaction region.
This will be referred to in the next subsection for
the $\alpha$-$\mu$ sticking probability $\omega_S^{0}$ .

\subsection{Initial sticking probability $\omega_S^0$}

The  absolute values of the fusion rates
$\lambda_{\rm f}^{\rm cont.}$ and
$\lambda_{\rm f}^{\rm bound}$ have been explicitly   calculated  
in the present three-body reaction calculation.
This requires a change in the way of  discussing the 
\mbox{$\alpha$-$\mu$} sticking, as will be emphasized later.

Now, it is possible to calculate the initial muon-sticking probability 
by the definition
\begin{eqnarray}
\omega_S^0=\frac{\lambda_{\rm f}^{\rm bound}}        
            {\lambda_{\rm f}^{\rm bound} + \lambda_{\rm f}^{\rm cont.}}
\label{eq:initial-stick}
\end{eqnarray}
that is  based on the original idea for $\omega_S^0$ (cf., for example,
Eq.~(192) in Ref.~~\cite{Froelich92}),
employing the nuclear interactions that reproduce the observation quantities 
in Figs.~\ref{fig:cal-cc-dt-an} and \ref{fig:sigma-an}.

Before discussing our calculation of $\omega_S^0$,
we review the essential point of previously reported studies on 
the \mbox{$\alpha$-$\mu$} sticking probability
referring to the review papers of $\mu$CF~\cite{Ponomarev90,Froelich92}.
Since the sudden approximation was used in the literature to define 
$\omega_S^0$, we first derive the same representation of their $\omega_S^0$
using our precise framework.
We start from our definition (\ref{eq:initial-stick}) and make the following
\mbox{approximations i) to v):}

i) In the \mbox{$T$-matrix} elements (\ref{eq:three-Tmat-0}) 
and (\ref{eq:three-Tmat-1}), the $\alpha n \mu$ outgoing wave 
$\Psi_{\frac{3}{2} M}^{(+)}(\alpha n \mu)$ is excluded from
the total wave function 
$\Psi_{\frac{3}{2} M}^{(+)}(E)$ of (\ref{eq:total-wf}).
Omitting the spin component, the wave function of $(dt\mu)_{J=v=0}$
is represented by  
{
$\Phi_0^{(dt\mu)}({\bf r}_3,{\bf R}_3)$ 
}
that was obtained, for example in Ref.~\cite{Kamimura89},  by
diagonalizing the $dt\mu$ Hamiltonian (\ref{eq:hamil-dtm}).
{
$\Phi_0^{(dt\mu)}({\bf r}_3,{\bf R}_3)$ is almost the same as 
$\mathring{\Phi}_{0}^{({\rm C})}({\bf r}_3, {\bf R}_3)
\!+\! {\widehat \Phi}_{0}^{({\rm N})}({\bf r}_3, {\bf R}_3)$ 
obtained using the linear equation (\ref{eq:Phi(N)-hat-eq})
(cf. Fig.~\ref{fig:density-dtmu-rs3}).
}

ii) In Eqs.~(\ref{eq:three-Tmat-0}) 
and (\ref{eq:three-Tmat-1}), the $dt$-$\alpha n$ transition 
interaction (\ref{eq:tensor-nonloc-34}), 
$V^{({\rm T})}_{\alpha n, dt}$, is replaced as
\begin{eqnarray} 
     \int d{\bf r}_3 V_{dt, \alpha n}^{({\rm T})}({\bf r}_3, {\bf r}_4)
    \to V_\delta \!\! \,\int d{\bf r}_3 \delta({\bf r}_3) 
     \,\delta({\bf r}_4),  
\label{eq:sudden-delta}
\end{eqnarray}
which is the essence of the sudden approximation.

iii) The momentum ${\bf K}_n$ of the plane wave 
$e^{ i {\bf K}_n \cdot {\bf R}_5 }$  is fixed to ${\bf K}$ given by
$\hbar^2 K^2 /2\mu_{R_5} = 17.589$ MeV. 

iv) The $T$-matrix elements and the fusion rates are given by 
\begin{eqnarray}
\label{eq:sudden-0}
\!\!\!\!\!\! &&T_{nlm} =  \langle \, 
   e^{ i {\bf K} \cdot {\bf R}_5 } \,\phi_{nlm}({\bf r}_5) 
          |\,  V_\delta \!\! \int \!{\rm d}{\bf r}_3 
          \delta({\bf r}_3)\delta({\bf r}_4)|  
     \,\Phi_0^{(dt\mu)}({\bf r}_3,{\bf R}_3) 
               \rangle_{{\bf r}_5,{\bf R}_5} ,   \nonumber\\
\!\!\!\!\!\!\!\!\!\!\!\! &&\qquad { =  V_\delta \, \langle \, 
   e^{ i {\bf K} \cdot {\bf R}_5 } \,\phi_{nlm}({\bf r}_5) \,
          |\, \delta({\bf r}_4)\, \Phi_0^{(dt\mu)}(0,{\bf R}_4)\, 
           \rangle_{{\bf r}_5, {\bf R}_5}, } \nonumber\\
\!\!\!\!\!\!\!\!\!\!\!\! &&\qquad  = V_\delta \, \langle \, 
   e^{ i {\bf q} \cdot {\bf r}_5 } \,\phi_{nlm}({\bf r}_5) 
          |\,\Phi_0^{(dt\mu)}(0,{\bf r}_5) 
           \rangle_{{\bf r}_5},  \quad  \\
\label{eq:sudden-1}
\!\!\!\!\!\!\!\!\!\!\!\! &&\lambda_{\rm f}^{\rm bound} = v
      \left( \frac{\mu_{R_5}}{2 \pi \hbar^2} \right)^2 
       \sum_{nlm} |T_{nlm}|^2 ,  \quad  \\  
\label{eq:sudden-2}
\!\!\!\!\!\!\!\!\!\!\!\! &&\lambda_{\rm f}  =  v
      \left( \frac{\mu_{R_5}}{2 \pi \hbar^2} \right)^2  V_\delta^2 \, 
 \langle \Phi_0^{(dt\mu)}(0,{\bf r}_5) 
         |\,\Phi_0^{(dt\mu)}(0,{\bf r}_5) 
         \rangle_{{\bf r}_5}, \quad  \;\;
\label{eq:sudden-3}
\end{eqnarray}
where ${\bf q}=\frac{m_\mu}{m_\alpha \! +\! m_\mu} {\bf K}$
and $v=\hbar K/\mu_{R_5}$. 
The completeness of the $\alpha$-$\mu$ basis functions 
\{$\phi_{nlm}({\bf r}_5), \phi_{lm}(k, {\bf r}_5)$\}
is used to derive $\lambda_{\rm f}$.
{
The use of $\delta({\bf r}_4)$ in Eq.~(\ref{eq:sudden-0})  is based on 
the approximation that the nuclear interactions can be regarded 
as a contact interaction because the interaction range 
is much smaller than the muonic molecular size. 
This choice of the interaction, however, imposes $S$-wave for 
the $\alpha$-$n$ relative motion denoted with ${\bf r}_4$, 
which contradicts the observed fact of $D$-wave.
}

v) Finally, the sticking probability $\omega_S^0$ defined by 
(\ref{eq:initial-stick})  is approximated by ${\widehat \omega}_S^0$ as
\begin{eqnarray}
{\widehat \omega}_S^0   =\frac{ \sum_{nlm} |\,\langle \, 
   e^{ i {\bf q} \cdot {\bf r}_5 } \,\phi_{nlm}({\bf r}_5) 
          |\,\Phi_0^{(dt\mu)}(0,{\bf r}_5) 
           \rangle_{{\bf r}_5}  |^2 }
         {  \langle \,\Phi_0^{(dt\mu)}(0,{\bf r}_5) 
         |\,\Phi_0^{(dt\mu)}(0,{\bf r}_5) 
         \rangle_{{\bf r}_5} }, 
\label{eq:sudden-stick}
\end{eqnarray}
wherein no calculation is performed on the absolute values of 
$\lambda_{\rm f}^{\rm bound}$, $\lambda_{\rm f}^{\rm cont.}$ 
and $\lambda_{\rm f}$, 
since the $dt$-$\alpha n$ coupling interaction 
$V_\delta \! \int \!{\rm d}{\bf r}_3 
\delta({\bf r}_3)\delta({\bf r}_4)$ is not appropriate for the purpose.

We see that Eq.~(\ref{eq:sudden-stick}) is  the same as 
the previous expression for
$\omega_S^0$ under the sudden approximation (for example, 
see Eq.~(207) of Ref.~\cite{Froelich92} and
Eq.~(36) of Ref.~\cite{Ponomarev90}).
Most of the literature calculations gave values in the region of 
\begin{equation}
{\widehat \omega}_S^0 \simeq 0.88-0.89 \%    
\label{eq:stick-88-89}
\end{equation}
{\it without} nuclear $d$-$t$ interactions
(see the lines for '\mbox{Theory:} Coulombic
problem' in Table 10 of the $\mu$CF review paper~\cite{Froelich92}).
Similarly, {\it with} the nuclear $d$-$t$, 
\begin{eqnarray}
{\widehat \omega}_S^0 \simeq 0.91-0.93\% 
\label{eq:stick-92-93}
\end{eqnarray}
were obtained by the optical-potential model~\cite{Kamimura89,
Bogdanova89} and the $R$-matrix method~\cite{Struensee88a,
Szalewicz90,Hale93,Hu1994,Cohen1996,Jeziorski91}
(In Refs.~\cite{Hu1994,Cohen1996,Jeziorski91},
an internuclear distance $a_{dt}=0.51$ fm was taken in place of
$r_3=a_{dt}=0$ in Eq.~(\ref{eq:sudden-stick})).
One of the present authors (M.K.)~\cite{Kamimura89} 
participated in those calculations.

However, it should be noted that there is a \mbox{serious} problem
in the discussion of ${\widehat \omega}_S^0$
in the case of (\ref{eq:stick-88-89}) with the Coulomb force only
for $\Phi_{dt\mu}({\bf r}_3,{\bf R}_3)$.
This problem occurs because attention was not paid to 
the absolute values of $\lambda_{\rm f}^{\rm bound}$ 
and $\lambda_{\rm f}^{\rm cont.}$. 
This is made clear in Table~\ref{table:fusion-rate-sticking} 
for $\lambda_{\rm f}^{\rm bound}$ and $\lambda_{\rm f}^{\rm cont.}$ 
together with their contributions 
from the three types of \mbox{$T$-matrix,} namely,
$T^{({\rm C})}, T^{({\rm N})}$ and $T^{(+)}$.
The column $|T^{({\rm C})}|^2$ is responsible for the calculation of 
(\ref{eq:stick-88-89}).
We see that the contribution from $T^{({\rm C})}$ to 
$\lambda_{\rm f}^{\rm cont.}$ and 
$\lambda_{\rm f}^{\rm bound}$ is much smaller than that
from $T^{({\rm N})}$;
this is known from Fig.~\ref{fig:density-dtmu-rs3}
since ${\mathring \rho}^{\rm (C)}(r_3)$ is much smaller
than $\rho^{\rm (N)}(r_3)$ in the nuclear interaction region.

Therefore, we say that such a calculation of ${\widehat \omega}_S^0$ 
using the minor components of $\lambda_{\rm f}^{\rm cont.}$ and 
$\lambda_{\rm f}^{\rm bound}$ is not so meaningful 
(although it is useful when comparing the accuracy of the employed 
calculation methods with each other). 
In this sense, we placed the symbol `---' in the column of  
$|T^{({\rm C})}|^2$ in the last line for $\omega_S^0$, and similarly in
the column of $|T^{({\rm +})}|^2$.
We also note that the statement 
``the additional effect of the nuclear force
to the sticking probability" is not appropriate since 
$T^{({\rm N})}$ dominantly contributes to 
$\lambda_{\rm f}$ (Table~\ref{table:fusion-rate-sticking}),
{\it not} additionally.

Our final result on the initial sticking probability is,
as shown in the full $T$-matrix column of 
Table~\ref{table:fusion-rate-sticking}, 
\begin{eqnarray}
\omega_S^0=0.857 \% \qquad \mbox{(present)}, 
\label{eq:cal-omega0}
\end{eqnarray}
which is reduced by $\sim 7\%$ from
the value in (\ref{eq:stick-92-93}). 
The origin of this reduction will be
discussed in the next-to-last paragraph of Sec.~V\,C.

Table~\ref{table:sticking-nl} lists the calculated result of  
the muon initial sticking probability $\omega_S^0$ (\%) 
of the $(dt\mu)_{J=v=0}$ state 
together with the $(\alpha \mu)_{nl}$-components.
The last \mbox{column} is shown , only for reference, and is from
our previous result in Ref.~\cite{Kamimura89} based on
the sudden approximation (\ref{eq:sudden-stick}), 
in which the $d$-$t$ nuclear interaction is included 
but the $\alpha$-$n$ channel is not considered.

\begin{table}[t!]
\caption{The initial $\alpha$-$\mu$ sticking probability $\omega_S^0$ (\%) 
of the  (dt$\mu)_{J=v=0}$ state
together with the $(\alpha \mu)_{nl}$-components.
The present wave function includes the $\alpha n \mu$ channel
with $D$-wave $\alpha$-$n$ relative motion. 
The last column is  from
our previous result~\cite{Kamimura89} 
which employed the \mbox{sudden} approximation (\ref{eq:sudden-stick});
the $d$-$t$ nuclear interaction was included 
but the $\alpha n \mu$ channel was not considered.
}
\label{table:sticking-nl}
\begin{center}
\begin{tabular}{ccccccc} 
\hline \hline
\noalign{\vskip 0.1 true cm} 
$J\!=\!v\!=\!0$   & $\quad$ & Present      &   & 
$\;$ Ref.~\cite{Kamimura89}\\
\noalign{\vskip 0.0 true cm} 
$\Psi_{\frac{3}{2} M}^{(+)}(\alpha n \mu)$   &  &  $\alpha$-$n \:D$ wave  & 
  &  No $\alpha$-$n$ wave \\
\noalign{\vskip -0.2 true cm} 
   & $\quad$ &    &     &  (sudden approx.) \\
\noalign{\vskip 0.2 true cm} 
\hline
\noalign{\vskip 0.1 true cm} 
$\lambda_{\rm f} ({\rm s}^{-1})$  &  &  $8.05\times 10^{12}$  &   
     &  $-$    \\
\noalign{\vskip 0.1 true cm} 
$\lambda_{\rm f}^{{\rm bound}} ({\rm s}^{-1})$  &  &  $6.90\times 10^{10}$ &  
      &  $-$    \\
\noalign{\vskip 0.1 true cm} 
\hline
\noalign{\vskip 0.15 true cm} 
$\omega_S^0 (\%)$  &  & 0.857 $\:$  &     & 0.9261   \\
\noalign{\vskip 0.15 true cm} 
$    1s$           &  &   0.6583 &    & 0.7141   \\
$    2s$           &  &   0.0950  &   & 0.1021   \\
$    2p$           &  &   0.0233  &   & 0.0248   \\
$    3s$           &  &   0.0289  &   & 0.0310   \\
$    3p$           &  &   0.0084  &   & 0.0089   \\
$    3d$           &  &   0.0002   &  & 0.0002   \\
$    4s$           &  &   0.0123  &   & 0.0132   \\
$    4p$           &  &   0.0038   &  & 0.0040   \\
$4d+4f $           &  &   0.0001  &   & 0.0001   \\
$    5s$           &  &   0.0063 &    & 0.0068   \\
\noalign{\vskip 0.2 true cm} 
all others         &  &   0.0204  &   & 0.0208   \\
\noalign{\vskip 0.1 true cm} 
\hline
\hline
\end{tabular}
\end{center}
\end{table}

\subsection{Effective sticking probability $\omega_S^{\rm eff}$}

Finally, we discuss the effective sticking probability $\omega_S^{\rm eff}$ 
that is defined as  
\begin{equation}
\omega_S^{\rm eff} = \omega_S^0 \,(1-R),
\end{equation}
where $R$ is the muon reactivation coefficient that expresses
the probability that the muon is shaken off 
from the $(\alpha \mu)_{nl}$ state during slowing down 
from the initial kinetic energy 3.5 MeV.
The effective sticking $\omega_S^{\rm eff}$ is the most crucial 
parameter in the $\mu$CF cycle because it sets a limit on 
the maximum possible fusion output per muon.

Regarding the sticking problem, we understand as follows 
from Secs.~8.4-8.6 of the $\mu$CF review paper~\cite{Froelich92} (1992): 
Using $R=0.287$ (for a low density $\phi=0.17$) 
of Ref.~\cite{Stodden90},
the theoretical values of $\omega_S^0 \simeq 0.91-0.93 \%$  
in (\ref{eq:stick-92-93}) result in
$\omega_S^{\rm eff} \sim 0.66\%$ , which is 10\% larger than the
experimental value $\omega_S^{\rm eff} =0.59 \pm 0.07\%$ ($\phi=0.175$) 
at PSI~\cite{Petitjean1990} (1991). 
Ref.~\cite{Froelich92} states that   
the 10\% difference is large enough to motivate
further studies, because it may be a signal that the sticking problem 
is not yet completely understood.

\begin{figure}[t!]
\begin{center}
\epsfig{file=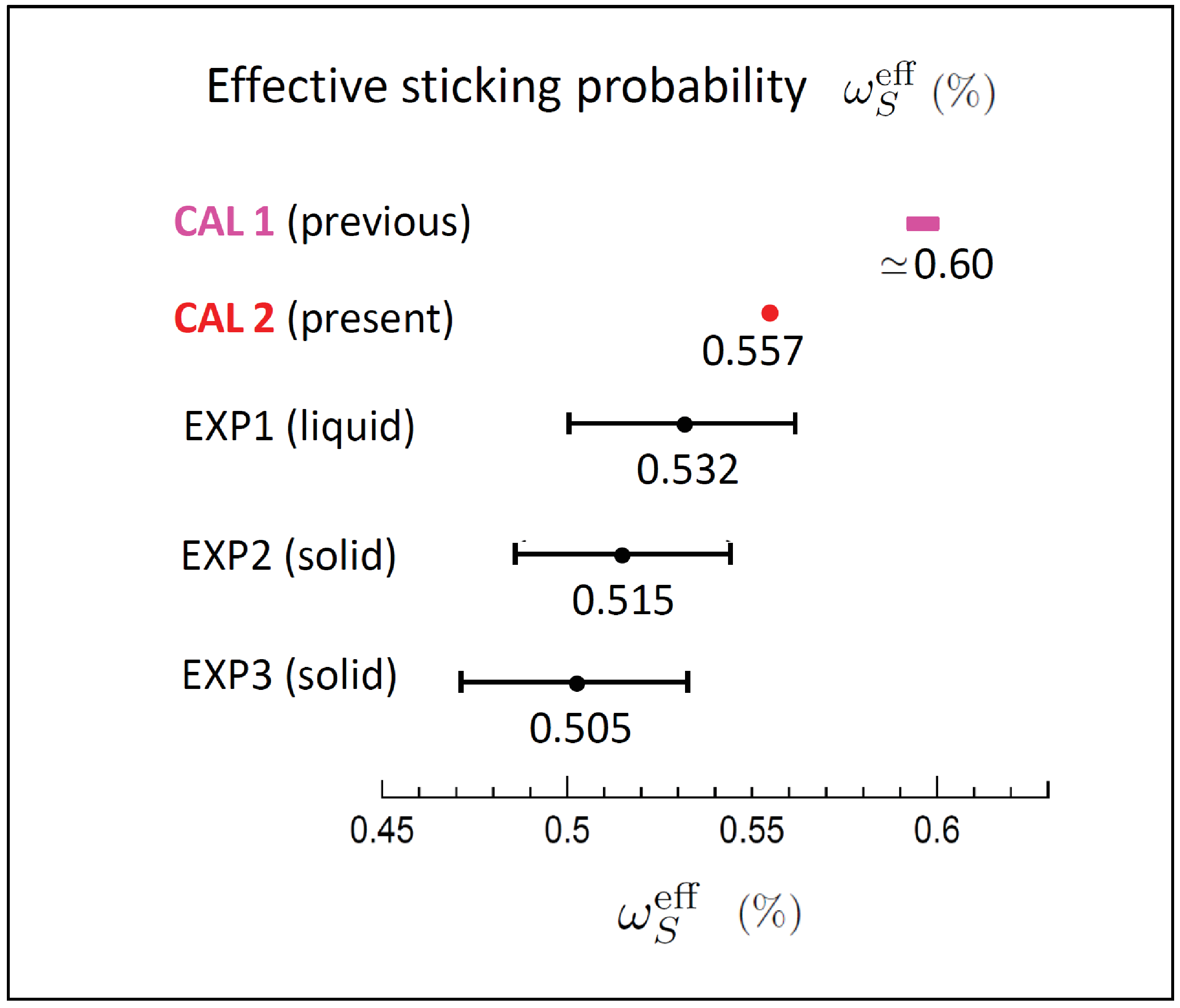,width=6.3cm,height=5.4cm}
\end{center}
\vskip -0.5cm
\caption{Comparison of the calculated effective sticking probability
$\omega_S^{\rm eff}$ with the observed values based on  from
the precise high-density experiments 
EXP1~\cite{Ishida2001}, EXP2~\cite{Ishida2001} 
and EXP3~\cite{Petitjean2001} in 2001. CAL1 is given by 
the previous work with $\omega_S^0 \simeq 0.91-0.93\%$~\cite{Kamimura89,
Bogdanova89,Struensee88a,
Szalewicz90,Hale93,Hu1994,Cohen1996,Jeziorski91},
and CAL2 is owing to the present result with $\omega_S^0=0.857$ \%
of (\ref{eq:cal-omega0}),  with the definition 
$\omega_S^{\rm eff}=\omega_S^0 (1- R)$ 
taking the reactivation coefficient as
$R=0.35$ for high densities~\cite{Struensee88b,Stodden90,Markushin,Rafelski}.
}
\label{fig:effective sticking}
\end{figure}

In 2001, the final (last) precise experimental data 
on $\omega_S^{\rm eff}$
were reported from RIKEN-RAL~\cite{Ishida2001} and
PSI~\cite{Petitjean2001} at high densities $(\phi=1.2 -1.4)$.
The results showed that 
\begin{eqnarray}
&& \omega_S^{\rm eff}=\begin{cases}
\;0.532\pm0.030 \% \quad 
\text{({\rm Liquid}~\cite{Ishida2001})}, \\     
\;0.515\pm0.030 \%\quad 
\text{({\rm Solid}~\cite{Ishida2001})},\;\; \\     
\;0.505\pm0.029 \%\quad 
\text{({\rm Liquid}~\cite{Petitjean2001})}. 
\end{cases}
\end{eqnarray}
We derive our $\omega_S^{\rm eff}$ 
with the use of $\omega_S^0=0.857$ \% 
in (\ref{eq:cal-omega0}) and 
\mbox{$R=0.35$}~\cite{Struensee88b,Stodden90,Markushin,Rafelski}
for high densities (density dependence seems 
to be very small for $\phi \sim 1.2-1.5$ in Table~VI of 
Ref.~\cite{Stodden90}). We obtain 
\begin{eqnarray}
\omega_S^{\rm eff}=0.557 \%\qquad \mbox{(present)},
\label{eq:cal-effective-sticking}
\end{eqnarray}
which can explain 
the observed values
illustrated in Fig.~\ref{fig:effective sticking},
whereas the ${\widehat \omega}_S^0 \simeq 0.91-0.93 \%$
by the previous work gives $\omega_S^{\rm eff} \simeq 0.60$\%.

To consider the origin of the change
to $\omega_S^0=0.857$\% in (\ref{eq:cal-omega0}) 
from  $0.91-0.93$\% in (\ref{eq:stick-92-93}),
we perform an additional calculation in which the \mbox{$\alpha$-$n$}
outgoing $D$-wave is replaced by $S$-wave.
This is performed only for a reference calculation since
this change contradicts the observation that the 
outgoing $\alpha$-$n$ channel with $I^\pi=\frac{3}{2}^+$
has $D$-wave angular momentum.

We consider the following nonlocal central-force
coupling, instead of the tensor force 
$V_{dt, \alpha n}^{({\rm T})}({\bf r}_3, {\bf r}_4)$ 
in Eq.~(\ref{eq:tensor-1}):
\begin{eqnarray}  
 V_{dt, \alpha n}^{({\rm S})}({\bf r}_3, {\bf r}_4)
            &\!\!\! =\!\!\!&v_0^{(S)}\, 
          e^{-\mu \,r_{34}^2- \mu' R_{34}^2}
\label{eq:scalar}
\end{eqnarray}
with $\mu=1/(1.0\, {\rm fm})^2$, $\mu'=1/(6.0\, {\rm fm})^2$
and  \mbox{$v_0^{(S)}= 0.81$} ${\rm MeV\, fm}^{-3}$  
with a slight change to $V_0=-38.05\,$ MeV in 
the \mbox{$d$-$t$} potential.  The quality of the fitting to the 
observed $S$-factors is
almost the same as the red curve in Fig.~\ref{fig:cal-cc-dt-an}.

We have obtained
$\lambda_{\rm f}\!= 8.87 \times  10^{11}\, {\rm s}^{-1}$ 
by solving the coupled-channels 
Schr\"{o}dinger equation (\ref{eq:schr-eq}) and  
$\lambda_{\rm f}= 7.56 \times 10^{11}\, {\rm s}^{-1}$ by calculating the 
$T$-matrix elements (\ref{eq:three-Tmat-0}) and (\ref{eq:three-Tmat-1}); 
this result is not unreasonable. 
As for the $\alpha$-$\mu$ sticking problem,
we see the change as follows:
\begin{eqnarray}
&&\!\!\!\!
\lambda_{\rm f, S{\rm -wave}}^{\rm bound}\!=7.09 \times\! 10^{9} {\rm s}^{-1} 
      \to 
\lambda_{\rm f, D{\rm -wave}}^{\rm bound}\!=6.90 \times \!10^{9} {\rm s}^{-1}\nonumber \\
&&\!\!\!\!
\lambda_{\rm f, S{\rm -wave}}^{\rm cont.}\!=7.49 \times\! 10^{11} {\rm s}^{-1}
    \!\!  \to 
\lambda_{\rm f, D{\rm -wave}}^{\rm cont.}\!=7.98 \times \!10^{11} {\rm s}^{-1} \nonumber,
\end{eqnarray}
which gives the change of 
$\omega_S^0 =\lambda_{\rm f}^{\rm bound}
/(\lambda_{\rm f}^{\rm bound} + \lambda_{\rm f}^{\rm cont.})$ 
from 0.938\% ($S$-wave) to 0.857\% ($D$-wave).
Furthermore, we see that the former number of $\omega_S^0$ is
close to that in (\ref{eq:stick-92-93}) by the sudden approximation.

In conclusion, we have much improved the sticking-probability calculation
by employing the $D$-wave $\alpha$-$n$ outgoing channel with the non-local 
tensor-force $dt$-$\alpha n$ coupling and by deriving the probability
based on the absolute values of 
the $\lambda_{\rm f}^{\rm bound}$ and $\lambda_{\rm f}^{\rm cont.}$.
The calculated result can reproduce the experimental value as mentioned above.

For more progress in the theoretical study of 
the effective sticking probability $\omega_S^{\rm eff}$,
development of studying the reactivation
coefficient $R$ is expected.  Our result on
the \mbox{{\it absolute}} values of the transition rates to the individual 
\mbox{$(\alpha \mu)$ bound} and continuum states in 
Table~\ref{table:sticking-nl} and Fig.~\ref{fig:k-continuum-rate}
will be useful.

\section{Momentum and Energy spectra   
of emitted muon}
%
In this section we calculate the momentum and energy spectra 
(in absolute values) of the emitted muon after 
the $dt\mu$ fusion (\ref{eq:mucf-reaction}).
It is often stated that `10-keV' muons are \mbox{emitted} 
during the reaction.
However, it should be noted that 10 keV is the `average' of the 
muon kinetic energy in the $({\rm He}\mu)_{1s}$ atom
wherein the muon momentum distribution has a long higher-momentum tail.  
If one considers the utilization of such muons, for example, 
as the source of an ultra-slow  negative muon 
beam~\cite{Nagamine1989,
NagamineMCF199091,
Strasser1993,Nagamine1996Hyp103,Strasser1996,Natori2020,okutsu,
YAMASHITA2021112580},
it is important to determine the momentum and energy spectra 
of the released muon.
{
In this Section, Set B is emplyed for the nuclear interactions.
}

For simplicity in expression, in this section, we refer to ${\bf K}, 
{\bf K}_i$ and ${\widetilde {\bf K}}_i$  as `muon momentum';
more precisely, it is the momentum of the relative motion 
between the muon and the $\alpha$-$n$ pair; and similarly for
`muon energy' $E, E_i$ and ${\widetilde E}_i$. 

\begin{figure}[t!]
\begin{center}
\epsfig{file=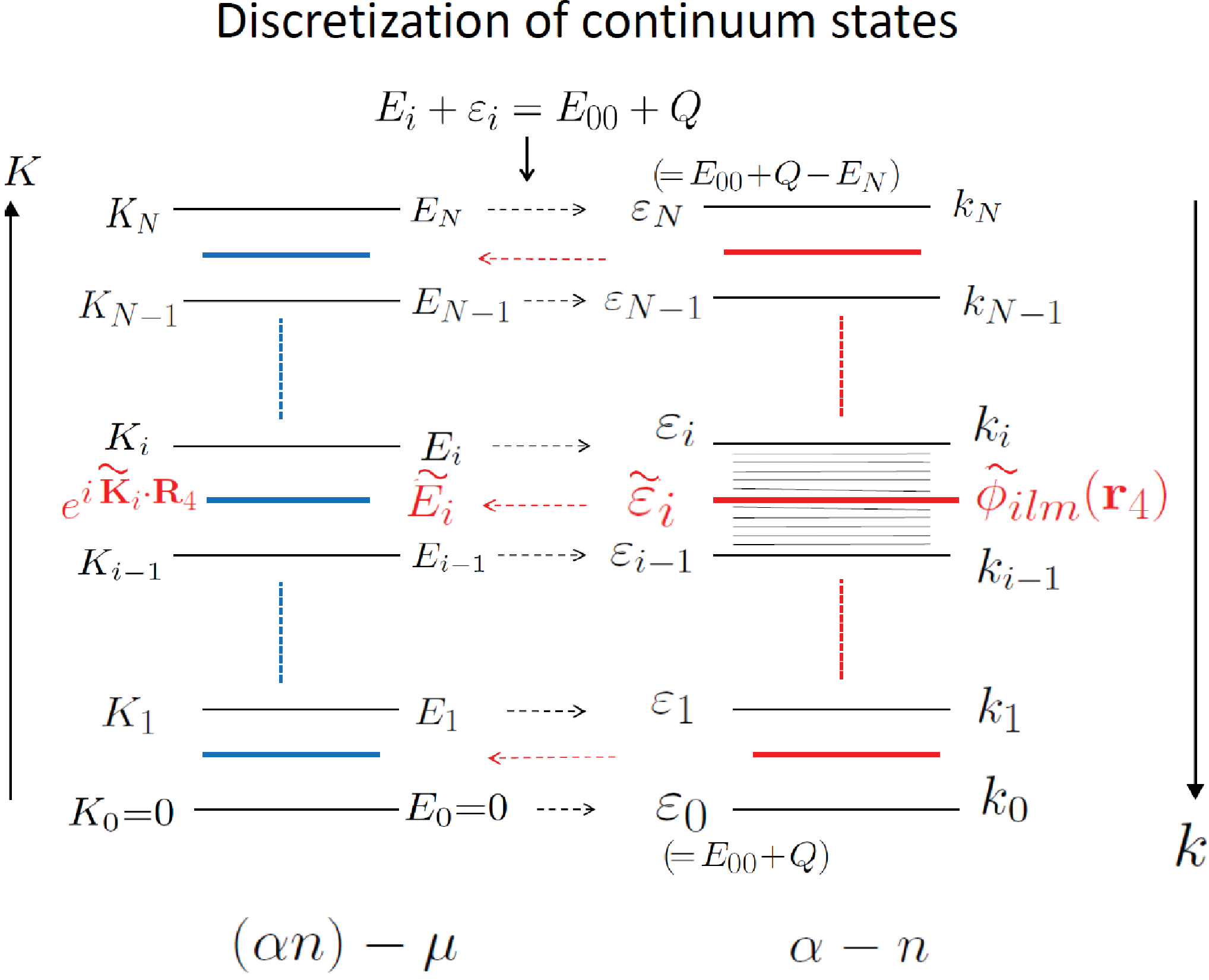,width=8.2cm,height=6.5cm}
\end{center}
\vskip -0.5cm
\caption{Schematic illustration for discretization of the 
momentum space $[K_0, K_N]$
of the ($\alpha n$)-$\mu$ relative motion 
along ${\bf R}_4$ (left half) and  that of 
the $\alpha$-$n$  relative motion 
$[k_N, k_0]$ 
along ${\bf r}_4$ (right half)
while maintaining 
$E_i + \varepsilon_i = E_{00} + Q$.
The resulting discretized \mbox{$\alpha$-$n$} continuum states 
${\widetilde \phi}_{ilm}({\bf r}_4) (i=1 - N)$ are shown by the red lines
and 
the associated muon plane waves 
$e^{ i {\widetilde {\bf K}}_i \cdot {\bf R}_4 }$ are
represented by the 
blue lines in the left half. \mbox{See text.}
}
\label{fig:risanka-muon-new-2}
\end{figure}

{
To derive the momentum (energy) spectra as a continuous
function of $K (E)$,  prcise discretization of 
the momentum space both in the ($\alpha n)$-$\mu$  and
the  $\alpha$-$n$ relative motions  is required while maintaining 
the energy conservation
of the energy-sum of such motions. Such a correlated discretization
is shown in Fig.~\ref{fig:risanka-muon-new-2}.
} 
We start by assuming an upper limit of the muon  
momentum $K_N$ at the top of the left end of the figure,
whereas the minimum momentum is $K_0=0$.
We then divide the $K$-space $[K_0, K_N]$
\mbox{into $N$ bins ($K_i, i=0 - N$)}
with equal intervals $\mathit{\Delta}K$.
Correspondingly, we divide  the \mbox{$k$-space} $[k_N, k_0]$
of the $\alpha$-$n$
relative motion on the right half of the figure into 
\mbox{$N$ bins ($k_i, i=0 - N$)}
with the energy \mbox{conservation kept as}
\begin{eqnarray}
 E_i + \varepsilon_{i}=E_{00}+Q,  \quad 
 E_i=\frac{\hbar^2}{2\mu_{R_4}}  K_i^2,    \quad 
 \varepsilon_i=\frac{\hbar^2}{2\mu_{r_4}}  k_i^2 ,
\label{eq:conserv-muonsp}
\end{eqnarray}
where $K_i$ increases with increasing $i$, but 
$k_i$ decreases with increasing $i$.
The bin width $\mathit{\Delta}\! K=K_i-K_{i-1}$ is constant in the 
left-half muon momentum space, whereas 
\mbox{$\mathit{\Delta}k_i=|k_i-k_{i-1}|$}
on the right half depends on $i$.
Now, ${\widetilde \phi}_{ilm}({\bf r}_4)$ is constructed 
using Eq.~(\ref{eq:bin}) but the $k$-integration runs from $k_i$ 
till $k_{i-1}$. The energy ${\widetilde \varepsilon}_i$ is given by
Eq.~(\ref{eq:average-energy}).
We take $N=200$ as the number of discretized 
momentum bins in Fig.~\ref{fig:risanka-muon-new-2}, 
setting $\hbar K_N=6.0 $\, MeV/$c \,(E_N=175$ keV).
{
This precise discretization is necessary for deriving the 
{\it continuous} muon $K (E)$-spectrum. 
}

{
\subsection{$T$-matrix calculation of fusion rate}
}

We calculate the muon spectra using 
the \mbox{$T$-matrix} procedure in Sec.~VI\,A
and the total wave function $\Psi_{\frac{3}{2} M}^{(+)}(E)$
that was already obtained in Sec.~III as the sum of three components
(cf. Eq.~(\ref{eq:total-wf})).
Correspondingly, as in Sec.~V\,A,
we divide the  $T$-matrix (\ref{eq:T-mat}) into 
three components on the Jacobi coordinates $({\bf r}_4, {\bf R}_4)$ 
in channel $c=4$ (cf. Fig.~\ref{fig:jacobi-dtmu}) as
\begin{eqnarray}
&& \!\!\!\!\!\! {T}^{({\rm C})}_{il, m m_s}\!\!  = \! \langle \, 
   e^{ i {\widetilde {\bf K}}_i \cdot {\bf R}_4 } \,{\widetilde \phi}_{ilm}({\bf r}_4) 
       \chi_{\frac{1}{2} m_s}\!(n)\,  
    |\,  V^{({\rm T})}_{\alpha n, dt} \,|   
 \mathring{\Psi}_{\frac{3}{2} M}^{({\rm C})}(dt\mu) \,\rangle ,\nonumber  \\ 
&& \!\!\!\!\!\! {T}^{({\rm N})}_{il, m m_s}\!\!  = \! \langle \, 
   e^{ i {\widetilde {\bf K}}_i \cdot {\bf R}_4 } \,{\widetilde \phi}_{ilm}({\bf r}_4) 
       \chi_{\frac{1}{2} m_s}\!(n)\,  
    |\,  V^{({\rm T})}_{\alpha n, dt} \,|   
  \Psi_{\frac{3}{2} M}^{({\rm N})}(dt\mu) \,\rangle , \nonumber \\ 
&& \!\!\!\!\!\! {T}^{(+)}_{il, m m_s}\!\!  = \! \langle \, 
           e^{ i {\widetilde {\bf K}}_i \cdot {\bf R}_4 } \, 
     {\widetilde \phi}_{ilm}({\bf r}_4) 
       \chi_{\frac{1}{2} m_s}\!(n)\,  
    |\,  V_{\alpha \mu} \,|   
 \Psi_{\frac{3}{2} M}^{(+)}(\alpha n \mu) \,\rangle , 
\label{eq:Tmat-muon-emit}
\end{eqnarray}
where ${\widetilde \phi}_{ilm}({\bf r}_4)\, (i=1 - N)$ is the 
discretized $\alpha$-$n$ continuum state
with energy $\widetilde{\varepsilon}_{i l}$,
and 
$e^{ i {\widetilde {\bf K}}_i \cdot {\bf R}_4 }$ is the associated 
$(\alpha n)$-$\mu$ plane wave, 
satisfying the energy conservation
\begin{eqnarray}
\qquad \quad  {\widetilde E}_i + {\widetilde \varepsilon}_{i}
=E_{00}+Q  \qquad (i=1 - N) .
\label{eq:conserv-muonsp2}
\end{eqnarray}
The third component $T^{(+)}$ of (\ref{eq:Tmat-muon-emit})
estimates the effect of the Coulomb
potential $V_{\alpha \mu}(r_5)$ on the $T$-matrix.
Here, we note that it is not necessary to use the Coulomb wave function,
instead of the plane wave,
in the bra-vector of the above $T$-matrix elements; 
the reason is explained in Appendix. 

The reaction rate (\ref{eq:reaction-rate-def})
for a muon emitted to the discretized continuum state
$(\alpha n)_{il}$-$\mu$ is written as
\begin{eqnarray}
\label{eq:reaction-rate-Sec6}
\!\!\! {r}_{il}={v}_{il}
      \left( \frac{\mu_{R_4}}{2 \pi \hbar^2} \right)^2  
      \sum _{m,m_s}   
   \int \big|  {T}^{({\rm C})}_{il,m m_s}
   +{T}^{({\rm N})}_{il,m m_s} 
     +{T}^{(+)}_{il,m m_s}    \big|^2 \,
     {\rm d}{\widehat {\widetilde {\bf K}}}_i ,
\end{eqnarray}  
where ${v}_{il}=\hbar {\widetilde K}_i/\mu_{R_4}$ is the velocity of the    
$(\alpha n)_{il}$-$\mu$ relative motion.
Since  ${r}_{il}$ does not depend on the 
$M$ \mbox{($z$-component} of the total angular momentum $\frac{3}{2}$),
it is not necessary to take the average with respect to $M$.
The sum of the transition rates
\begin{eqnarray}
\lambda_{\rm f}=\sum_{il}\; {r}_{il}  
\label{eq:fusion-rate-muon}
\end{eqnarray}
is the fusion rate of the reaction~(\ref{eq:mucf-reaction})
using the $T$-matrix based on channel $c=4$.
This $\lambda_{\rm f}$  is  
compared with the $\lambda_{\rm f}$ 
obtained in Secs.~II\,B and V\,C using different prescriptions.

To investigate the role of the three types of 
\mbox{$T$-matrix} elements in Eq.~(\ref{eq:reaction-rate-Sec6}),  
we calculate the individual reaction rates
\begin{eqnarray}
\!\!\!\!\! &&\lambda_{\rm f}^{\rm (C)}=\sum_{il}\; {r}_{il}^{\rm (C)}, \quad
{r}_{il}^{({\rm C})}= {v}_{il}
      \left( \frac{\mu_{R_4}}{2 \pi \hbar^2} \right)^2  
     \sum _{m,m_s}   
   \int \big|  {T}^{({\rm C})}_{il,m m_s}
    \big|^2 \,
     {\rm d}{\widehat {\widetilde {\bf K}}}_i \,,  \nonumber \\
\!\!\!\!\!&& \lambda_{\rm f}^{\rm (N)}=\sum_{il}\; {r}_{il}^{\rm (N)}, \quad
             {r}_{il}^{({\rm N})}= {v}_{il}
      \left( \frac{\mu_{R_4}}{2 \pi \hbar^2} \right)^2  
     \sum _{m,m_s}   
   \int \big|  {T}^{({\rm N})}_{il,m m_s}
    \big|^2 \,
     {\rm d}{\widehat {\widetilde {\bf K}}}_i \,,   \nonumber \\
\!\!\!\!\! &&\lambda_{\rm f}^{\rm (+)}=\sum_{il}\; {r}_{il}^{\rm (+)}, \quad
        {r}_{il}^{(+)}= {v}_{il}
      \left( \frac{\mu_{R_4}}{2 \pi \hbar^2} \right)^2  
     \sum _{m,m_s}   
   \int \big|  {T}^{({\rm +})}_{il,m m_s}
    \big|^2 \,
     {\rm d}{\widehat {\widetilde {\bf K}}}_i \,. \nonumber \\ 
\label{eq:reaction-rate3}
\end{eqnarray}  
In the calculation of the
$\lambda_{\rm f}^{({\rm C})}$ and $ \lambda_{\rm f}^{({\rm N})}$, 
the contribution from the final states ${\widetilde \phi}_{ilm}({\bf r}_4)$
with $l \neq 2$ is negligible under the  $dt$-$\alpha n$ tensor coupling
interaction.
In $\lambda_{\rm f}^{({\rm +})}$ for the $\alpha$-$\mu$ Coulomb force effect,
the contribution from $l>4$ is negligible.

The calculated fusion rate $\lambda_{\rm f}$ and the individual contributions 
$\lambda_{\rm f}^{\rm (C)}, \lambda_{\rm f}^{\rm (N)}$
and $\lambda_{\rm f}^{\rm (+)}$ are listed 
in Table~\ref{table:fusion-rate-muon-1}.
Finally, we obtain the full fusion rate as
\begin{equation}
       \lambda_{\rm f}=1.15 \times 10^{12} \,{\rm s}^{-1}.
\label{eq:lambda115}
\end{equation}
From $\lambda_{\rm f}^{({\rm C})}$ and $\lambda_{\rm f}^{({\rm N})}$ 
in Table~\ref{table:fusion-rate-muon-1}, 
it is observed that a fusion reaction occurs
mostly from $\Psi_{\frac{3}{2} M}^{({\rm N})}(dt\mu)$,
whereas the contribution from  $\mathring{\Psi}_{\frac{3}{2} M}^{({\rm C})}(dt\mu)$ 
is minor; this was already expected based on Fig.~\ref{fig:density-dtmu-rs3}
since ${\mathring \rho}^{\rm (C)}(r_3)$ is much smaller
than $\rho^{\rm (N)}(r_3)$ in the nuclear interaction region.

The term $T^{({\rm +})}$ describes the effect of the 
Coulomb force that acts on the $\alpha$ and $\mu$
in the outgoing $\alpha n \mu$ channel. 
{
This force
was omitted when solving the Schr\"{o}dinger equation 
(\ref{eq:schr-eq}); 
however, this approximation  is corrected 
via the \mbox{$T$-matrix} calculation presented in this subsection.
The correction was obtained as $\lambda_{\rm f} - \lambda_{\rm f}^{({\rm N})}
=0.11 \times 10^{12}\, {\rm s}^{-1}$.
Therefore,  the final result of the fusion rate in this study
is expressed as Eq.~(\ref{eq:lambda115}), $1.15 \times 10^{12}\, {\rm s}^{-1}$.
}   

\begin{table}[t!]  
\caption{Fusion rate 
$\lambda_{\rm f}$ (in units of ${\rm s}^{-1}$) calculated 
by (\ref{eq:fusion-rate-muon}) and individual rates 
$\lambda_{\rm f}^{({\rm C})}$, $\lambda_{\rm f}^{({\rm N})}$,
and $\lambda_{\rm f}^{({\rm +})}$  by 
(\ref{eq:reaction-rate3}) on the Jacobi-coordinate channel $c=4$.
The second line shows the type of used $T$-matrix.
}
\begin{center}
\begin{tabular}{ccccccccc} 
\hline \hline
\noalign{\vskip 0.2 true cm} 
 $ \lambda_{\rm f}$    & 
      &   $ \lambda_{\rm f}^{({\rm C})}$    &  $\;\;$
         & $ \lambda_{\rm f}^{({\rm N})}$&  $\;\;$  & 
        $ \lambda_{\rm f}^{({\rm +})}$    \\ 
\noalign{\vskip 0.1true cm} 
 $|T^{({\rm C})}\!+\!T^{({\rm N})}\!+\!T^{(+)}|^2$   
           &     &    $|T^{({\rm C})}|^2$    &  
         & $|T^{({\rm N})}|^2$   &    & $|T^{({\rm +})}|^2$     \\ 
\noalign{\vskip 0.1true cm} 
\hline 
\noalign{\vskip 0.1 true cm} 
\noalign{\vskip 0.1 true cm} 
    $1.15 \times 10^{12} $  &  $\;\;\;$   
     &  $4.58 \times 10^{8} $ &  $\;\;\;$ &   $1.04 \times 10^{12} $ &  $\;\;\;$ 
     &   $1.80 \times 10^{11} $    \\ 
\noalign{\vskip 0.2 true cm} 
\hline
\hline
\noalign{\vskip -0.3 true cm} 
\end{tabular}
\label{table:fusion-rate-muon-1}
\end{center}
\end{table}

It should be noted that, in some cases, the calculation of $T$-matrix 
elements for the Coulomb force between the continuum states 
is hindered by a problem 
in the integration up to the infinity. 
However, this issue is circumvented in the present work 
since we employ the discretization of the continuum states 
then smooth it as done in Eqs.~(\ref{eq:muon-K-spec-1}) 
and (\ref{eq:muon-K-spec-2}).

{
Here, it is to be emphasized that
both the solution of  the Schr\"{o}dinger equation
and the calculation of  $T$-matrix elements
are  very accurate in the case wherein 
the Coulomb force $V_{\alpha \mu}^{({\rm C})}(r_5)$ is
omitted. The reason is as follows:
As shown in Sec.~IV, provided that the total wave function
 $\Psi_{\frac{3}{2} M}^{(+)}(E)$  obtained by solving the
Schr\"{o}dinger equation is \mbox{{\it exact},} 
the asymptotic behavior exhibited  by the $T$-matrix calculation
is the same as that given by 
$\Psi_{\frac{3}{2} M}^{(+)}(E)$.
In the present case, 
the fusion rate $\lambda_{\rm f}$ was obtained as 
$1.04 \times 10^{12}\,{\rm s}^{-1}$
by the $T$-matrix calculation 
(cf. the case
$|T^{({\rm C})}\!+\!T^{({\rm N})}|^2 \simeq 
|T^{({\rm N})}|^2$   in Table~\ref{table:fusion-rate-muon-1})
and as $1.03 \times 10^{12}\,{\rm s}^{-1}$
using the $S$-matrix of $\Psi_{\frac{3}{2} M}^{(+)}(E)$ 
(cf. Eq.(\ref{eq:lambda-rate})).
}

\vskip 0.5cm
\subsection{Muon spectrum}

The aim of this Sec.~VI is to calculate
the muon momentum and energy spectra  
as continuous functions of $K$ and the kinetic energy 
$E$, respectively.
Here, $K (E)$ is the \mbox{momentum} (kinetic energy) 
of the $(\alpha n)$-$\mu$ relative motion associated with ${\bf R}_4$, 
whereas the muon momentum $K_\mu$ (kinetic energy $E_\mu$) measured 
from the center of mass of the $\alpha n \mu$ system is given by
\begin{equation}
   K_\mu = K, \quad \quad E_\mu= \gamma E ,    \quad 
\label{eq:muon-motion}
\end{equation}
with $\gamma =\frac{m_\alpha + m_n}{m_\alpha + m_n + m_\mu}=0.9779$.
We note that the center of mass of the $\alpha n \mu$  is almost at rest
in the laboratory system since the $(dt\mu)$ molecule 
is also almost at rest at the fusion 
(\ref{eq:mucf-reaction}). 

The momentum spectra $r(K)$ and $r^{({\rm N})}(K)$ 
can be obtained, 
by smoothing ${r}_{il}$ and
${r}_{il}^{({\rm N})}$,  respectively, as
\begin{eqnarray}
&&\!\!\!\! \lambda_{\rm f}= \sum_{il} 
\Big( \frac{{r}_{il}}{\mathit{\Delta}\! K} \Big)\, \mathit{\Delta}\!K
 \stackrel{\mathit{\Delta}\!K \to 0}{\longrightarrow} 
  \int_0^{K_N} \! r(K)\, {\rm d} K , 
\label{eq:muon-K-spec-1}      \\
&&\!\!\!\! \lambda_{\rm f}^{({\rm N})}= \sum_{il} 
\Big( \frac{{r}^{({\rm N})}_{il}}{\mathit{\Delta}\! K} \Big)\, 
\mathit{\Delta}\! K
 \stackrel{\mathit{\Delta}\!K \to 0}{\longrightarrow} 
  \int_0^{K_N} \! r^{({\rm N})}(K)\, {\rm d} K , \qquad
\label{eq:muon-K-spec-2}
\end{eqnarray}  
and similarly for $r^{({\rm C})}(K)$ and $r^{({\rm +})}(K)$.

\begin{figure}[t!]
\begin{center}
\epsfig{file=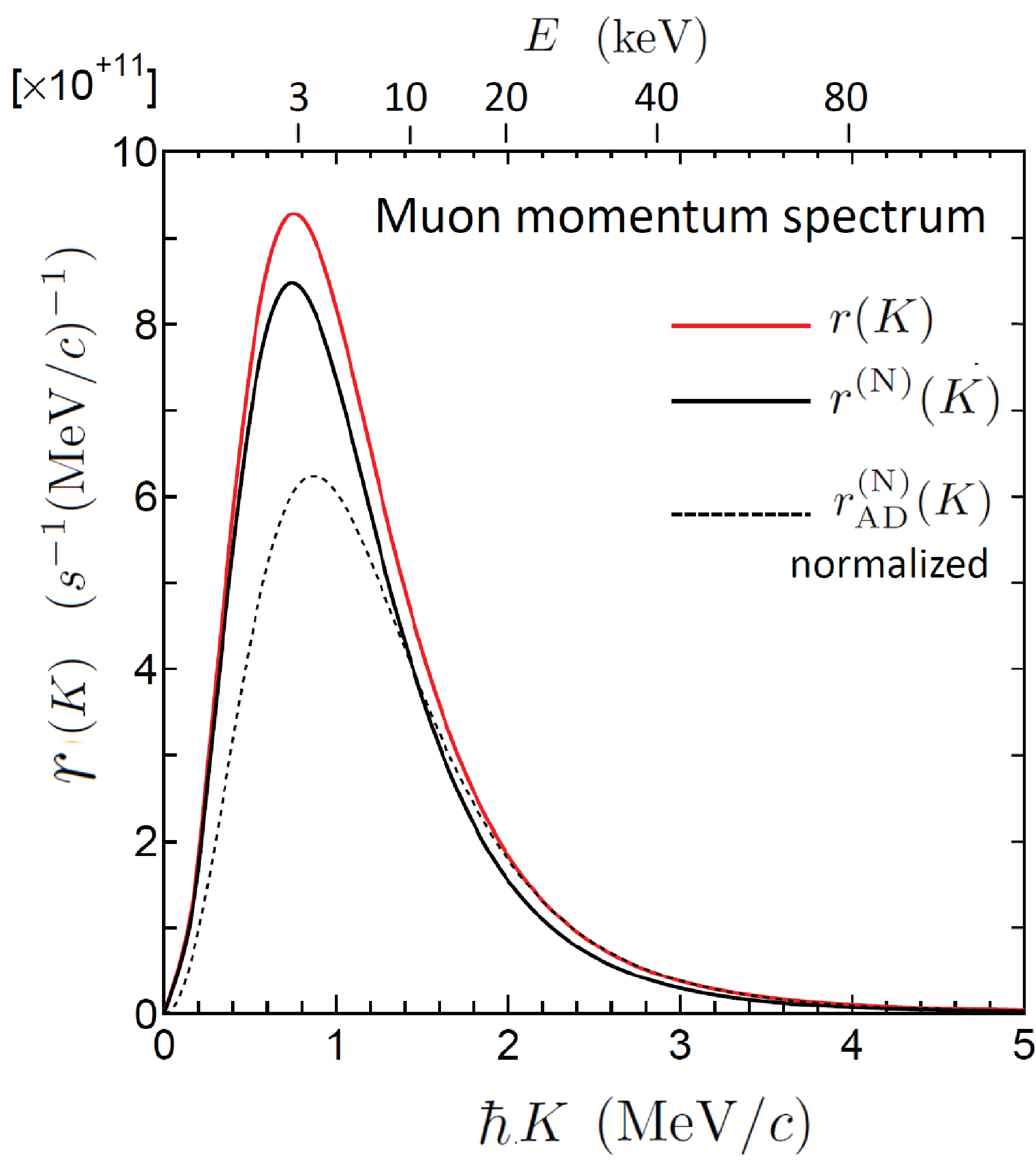,width=7.0cm,height=7.5cm}
\end{center}
\vskip -0.5cm
\caption{Momentum spectrum of the muon emitted by the $dt\mu$ fusion.
The red and black curves denote ${r}(K)$ and ${r}^{({\rm N})}(K)$
defined in (\ref{eq:muon-K-spec-1}) and (\ref{eq:muon-K-spec-2}), respectively.
The dotted curve shows $r^{({\rm N})}_{\rm AD}(K)$ by the adiabatic
approximation for $r^{({\rm N})}(K)$ (see text); 
$r^{({\rm N})}_{\rm AD}(K)$ is normalized to $r^{({\rm N})}(K)$ 
to have the same $K$-integrated values, $\lambda_{\rm f}^{\rm (N)}
=1.04 \times 10^{12}\, {\rm s}^{-1}$.
}
\label{fig:muon-spectrum}
\end{figure}
\begin{figure}[t!]
\begin{center}
\epsfig{file=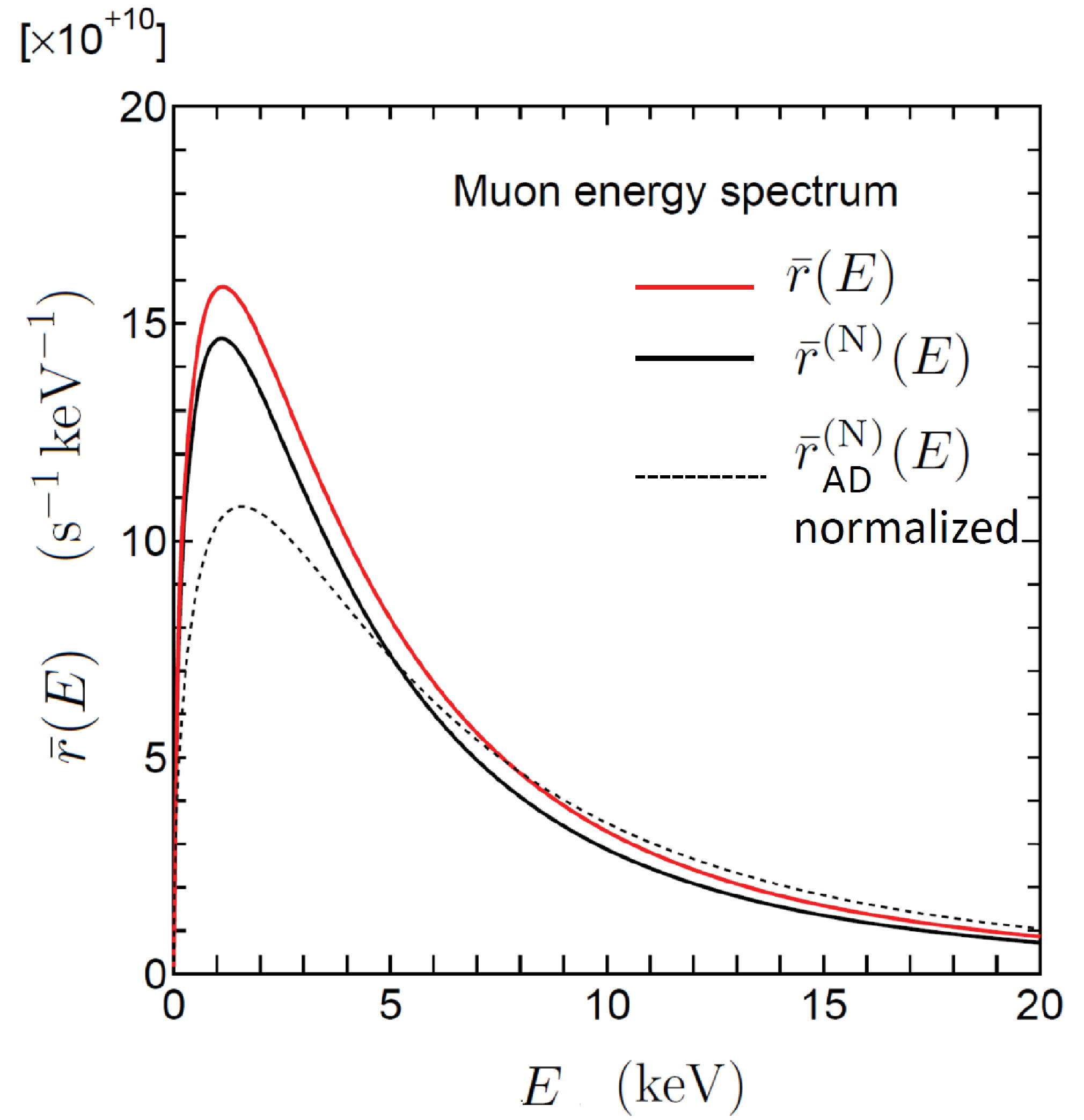,width=7.2cm,height=7.0cm}
\end{center}
\vskip -0.5cm
\caption{Energy spectrum of the muon emitted during the $dt\mu$ fusion.
The red and black curves denote ${\bar r}(E)$ and ${\bar r}^{({\rm N})}(E)$
defined in (\ref{eq:T-spect-1}) and (\ref{eq:T-spect-2}), respectively.
The peak position is at $E=1.1$ keV in the two cases.  
The dotted curve shows ${\bar r}^{({\rm N})}_{\rm AD}(E)$ obtained 
using the adiabatic
approximation for ${\bar r}^{({\rm N})}(E)$ (see text); 
${\bar r}^{({\rm N})}_{\rm AD}(E)$ is normalized to ${\bar r}^{({\rm N})}(E)$ 
to have the same \mbox{$E$-integrated} values, $\lambda_{\rm f}^{\rm (N)}
=1.04 \times 10^{12}\, {\rm s}^{-1}$.
}
\label{fig:muon-E-spec-linear}
\end{figure}

The energy spectra ${\bar r}(E)$ and 
${\bar r}^{({\rm N})}(E)$ are derived, with  the use of 
$E=\hbar^2 K^2/2\mu_{{\rm R}_4}$, by
\begin{eqnarray}
{\bar r}(E) \,{\rm d}E =\!\!\!\!&& r(K)\, {\rm d} K, 
\label{eq:T-spect-1}  \\
{\bar r}^{({\rm N})}(E) \,{\rm d}E =
\!\!\!\!&& r^{({\rm N})}(K)\, {\rm d} K,  
\label{eq:T-spect-2}  
\end{eqnarray}
and similarly for ${\bar r}^{({\rm C})}(E)$ 
and ${\bar r}^{({\rm +})}(E)$.

The calculated momentum spectrum $r(K)$ is illustrated 
in Fig.~\ref{fig:muon-spectrum} by the red curve in units of 
${\rm s}^{-1} ({\rm MeV} /c)^{-1}$, whereas
$r^{({\rm N})}(K)$ is represented by the black curve. 
The energy spectrum ${\bar r}(E)$ is shown
in Fig.~\ref{fig:muon-E-spec-linear} by the red curve in units of 
${\rm s}^{-1} ({\rm keV})^{-1}$, whereas
${\bar r}^{({\rm N})}(E)$ is represented by the black curve.

\begin{figure}[b!]
\begin{center}
\epsfig{file=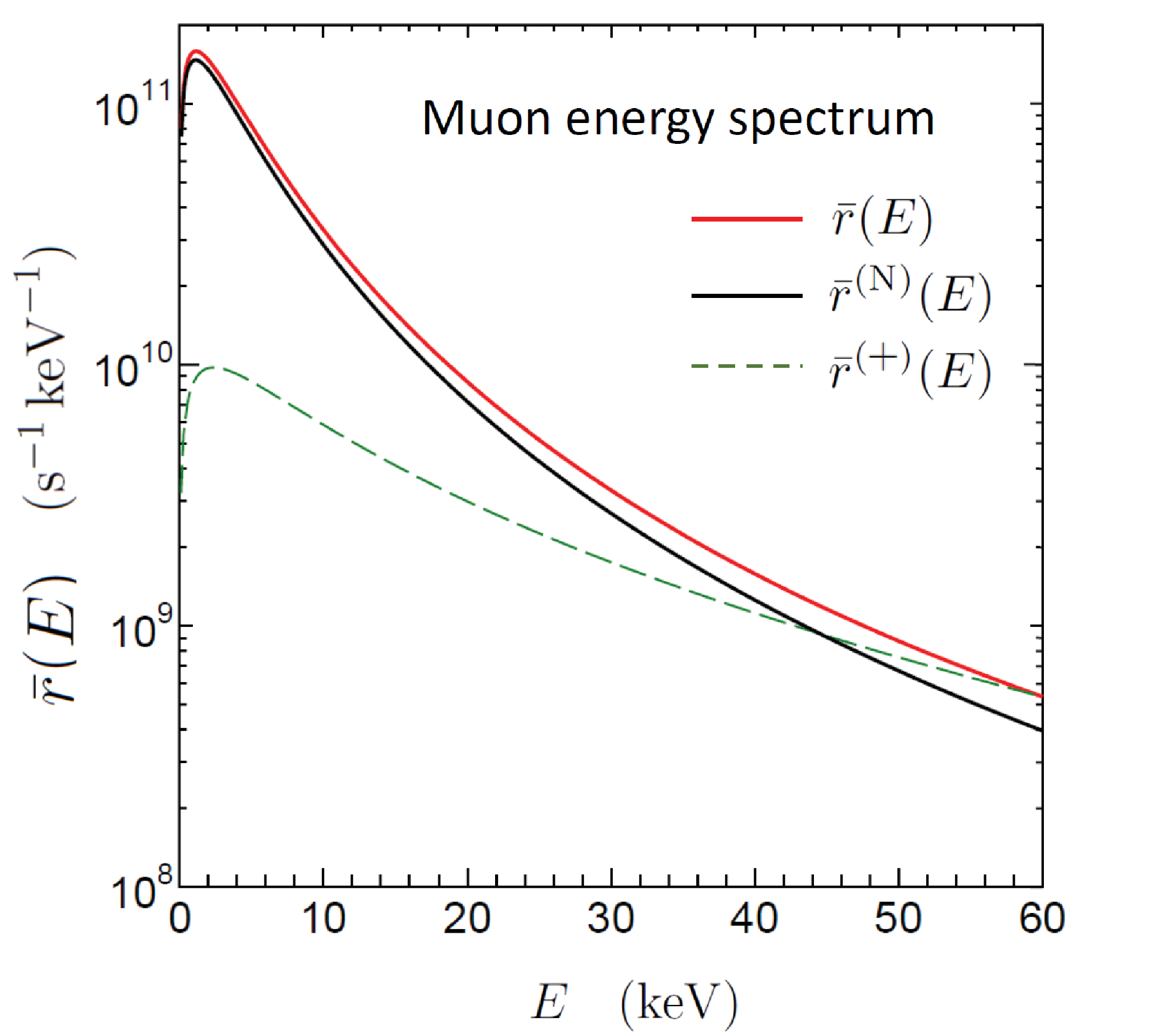,width=7.6cm,height=6.8cm}
\end{center}
\vskip -0.5cm
\caption{Energy spectrum of the muon emitted during the $dt\mu$ fusion
in log scale.
The red and black curves denote ${\bar r}(E)$ and 
${\bar r}^{({\rm N})}(E)$
defined in (\ref{eq:T-spect-1}) and (\ref{eq:T-spect-2}), respectively.
The dotted green curve represents ${\bar r}^{({\rm +})}(E)$
when only $|{T}^{(+)}|^2$ is used.
}
\label{fig:muon-E-spec-log}
\end{figure}

The difference between the red and black curves 
in Figs.~\ref{fig:muon-spectrum} and \ref{fig:muon-E-spec-linear}
originates from the $\alpha$-$\mu$
Coulomb-force contribution $T^{({\rm +})}$
in Eq.~(\ref{eq:reaction-rate-Sec6}). 
The contribution from $T^{({\rm C})}$ is minor.
The effect of $T^{({\rm +})}$ is small 
at low energies but becomes relatively large
at high energies, which is seen in Fig.~\ref{fig:muon-E-spec-log} 
for the log scale,
in the dotted green curve derived based on only $|T^{(+)}|^2$.

\begin{table}[b!]  
\caption{Property of the energy spectrum of muon emitted from 
$(dt\mu)_{J=v=0} \to \alpha +n +\mu$ given by the present calculation,
${\bar r}(E)$ and ${\bar r}^{({\rm N})}(E)$,
and the adiabatic approximation ${\bar r}^{({\rm N})}_{\rm AD}(E)$
which gives no absolute value (cf. Fig.~\ref{fig:muon-E-spec-linear}).
}
\begin{center}
\begin{tabular}{lcccccc} 
\hline \hline
\noalign{\vskip 0.2 true cm} 
Muon energy spectrum   & $ \;$ &  Peak  &   &
 Average    &  & Peak   \\ 
\noalign{\vskip -0.1true cm} 
  &  &   energy &   &  energy   &  &  strength  \\ 
\noalign{\vskip 0.01true cm} 
  &  &  (keV) & $\;$  &  (keV)   &  & 
    $({\rm s}\cdot{\rm keV})^{-1}$  \\ 
\noalign{\vskip 0.1true cm} 
\hline
\noalign{\vskip 0.2true cm} 
Present, ${\bar r}(E)$    &  &    1.1 &   &  9.5   &  & $1.60 \times 10^{11}$     \\ 
\noalign{\vskip 0.1true cm} 
Present, ${\bar r}^{({\rm N})}(E)$    &  &    1.1 &  &  8.5   &  & $1.47 \times 10^{11}$     \\ 
\noalign{\vskip 0.1true cm} 
\noalign{\vskip 0.1true cm} 
  Adiabatic, ${\bar r}^{({\rm N})}_{\rm AD}(E)$  &  &  1.6 &   &  10.9 & &  $-$  $\;\;$   \\ 
\noalign{\vskip 0.2 true cm} 
\hline
\hline
\noalign{\vskip -0.3 true cm} 
\end{tabular}
\label{table:muon-spec}
\end{center}
\end{table}

As shown in Fig.~\ref{fig:muon-E-spec-linear} and 
Table~\ref{table:muon-spec}, the peak of 
the energy spectrum is located at
$E=1.1$ keV  both for ${\bar r}(E)$ and ${\bar r}^{\rm (N)}(E)$. 
Since the spectrum has a long high-energy tail,
the average energy is $9.5$ keV $(8.5 $ keV)
for ${\bar r}(E)\, ({\bar r}^{\rm (N)}(E))$.
Therefore, `muons with \mbox{1-keV} peak energy 
and 10-keV average energy' are emitted 
by the $dt\mu$ fusion. 
\mbox{This result} (more precisely, 
Figs.~\ref{fig:muon-spectrum} and \ref{fig:muon-E-spec-linear})
will be useful for the ongoing experimental project to
realize an ultra-slow  negative muon beam 
using the $\mu$CF~\cite{Nagamine1989,NagamineMCF199091,
Strasser1993,Strasser1996,Nagamine1996Hyp103,Natori2020}
(cf. Type II of Sec.~I). 

When the authors of Refs.\cite{Nagamine1989,NagamineMCF199091,
Nagamine1996Hyp103} 
proposed the solid \mbox{D-T} layer system that cools the 
incident muon beam by utilizing the $\mu$CF, 
they used the calculated muon energy spectrum in 
\mbox{Fig.~1 of Ref.~\cite{Mueller},} wherein
the spectrum was represented by a shape that was
set to unity at $E=0$.  
However, the definition of this energy spectrum 
is different from our energy spectrum 
${\bar r}(E) \,(=0$ at $E=0$) that is
represeted in {\it absolute} value and $\int_0^\infty \bar{r}(E) {\rm d}E$
gives the fusion rate $\lambda_{\rm f}=1.15 \times 10^{12} {\rm s}^{-1}$.
The role of the $\alpha$-$\mu$ Coulomb force,  which 
was discussed in Ref.~\cite{Mueller} 
by using their convoy-muon approximation,  
is properly included in our formulation
via the $T$-matrix $T^{(+)}$.

\begin{figure}[t!]
\begin{center}
\epsfig{file=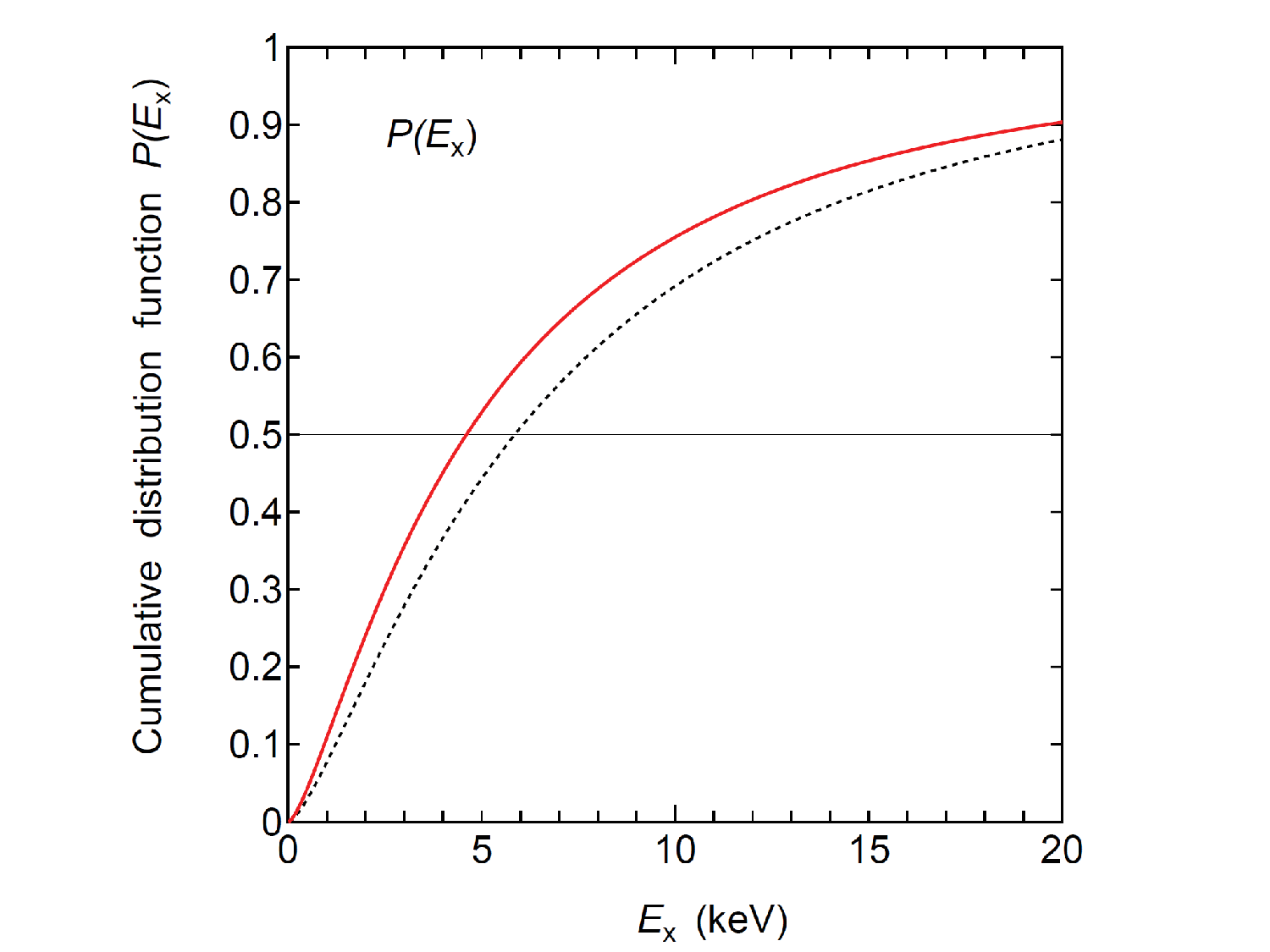,width=9.0cm,height=6.5cm}
\end{center}
\vskip -0.5cm
\caption{
Cumulative distribution function $P(E_{\rm x})$ 
associated with the muon energy spectrum ${\bar r}(E)$.
$P(E_{\rm x})$ is defined 
by Eq.~(\ref{eq:muon-prob}). 
The red curve is derived by integrating ${\bar r}(E)$
of Fig.~\ref{fig:muon-E-spec-linear}, whereas the dotted curve is 
given by using ${\bar r}^{({\rm N})}_{\rm AD}(E)$ instead of  
${\bar r}(E)$. 
}
\label{fig:muon-probability}
\end{figure}

For the sake of the observation of the muon energy spectrum,
we present 
a {\it cumulative distribution function}, 
$P(E_{\rm x})$, associated with the muon energy spectrum ${\bar r}(E)$,
defined by
\begin{eqnarray}
P(E_{\rm x}) = \frac{\int_0^{E_{\rm X}}  \, {\bar r}(E)\, {\rm d}E}
        {\int_0^\infty \, {\bar r}(E)\, {\rm d}E} \leq 1\:,
\label{eq:muon-prob}
\end{eqnarray}
which is illustrated by the red curve in Fig.~\ref{fig:muon-probability}. 
The dotted black curve is for $P(E_{\rm x})$ calculated with the adiabatic
approximation, namely using ${\bar r}^{({\rm N})}_{\rm AD}(E)$
instead of ${\bar r}(E)$ in Eq.~(\ref{eq:muon-prob}). Here,
absolute value of the energy spectrum is not concerned.

The red curve indicates
that 24 \% of the emitted muon is in the region
$0 < E < 2$ keV and 35\% is in $0 < E < 3$ keV, and hence
11\% is from $2 < E < 3$ keV, whereas
the muon having $0< E < 4.7$ keV is 50\% and
the one with $0< E < 10$ keV amounts to 75\%. 
The curve reaches 99\% when $E_{\rm x}=80$ keV.

\vskip 0.1cm
It is found that the `shape' of the two black curves for 
$r^{({\rm N})}(K)$ and ${\bar r}^{({\rm N})}(E)$ 
in Figs.~\ref{fig:muon-spectrum} and~\ref{fig:muon-E-spec-linear},
respectively,
are well simulated by simple functions as
\begin{eqnarray}
&&r^{\rm (N)}(K)  \propto \frac{K^2}{(1+ K^2 a^2)^4}, \quad 
\label{eq:rNK-simulate} \\
&&{\bar r}^{\rm (N)}(E)  \propto \frac{K}{(1+ K^2 a^2)^4}, 
\quad E=\hbar^2 K^2/2\mu_{{\rm R}_4} \qquad
\label{eq:rNE-simulate} 
\end{eqnarray}
with $a=154$ fm (this number appeared in Fig.~\ref{fig:psi-1s-1}).
The reason is as follows:
Using the property (\ref{eq:Phi(N)hat-separation}) 
of ${\widehat \Phi}_0^{({\rm N})}({\bf r}_3, {\bf R}_3)$,
we can represent the $T$-matrix (\ref{eq:Tmat-muon-emit}), 
without the spin part, as
\begin{eqnarray}
 \!\!\!\!\! T_{il,m}^{({\rm N})}\!=\! \langle  
               {\widetilde \phi}_{ilm}({\bf r}_4) | 
               V^{({\rm T})}({\bf r}_4,{\bf r}_3) 
               |\, \varphi^{({\rm N})}_0({\bf r}_3) 
               \rangle 
            \langle e^{ i {\bf {\widetilde K}_i} \cdot {\bf R}_4 }
              \,|\, \psi^{({\rm N})}_0({\bf R}_4)  \rangle .\; 
\label{eq:shape-T}
\end{eqnarray}  
Taking $E_N \approx 0.2 \, {\rm MeV} \ll Q$ and the relations 
(cf. Fig.~\ref{fig:risanka-muon-new-2})
\begin{equation}
  Q- E_N \leq   \varepsilon_i \leq Q \;\; \;
({\rm namely,} \; \varepsilon_i \approx Q,\: k_i \approx k_0) 
\end{equation}
together with  Eq.~(\ref{eq:conserv-muonsp}), 
we can derive 
\begin{eqnarray}
\mathit{\Delta} k_i \approx \frac{\mu_{r_4}}{\mu_{R_4}}
\frac{K_i}{k_0} \mathit{\Delta}\! K \propto{K_i}.
\end{eqnarray}
We then obtain, in the interaction region of 
$V^{({\rm T})}({\bf r}_4,{\bf r}_3)$, 
\begin{eqnarray}
\!\!\!\!\!\!\!
{\widetilde \phi}_{ilm}({\bf r}_4)\!\!\!\!\!&&=
\frac{1}{\sqrt{\mathit{\Delta} k_i}}
\int_{k_{i}}^{k_{i-1}}
\!\! \phi_{lm}(k,{\bf r}_4)\,{\rm d}k
  \approx \! \sqrt{\mathit{\Delta} k_i} \,\phi_{lm}(k_i,{\bf r}_4)  \nonumber\\
\!\!\!\!&&\approx \sqrt{\mathit{\Delta} k_i} \,\phi_{lm}(k_0,{\bf r}_4) 
\propto \! \sqrt{K_i} \,\phi_{lm}(k_0,{\bf r}_4).
\label{eq:propo-Ki}
\end{eqnarray}
Substituting this ${\widetilde \phi}_{ilm}({\bf r}_4)$  
into Eq.~(\ref{eq:shape-T}) and smoothing ${\bf K}_i$ and 
${\bf {\widetilde K}_i}$  
to ${\bf K}$, we finally obtain 
(note $v_{il} \propto {\widetilde K}_i$)
\begin{eqnarray}
r^{({\rm N})}(K) \propto  K^2 \!\! \int \!\! \,
 \big| \langle \,e^{ i {\bf K} \cdot {\bf R}_4 }
               \,| \,\psi^{({\rm N})}_0({\bf R}_4) 
               \rangle \big|^2 {\rm d} {\widehat {\bf K}} .
\label{eq:rNK-sim-integ}
\end{eqnarray}
As shown in Fig.~\ref{fig:psi-1s-1} (note ${\bf R}_4={\bf R}_3$), 
$\psi_0^{({\rm N})}(R_4)$
is well represented by $\propto e^{-R_4/a}$ with $a=154$ fm. 
Putting this function form
into Eq.~(\ref{eq:rNK-sim-integ}), 
we immediately obtain Eq.~(\ref{eq:rNK-simulate}),
from which we have Eq.~(\ref{eq:rNE-simulate})
with Eq.~(\ref{eq:T-spect-2}).
We found that both of the simulated functions well reproduce 
the corresponding black solid curves in Figs.~\ref{fig:muon-spectrum} 
and  \ref{fig:muon-E-spec-linear} within the width of the curves
under the normalization at the peaks. 

Finally, we discuss  the muon momentum and energy spectra
if we take the adiabatic approximation for the \mbox{$d$-$t$} 
relative motion just before  the fusion reaction occurs.
In this case, the wave function of the $(dt)$-$\mu$ relative motion
is simply given by $\propto e^{-R_4/a_0}$ with $a_0=131$ fm
(namely, the \mbox{$1s$ wave function} of the ${\rm He}\mu$ atom
as seen in Fig.~\ref{fig:psi-1s-1}), 
which has the mean kinetic energy of 10.9 keV.
Based on the preceding discussion, 
the `shape' of the muon momentum spectrum, $r^{({\rm N})}_{\rm AD}(K)$,
is given by Eq.~(\ref{eq:rNK-simulate}) with $a=a_0$;
similarly for the muon energy spectrum, ${\bar r}^{({\rm N})}_{\rm AD}(E)$,
given by Eq.~(\ref{eq:rNE-simulate}).
The spectra are illustrated in Figs.~\ref{fig:muon-spectrum} 
and  \ref{fig:muon-E-spec-linear} by the dotted curves that are
normalized  as explained in the figure captions. 
It should be noted that, in both figures, the peak of
the dotted curve has higher energy and broader width 
than the solid black curve (cf. Table~\ref{table:muon-spec}).

\section{Summary}

Recently, the study of $\mu$CF has regained
significant interest owing 
to several new developments and applications as explained 
in Introduction. In this regards, 
we have comprehensively studied the fusion reaction
$(dt\mu)_{J=v=0}  \to \alpha + \mu + n \,$ and $\; (\alpha \mu)_{nl} + n$,
by employing the $dt\mu$- and $\alpha n \mu$-channel coupled three-body model.
For the first time, we have solved  the coupled-channels 
Schr\"{o}dinger equation 
(\ref{eq:schr-eq}) under
the boundary condition whereby the muonic 
molecular bound state $(dt\mu)_{J=v=0}$
is the initial state and the outgoing wave in the $\alpha n \mu$ channel.
The total wave function (\ref{eq:total-wf})  
is composed of the three components  
$\mathring{\Psi}^{({\rm C})}(dt\mu)  
   + \Psi^{({\rm N})}(dt\mu)  
   + \Psi^{(+)}(\alpha n \mu)$. Here, 
$\mathring{\Psi}^{({\rm C})}(dt\mu)$ is the {\it given} function 
employed to describe nonadiabatically the 
$(dt\mu)_{J=v=0}$ 
state with only the Coulomb force, and is treated as the source term in the 
Schr\"{o}dinger equation. $\Psi^{({\rm N})}(dt\mu)$ is the additional
$dt\mu$ wave function required to correlate with the nuclear interactions.
$\Psi^{(+)}(\alpha n \mu)$ is the outgoing  wave function
of the $\alpha n \mu$ channel. 

We take the $d$-$t$ and $\alpha$-$n$ nuclear potentials
together with the  nonlocal {\it tensor} force  to couple 
the $S$-wave $d$-$t$ channel and the $D$-wave $\alpha$-$n$ channel.
They were then determined to reproduce the observed 
low-energy \mbox{$S$-factor} 
of the reaction \mbox{$d+t \to \alpha + n + $ 17.6 MeV}  
(Fig.~\ref{fig:cal-cc-dt-an}).
Use of the determined interactions simultaneously 
accounted for the $\alpha+n$ total cross section 
(Fig.~\ref{fig:sigma-an}). 
Applying the obtained total wave function 
to the $T$-matrix framework based on the Lippmann-Schwinger equation,
we have investigated the reaction rates  going to the individual 
$\alpha$-$\mu$ bound states and the continuum states together with
the $\alpha$-$\mu$ sticking probability.  We also studied
the momentum and energy spectra  of the muon emitted 
via the $\mu$CF.

The main conclusions are summarized as follows.

i) From the calculated $S$-matrix of the outgoing wave, we have derived
the fusion rate $\lambda_{\rm f}=1.03 \times 10^{12} {\rm s}^{-1}$.
This is consistent with the previously obtained values,
for example, by utilizing the \mbox{$d$-$t$} optical-potential 
model~\cite{Kamimura89,Bogdanova89}
and the $R$-matrix method~\cite{Struensee88a,Szalewicz90,
Hale93,Hu1994,Cohen1996,Jeziorski91}.
{
As the nuclear interactions employed in this work are
phenomenological ones, we examined three different interactions, Sets A, B, and C 
(Table~I).  
We have found that the calculated fusion rates $\lambda_{\rm f}$  \mbox{(Table~II)}
are independent of the details of the 
employed interactions that
reproduced the observed data in Figs.~\ref{fig:cal-cc-dt-an}
and \ref{fig:sigma-an}.  Set B is employed  for other 
calculations in this work.
}

ii) By performing the $T$-matrix calculation 
on the Jacobi-coordinate channel $c=5$ (Fig.~\ref{fig:jacobi-dtmu})
with the use of the  total wave function obtained in the above item i), 
we have calculated for the first time the {\it absolute} values of
the reaction (\ref{eq:mucf-reaction})
going to the $\alpha$-$\mu$ bound  and continuum states.
Using those values we obtain 
the fusion rate $\lambda_{\rm f}^{\rm bound}=6.90 \times 10^9 {\rm s}^{-1}$
to the $(\alpha \mu)_{\rm bound} + n$ states and
$\lambda_{\rm f}^{\rm cont.}=7.98 \times 10^{11} {\rm s}^{-1}$ to the 
$(\alpha \mu)_{\rm cont.} + n$ states, giving their sum as 
$\lambda_{\rm f}=8.05 \times 10^{11} {\rm s}^{-1}$. 
According to the {\it original} definition of sticking probability 
$\omega_S^0=\lambda_{\rm f}^{\rm bound}/
(\lambda_{\rm f}^{\rm cont.} + \lambda_{\rm f}^{\rm bound})$,
we obtain $\omega_S^0=0.857\%$.
This is smaller by \mbox{$\sim\!$ 7\%} than the literature result
$\omega_S^0 \simeq 0.91-0.93\%$ based on the sudden approximation
including the nuclear $d$-$t$ potential.
{Here, it is to be emphasized that 
we have much improved the sticking-probability calculation
by employing the $D$-wave $\alpha$-$n$ outgoing channel with the non-local 
tensor-force $dt$-$\alpha n$ coupling
and by deriving the probability
based on the aboslute values of 
the $\lambda_{\rm f}^{\rm bound}$ and $\lambda_{\rm f}^{\rm cont.}$.
}

iii) The value of $\omega_S^0=0.857\%$  corresponds, 
with the reactivation coefficient 
$R=0.35$~\cite{Struensee88b,Stodden90,Markushin,Rafelski},
to $\omega_S^{\rm eff}=(1-R)\, \omega_S^0=0.557\%$
which can explain the experimental 
data (Fig.~\ref{fig:effective sticking}).
For further progress on the study of $\omega_S^{\rm eff}$,
\mbox{development} in the calculation of $R$ is expected;  
our result on
the absolute values of the transition rates to the individual 
$\alpha$-$\mu$ bound  and continuum 
states (Fig.~\ref{fig:k-continuum-rate} and Table~IV) will be useful.

iv) In the  $T$-matrix calculation of 
$\lambda_{\rm f}^{\rm cont.}$,  $\lambda_{\rm f}^{\rm bound}$ 
and their sum $\lambda_{\rm f}$, 
we have found that $\Psi^{({\rm N})}(dt\mu)$ dominantly contributed 
to the fusion rates, whereas  $\mathring{\Psi}^{({\rm C})}(dt\mu)$ 
and $\Psi^{({\rm +})}(\alpha n \mu)$ play
a minor role (Table~\ref{table:fusion-rate-sticking}).
We then conclude that the calculation of
the initial sticking $\omega_S^0$ using
$\mathring{\Psi}^{({\rm C})}(dt\mu)$ only 
is not meaningful and that 
the statement ``the additional effect of the nuclear force
to the sticking probability" is not appropriate since 
$\Psi^{({\rm N})}(dt\mu)$ dominantly contributes to 
the fusion rate $\lambda_{\rm f}$.

v) We have performed another $T$-matrix calculation \mbox{to derive} 
{\it absolute} values for the momentum and energy spectra
of the muon emitted during the fusion process
\mbox{(Figs.~\ref{fig:muon-spectrum} and \ref{fig:muon-E-spec-linear}).}
The most important conclusion is that the `peak' energy
of the muon energy spectrum is 1.1 keV, 
whereas the `mean' energy is  9.5 keV (Table~\ref{table:muon-spec}) owing to  
the long higher-energy tail. 
This result will be useful to the new ongoing experimental project to 
realize an ultra-slow negative muon beam by utilizing the 
fusion reactions in the $dt\mu$ molecule as well as in the $dd\mu$ one,
and for a variety of applications e.g. a scanning negative muon microscope 
and an injection source for the muon collider.
The \mbox{$T$-matrix} calculation for the channel 
$c=4$ (Fig.~\ref{fig:jacobi-dtmu})
gives $\lambda_{\rm f}=1.15 \times 10^{12}\, {\rm s}^{-1}$.
{
We have examined the fusion rate $\lambda_{\rm f}$ 
and concluded that this value with the correction owing to the 
$\alpha$-$\mu$ Coulomb force is the final result in this study (sec. VI\,A).  
}

vi) As mentioned above,
we have reported three numbers for the fusion 
rate $\lambda_{\rm f}$ in items i), ii), and v), which 
are calculated using very different \mbox{methods.}
The values are consistent with each other but not {\it equal}. 
This is because the solution of the Schr\"{o}dinger 
equation (\ref{eq:schr-eq}) 
used in the $T$-matrix calculations is not  
\mbox{{\it exact}}. However, before this situation will  be improved,
we shall proceed, in the next coming paper, to the detailed study
of nuclear fusion  reactions in the $dd\mu$ molecule
because of its urgent importance, as indicated in v).

\vspace {-0.5cm}
\section*{Acknowledgements}
The authors would like to thank Prof. K. Nagamine
for his valuable discussions on 
the recent developments in the $\mu$CF experiments.
We are grateful to Prof. K. Ogata and Dr. T. Matsumoto
for their helpful discussions  on the nuclear reaction mechanisms.
We are also thankful to Dr. K. Ishida for helpful discussions on
the observation of the $\alpha$-$\mu$ sticking.
This work is supported by 
the Grant-in-Aid for Scientific Research on Innovative Areas, 
``Toward new frontiers: Encounter and synergy of state-of-the-art 
astronomical detectors and exotic quantum beams", 
JSPS KAKENHI Grant No.JP18H05461.
The computation was conducted on the ITO \mbox{supercomputer}
at Kyushu University.

\section*{{\bf Appendix}}

In Sec.~VI for the muon spectrum emitted by the $\mu$CF,
we employed the $T$-matrix (\ref{eq:Tmat-muon-emit}).
Here, we explain that it is not necessary to use the Coulomb wave function,
instead of the plane wave, in the bra-vector of 
the $T$-matrix elements. 
 
Taking the $\alpha$-$\mu$ Coulomb potential into account,  
we examined the  following $T$-matrix elements:

\begin{eqnarray}
&& \!\!\!\!\!\! {T}^{({\rm C})}_{iLl,M m m_s}\!\!  = \! \langle \, 
   f_{ilLM}({\widetilde K}_i, {\bf R}_4) \,{\widetilde \phi}_{ilm}({\bf r}_4) 
       \chi_{\frac{1}{2} m_s}\!(n)\,  
    |\,  V^{({\rm T})}_{\alpha n, dt} \,|\,   
 \mathring{\Psi}_{\frac{3}{2} M}^{({\rm C})}(dt\mu) \,\rangle ,\nonumber  \\ 
&& \!\!\!\!\!\! {T}^{({\rm N})}_{iLl,M m m_s}\!\!  = \! \langle \, 
   f_{ilLM}({\widetilde K}_i, {\bf R}_4)  \,{\widetilde \phi}_{ilm}({\bf r}_4) 
       \chi_{\frac{1}{2} m_s}\!(n)\,  
    |\,  V^{({\rm T})}_{\alpha n, dt} \,| \,  
  \Psi_{\frac{3}{2} M}^{({\rm N})}(dt\mu) \,\rangle , \nonumber \\ 
&& \!\!\!\!\!\! {T}^{(+)}_{iLl,M m m_s}\!\!  = \! \langle \, 
           f_{ilLM}({\widetilde K}_i, {\bf R}_4)  \, 
     {\widetilde \phi}_{ilm}({\bf r}_4) 
       \chi_{\frac{1}{2} m_s}\!(n)\,  
     |\,  V_{\alpha \mu}(R_5)- U_{il}(R_4) \,|  \nonumber \\  
 &&    \hskip 1.0cm
 \times \, \Psi_{\frac{3}{2} M}^{(+)}(\alpha n \mu) \,\rangle , \nonumber 
\label{eq:Tmat-muon-emit-coul}
\end{eqnarray}
where ${\widetilde \phi}_{ilm}({\bf r}_4)\, (i=1 - N)$ is the 
discretized $\alpha$-$n$ continuum 
state with energy $\widetilde{\varepsilon}_{i l}$,
and  
$f_{ilLM}({\widetilde K}_i, {\bf R}_4)$ is the associated 
\mbox{$(\alpha n)$-$\mu$} Coulomb wave function which
satisfies the energy conservation
\begin{eqnarray}
\quad  {\widetilde E}_i + {\widetilde \varepsilon}_{i}
=E_{00}+Q, \quad {\widetilde E}_i=\frac{\hbar^2}{2\mu_{R_4}}
  {\widetilde K}_i^2   \quad (i=1 - N).  \nonumber
\label{eq:conserv-muonsp2}
\end{eqnarray}
$ f_{ilLM}({\widetilde K}_i, {\bf R}_4)$ is obtained 
as the regular solution of 
\begin{eqnarray}
\big( T_{{\bf R}_4} + U_{il}(R_4) -{\widetilde E}_i \big)\,
f_{ilLM}({\widetilde K}_i)=0,  \nonumber
\label{eq:Coulomb-f}
\end{eqnarray}
where the potential $U_{il}(R_4)$ is derived by folding the $\alpha$-$\mu$
Coulomb potential into the density of the $i$th discretized state
${\widetilde \phi}_{ilm}({\bf r}_4)$ of the $\alpha$-$n$ momentum space. 
\begin{eqnarray}
U_{il}(R_4)= \langle \, {\widetilde \phi}_{ilm}({\bf r}_4) \, |\,
-\frac{2e^2}{r_5} \,|\, {\widetilde \phi}_{ilm}({\bf r}_4)\, .
\rangle_{{\bf r}_4}  \nonumber
\label{eq:folding-pot-R4}
\end{eqnarray}
To understand the behavior of 
$f_{ilLM}({\widetilde K}_i, {\bf R}_4)$ and $U_{il}(R_4)$ both in 
the asymptotic region and 
the muon's amplitude ($\psi_0^{({\rm N})}(R_{3,4})$) region 
in the total wave function $\Psi_{\frac{3}{2} M}^{(+)}(E)$,
we explain by using  Fig.~\ref{fig:risanka-muon-new-2},
the discretization of the $K$-space of the $(\alpha n)$-$\mu$ motion
and associated $k$-space in the $\alpha$-$n$ motion.

\begin{figure}[b!]
\begin{center}
\epsfig{file=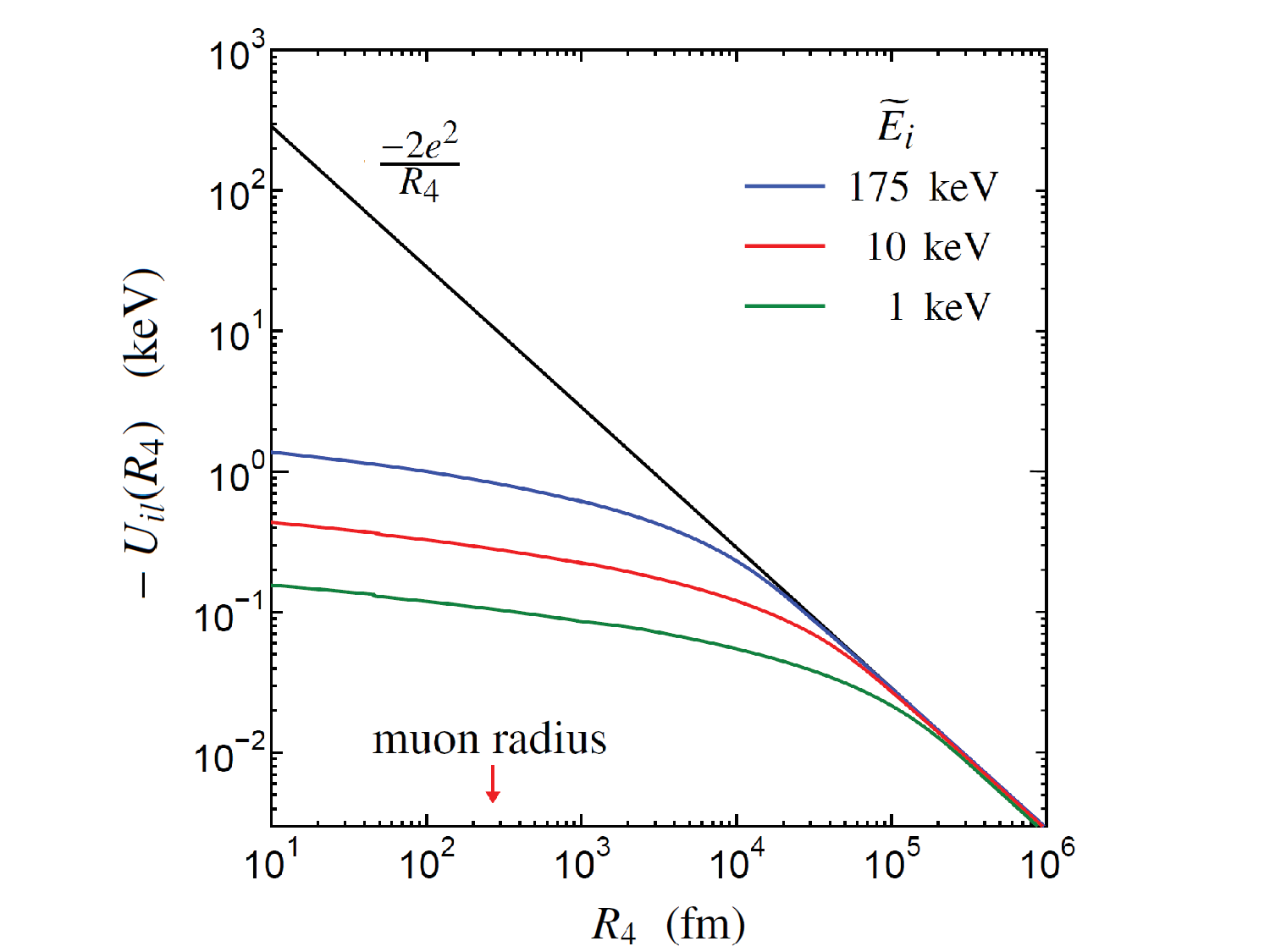,width=8.2cm,height=5.8cm}
\end{center}
\vskip -0.5cm
\caption{Folding potentials $U_{il}(R_4)$
used to determine the Coulomb
wave function $f_{ilLM}({\widetilde K}_i, {\bf R}_4)$ with the muon energy
${\widetilde E}_i$.
The green, red, and blue curves illustrate $U_{il}(R_4)$ 
for the cases of $l=2$ and $i=$ 16, 49, and 200, which correspond to
the muon energies ${\widetilde E}_i \simeq $ 1, 10, and 175 keV, respectively.
The black line denotes the pure Coulomb potential $-2e^2/R_4$.
}
\label{fig:folding-pot}
\end{figure}

Figure~\ref{fig:folding-pot} illustrates  
the folding potentials for the muon energies $E_i=$ 1, 10, and 175 keV
($i=$ 16, 49, and 200, respectively); note that 1 keV almost corresponds to the
the peak energy of the muon spectrum Fig.~\ref{fig:muon-E-spec-linear}) 
and 10 keV is almost the mean energy of the emitted muon. 
The potentials asymptotically converge to the pure Coulomb potential
$-2e^2/R_4$, but we note that they are very shallow in the 
region of muon amplitude $\psi_0^{({\rm N})}(R_{3,4})$  
($R_4 \lesssim  10^3$ fm) of the total wave function 
$\Psi_{\frac{3}{2} M}^{(+)}(E)$ which
appears as the {\it ket} wave function in the third member of the
$T$-matrix (\ref{eq:Tmat-muon-emit}).

If the discretization is made more precise by using larger $N$ values,
the attractive folding potentials become shallower. 
Therefore, in actual calculations 
in  Secs.~VI A and B, we can neglect 
$U_{il}(R_4)$ and replace $f_{ilLM}({\widetilde K}_i, {\bf R}_4)$ by the 
plane wave $e^{ i {\widetilde {\bf K}}_i \cdot {\bf R}_4 }$.
%


\end{document}